\documentclass[a4paper,titlepage,11pt]{article}
\pdfoutput=1
\usepackage[utf8]{inputenc}
\usepackage{geometry}
\usepackage[english]{babel}
\usepackage{amsmath, amsfonts, graphicx}
\usepackage{mathtools}
\usepackage{graphicx}
\usepackage{float}
\usepackage{pstricks}
\usepackage{hyperref}
\usepackage[labelfont=bf,font={small}]{caption}
\usepackage{pgfplots}
\usepackage{tikz}
\usepackage{axodraw4j}
\usetikzlibrary{positioning, arrows}
\usetikzlibrary{external}
\usepackage{footmisc}
\usepackage{multirow}
\usepackage{url}
\usepackage[breakable, theorems, skins]{tcolorbox}
\usepackage{enumerate}
\usepackage{physics}
\usepackage{bm}
\usepackage{epstopdf}

\numberwithin{equation}{section}

\def\nspc{{\hspace{-2pt}}}

  \newcommand{\sj}[6]{ \begin{Bmatrix}
   #1 & #2 & #3 \\
   #4 & #5 & #6 
  \end{Bmatrix}}
	\newcommand{\sltr}{\text{SL}(2,\mathbb{R})}
		\newcommand{\slnr}{\text{SL}{}^n(2,\mathbb{R})}
	\newcommand{\threej}[6]{\left(\mbox{\small$\!\begin{array}{ccc} #1 \! & \!\! #3 \! & \!\! #5 \nspc \\[-1mm]  #2 \!  & \!\! #4 \!  & \!\! #6 \nspc \end{array}\!$}\!\right)}
		\newcommand{\gphi}{g^{-1}\partial_\phi g}

\newcommand{\gtau}{g^{-1}\partial_\tau g}

\begin{document}

\begin{titlepage}

\setcounter{page}{1} \baselineskip=15.5pt \thispagestyle{empty}

\vfil

${}$
\vspace{1cm}

\begin{center}

\def\thefootnote{\fnsymbol{footnote}}
\begin{changemargin}{0.05cm}{0.05cm} 
\begin{center}
{\Large \bf Defects in Jackiw-Teitelboim Quantum Gravity}
\end{center} 
\end{changemargin}

~\\[1cm]
{Thomas G. Mertens${}^{\rm a}$\footnote{\href{mailto:thomas.mertens@ugent.be}{\protect\path{thomas.mertens@ugent.be}}} and Gustavo J. Turiaci${}^{\rm b}$\footnote{\href{mailto:turiaci@ucsb.edu}{\protect\path{turiaci@ucsb.edu}}}}
\\[0.3cm]
\vspace{0.7cm}
{\normalsize { \sl ${}^{\rm a}$Department of Physics and Astronomy,
\\[1.0mm]
Ghent University, Krijgslaan, 281-S9, 9000 Gent, Belgium}} \\[3mm]
{\normalsize { \sl ${}^{\rm b}$
Physics Department, University of California, Santa Barbara, CA, 93106 USA}}

\end{center}


 \vspace{0.2cm}
\begin{changemargin}{01cm}{1cm} 
{\small  \noindent 
\begin{center} 
\textbf{Abstract}
\end{center} }
We classify and study defects in 2d Jackiw-Teitelboim gravity. We show these are holographically described by a deformation of the Schwarzian theory where the reparametrization mode is integrated over different coadjoint orbits of the Virasoro group. We show that the quantization of each coadjoint orbit is connected to 2d Liouville CFT between branes with insertions of Verlinde loop operators. We also propose an interpretation for the exceptional orbits. We use this perspective to solve these deformations of the Schwarzian theory, computing their partition function and correlators. In the process, we define two geometric observables: the horizon area operator $\Phi_h$ and the geodesic length operator $L(\gamma)$. We show this procedure is structurally related to the deformation of the particle-on-a-group quantum mechanics by the addition of a chemical potential. As an example, we solve the low-energy theory of complex SYK with a U(1) symmetry and generalize to the non-abelian case. 

\end{changemargin}
 \vspace{1.3cm}
\vfil
\begin{flushleft}
\today
\end{flushleft}
\vspace{1cm}
\end{titlepage}

\newpage
\tableofcontents

\setcounter{footnote}{0}

\section{Introduction and Summary} 

There has been a recent explosion of interest in the so-called Sachdev-Ye-Kitaev models, see e.g. \cite{KitaevTalks,Sachdev:1992fk,Polchinski:2016xgd,Jevicki:2016bwu,Maldacena:2016hyu,Jevicki:2016ito,Cotler:2016fpe, Mandal:2017thl, wittenstanford,Turiaci:2017zwd,Gross:2017hcz,Gross:2017aos,Das:2017pif,Das:2017wae,Berkooz:2018qkz,gurari,Berkooz:2018jqr}. These 0+1d quantum models of all-to-all interacting fermions contain 1+1d gravitational features such as black holes and chaotic behavior. The low-energy dynamics of these models is described by a 1+1d model of dilaton gravity in AdS$_2$, the Jackiw-Teitelboim (JT) model \cite{JT}. Within this context, this model was picked up in \cite{Almheiri:2014cka,Jensen:2016pah, Maldacena:2016upp, Engelsoy:2016xyb} and studied intensively since then. With suitable boundary conditions, the model is described by a Schwarzian action:
\begin{equation}
\label{SSch}
S[f] = -C\int_0^\beta d\tau \, \left\{f,\tau\right\}, \qquad \left\{f,\tau \right\} \equiv \frac{f'''}{f'} - \frac{3}{2}\left(\frac{f''}{f'}\right)^2,
\end{equation}
where a prime denotes derivative with respect to Euclidean time $\tau$. The Schwarzian coupling constant $C$ has dimensions of length, and the semi-classical regime is reached by taking large $C/\beta$.\footnote{We will write most expressions for generic $C$, but take $C=1/2$ for some applications.}  

\noindent Structurally, the Schwarzian action can be viewed as a geometric action \cite{wittenstanford,Alekseev:1988ce, Alekseev:1990mp} and is deeply linked to Virasoro representation theory. For SYK purposes, one usually considers the vacuum orbit $\mathcal{M}_f = {\rm Diff}(S^1)/\sltr$ as integration manifold for the reparametrization $f$. It has recently become apparent that also other Virasoro orbits describe interesting deformed Schwarzian systems \cite{Blommaert:2018iqz, Anninos:2018svg, sss2, Maldacena:2019cbz}, see also \cite{Turiaci:2019nwa}. It is our goal here to systematically study and classify these deformations and in particular identify their JT gravitational content. 
\\~\\
JT gravity can be viewed as a constrained $\sltr$ BF gauge theory \cite{Jackiw:1992bw}, and as usual it is simpler to first understand the BF with compact gauge group story before adding the additional difficulties that gravity imposes. This will also be directly relevant for the complex SYK model, a generalization that uses complex instead of Majorana fermions \cite{Sachdev:2015efa}.	
\\~\\
Our main players in this work are the two dimensional BF model with compact gauge group $G$, described by the action:
\begin{equation}
\label{BFaction}
S_{\text{BF}}[\chi, A] = \int \text{Tr}\chi F + \frac{1}{2}\oint d\tau \text{Tr}\chi A_0,
\end{equation}
and the JT gravity model, given by:
\begin{equation}
\label{JTaction}
S_{\text{JT}}[\Phi,g] = \frac{1}{16\pi G}\left[\int d^2x \sqrt{g}\Phi(R+2) - 2\int d\tau \Phi K\right].
\end{equation}

\noindent Let us summarize our results. In previous work, diagrammatic rules were derived to structure the generic amplitude of bilocal operators within JT gravity \cite{Mertens:2017mtv} and BF theory \cite{Mertens:2018fds} on a disk-shaped region.\footnote{See also \cite{Lam:2018pvp,Blommaert:2018oro} and for a different approach see \cite{Kitaev:2018wpr, zhenbin}.} The rules are stated in those references and we will not repeat them here. As shown in these papers, for the compact group BF model, the regions of the disk are labeled by a discrete parameter, the irreducible representation $R$, whereas for the non-compact JT model, the region is labeled by a continuous parameter $k$ (from the 2d Liouville CFT perspective this is related to the Liouville momentum, while from a BF perspective this parameter labels continuous representations of $\sltr$). In each of these regions, the Hamiltonian is given by the Casimir of the representation. \\
In this work, we add a new rule to this dictionary by inserting a defect  (depicted by a cross in the diagram) into the bulk geometry. Some examples of diagrams with defect insertions are:
\begin{equation}
{\rm BF~theory:}~~~\begin{tikzpicture}[scale=0.8, baseline={([yshift=-0.13cm]current bounding box.center)}]
\draw[thick] (0,0) circle (1.5);
\draw[thick] (1.06,1.06) arc (300:240:2.16);
\draw[thick] (-1.06,-1.06) arc (120:60:2.16);
\draw[fill,black] (-1.06,-1.06) circle (0.1);
\draw[ultra thick] (-0.2,1.3)--(0.2,0.9);
\draw[ultra thick] (-0.2,0.9)--(0.2,1.3);
\draw[ultra thick] (0,1.4)--(0,1.6);
\draw[thick] (-1.5,0)--(1.5,0);
\draw (-0.6,1.1) node {\small \color{red} $R$};
\draw[fill,black] (1.06,-1.06) circle (0.1);
\draw[fill,black] (-1.06,1.06) circle (0.1);
\draw[fill,black] (1.06,1.06) circle (0.1);
\draw[fill,black] (1.5,0) circle (0.1);
\draw[fill,black] (-1.5,0) circle (0.1);
\draw[thick] (-1.4,0.55)--(0,-1.5);
\draw[fill,black] (-1.4,0.55) circle (0.1);
\draw[fill,black] (0,-1.5) circle (0.1);
\end{tikzpicture}
\label{Dcomp}
\qquad 
{\rm JT~gravity:}~~~\begin{tikzpicture}[scale=0.8, baseline={([yshift=0cm]current bounding box.center)}]
\draw[thick] (0,0) circle (1.5);
\draw[thick] (1.06,1.06) arc (300:240:2.16);
\draw[thick] (-1.06,-1.06) arc (120:60:2.16);
\draw[fill,black] (-1.06,-1.06) circle (0.1);
\draw[ultra thick] (-0.2,0.4)--(0.2,0);
\draw[ultra thick] (-0.2,0)--(0.2,0.4);
\draw[thick] (-1.4,0.539)--(1.4,0.539);
\draw[thick] (-1.4,-0.539)--(1.4,-0.539);
\draw[thick] (0,1.5)--(1.5,0);
\draw[fill,black] (0,1.5) circle (0.1);
\draw (-0.7,0) node {\small \color{red} $k$};
\draw[fill,black] (-1.4,-0.539) circle (0.1);
\draw[fill,black] (-1.4,0.539) circle (0.1);
\draw[fill,black] (1.4,-0.539) circle (0.1);
\draw[fill,black] (1.4,0.539) circle (0.1);
\draw[fill,black] (1.06,-1.06) circle (0.1);
\draw[fill,black] (-1.06,1.06) circle (0.1);
\draw[fill,black] (1.06,1.06) circle (0.1);
\draw[fill,black] (1.5,0) circle (0.1);
\end{tikzpicture}
\end{equation}

\noindent From the point of view of the BF theory with group $G$ bulk, these defects correspond to the insertion of a ``magnetic monopole'' implementing a non-trivial monodromy of the gauge field around it, labeled by the holonomy $z\in G$. We will refer to those as chemical potential defects, since from a boundary perspective they have the effect of turning on a chemical potential. \\
From the point of view of JT gravity, these defects correspond to inserting defects inside the bulk, either microscopic punctures (conical defects labeled by the angle $\theta$) or macroscopic tubes (wormholes labeled by the neck radius $\lambda$). In the Schwarzian theory perspective, they have the effect of changing the integration manifold from $ {\rm Diff}(S^1)/\sltr$ to other Virasoro orbits. 
\\~\\
For the case of BF theory with a compact group $G$, in this paper we will derive that the effect of these deformations by the defects is to insert the following terms inside the partition function or correlation functions
\begin{equation}
D_\mu(R) = \frac{\chi_R(z)}{\text{dim R}}, \qquad z = e^{i\beta \mu}, \quad \mu = \mu^a \lambda_a,
\end{equation}
where $\lambda_a$ ($a=1,\ldots, {\rm rank}$) are generators spanning the Cartan subalgebra of $G$, and $\mu^a$ the associated chemical potentials. $\chi_R(z)$ denotes the character of the group element $z$ evaluated in the representation $R$. 

 The representation $R$ of the insertion should be the one in the region of the disk where the Boltzmann factor $e^{-\beta H}$ is present \footnote{This guarantees that we are computing ${\rm Tr}[ e^{-\beta H - \mu Q} \ldots ]$, with the dots denoting operator insertions. Putting the defect in different regions would amount to compute a correlator with a different ordering ${\rm Tr}[e^{-\beta H} \ldots e^{- \mu Q} \ldots ]$.}. (For a reader unfamiliar with the diagrammatic rules, explicit formulas for the case of $SU(2)$ are given below in equations \eqref{sutwo} (partition function) and \eqref{2ptsu2} (two point function)). The U(1) case is particularly relevant in order to solve the low-energy description of complex SYK, with the parameter $\mu$ corresponding to the chemical potential. The non-abelian case generalizes this solution to other possible SYK-type models with non-abelian global symmetries.
\\~\\
For JT gravity, the presence of defects has the effect of changing the integration manifold from $ {\rm Diff}(S^1)/\sltr$ to other Virasoro orbits $\mathcal{M}_H= {\rm Diff}(S^1)/H$, with $H$ being the orbit stabilizer groups listed below. The action also changes depending on $H$ to
\begin{equation}\label{eqintroaction}
S_H[f]=-C \int_0^\beta d\tau \{ F \circ_H f(\tau), \tau\},
\end{equation}
where $F(\tau)$ is a new variable given in terms of $f(\tau)$ in an orbit dependent way as $F \circ_H f(\tau)$, with a specific functional relation given below. The variable $f$ is a reparametrization and therefore always has boundary conditions $f(\tau+\beta)=f(\tau)+\beta$. The path integral measure over $f$ is the natural one derived from the symplectic structure of $\mathcal{M}_H$ \cite{Alekseev:1988ce, Alekseev:1990mp}. The partition functions and correlators are the same as in \cite{Mertens:2017mtv} but with insertions implementing the presence of the bulk defects. For each $H$, the defect insertions are
\begin{itemize}
\item Elliptic $H=U(1)_\theta$
\begin{equation}\label{intro:ell1}
F\circ_\theta f = \tan \frac{\pi}{\beta} \theta f,~~~D_{\text{U(1)}_\theta}(k) = \frac{\cosh( 2 \pi \theta k)}{k\sinh(2 \pi k)}.
\end{equation}
\item Elliptic Exceptional $H=\slnr$
\begin{equation}
F\circ_n f = \tan \frac{\pi}{\beta} n f,~~~D_{\slnr}(k) = \frac{\sinh( 2 \pi n k)}{\sinh(2 \pi k)}.
\end{equation}
\item Hyperbolic $H=U(1)_\lambda$
\begin{equation}
F \circ_\lambda f = \tanh \frac{\pi}{\beta}\lambda f,~~~D_{\text{U(1)}_\lambda}(k) = \frac{\cos( 2 \pi \lambda k)}{k\sinh(2 \pi k)}.
\end{equation}
\item Parabolic $H=U(1)_0$ 
\begin{equation}
\label{intro:par1}
F\circ_0 f =  f,~~~D_{\text{U(1)}_0}(k) = \frac{1}{k\sinh(2 \pi k)}.
\end{equation}
\end{itemize} 
For example, the partition function for each deformation is $Z_H = \int d\mu(k) D_H(k) e^{-\beta \frac{k^2}{2C}}$, with the usual density of states $d\mu(k)=2k\sinh(2\pi k)$ and the orbit-dependent defects $D_H(k)$ given above. For $H=\slnr$ with $n=1$, $D_H(k)=1$ and one recovers the usual Schwarzian partition function. 

Even though the action \eqref{eqintroaction} looks the same in terms of the `uniformizing' coordinate $F(\tau)$ for all orbits, the monodromy of $F$ along the thermal circle depends on $H$ through the functions $F \circ_H f$ listed above. This makes it analogous to turning on a chemical potential for the compact group case, which can be achieved by adding a non-trivial holonomy around the thermal circle. 

From the 2D Liouville CFT perspective on the Schwarzian theory derived in \cite{Mertens:2017mtv}, we will see, for the orbits \eqref{intro:ell1} to \eqref{intro:par1}, how these deformations are related to the insertions of Verlinde loop operators. 
\\~\\
So far we ignored another class of coadjoint orbits, called special or exceptional:
\begin{itemize}
\item Special/Exceptional $H=\mathcal{T}_{n,\lambda}$
\begin{equation}
\label{spexc}
F \circ_{n,\lambda} f = e^{\lambda f}\left(\tan\frac{nf}{2} + \frac{\lambda}{2n}\right),~~~D_{n,\lambda}(k) = \frac{\sinh(2\pi nk)}{\sinh(2\pi k)}\frac{\sinh(2\pi \lambda k)}{k}.
\end{equation}
\end{itemize}
The precise interpretation of the special/exceptional orbits in \eqref{spexc} is more complicated and, exploiting the connection with 2D CFT, we propose that they are again related to a certain type of Verlinde loops (see section \ref{sec:excep} for details). We then use this connection to solve the theory by finding the corresponding defect insertion. 
\\~\\
Finally, as an aside, we will define an operator in section \ref{sect:areaop}, which we will show has the effect of adding the following defect insertion
\begin{equation}
D_A(k) = 2\pi k \coth 2\pi k -1.
\end{equation}
We will see the expectation value of this operator gives the horizon area. 
\\~\\
This work is structured as follows. Section \ref{sec:compact} discusses the low-energy action of complex SYK and the solution for its correlation functions. We generalize to the non-abelian case and relate the chemical potential to a defect insertion in the diagrammatic language, as in \eqref{Dcomp}. Defects in JT gravity are studied afterwards, and can be read independently. First a general discussion on the relation between defects and orbits is given in Section \ref{sec:orbits}. Sections \ref{sect:ell}, \ref{sect:hyper} and \ref{sect:para} then focus on the specific types of defect one can study (elliptic, hyperbolic and parabolic). Next to deriving the above formulas, we will provide a geometric interpretation of these, both as vertical Wilson loops in Chern-Simons theory, and as Verlinde loops in the ZZ-ZZ system related to Schwarzian QM.
A short discussion is made on exceptional orbits in section \ref{sec:excep} and some prospects for future work are given in section \ref{sec:concl}. Supplementary material on orbits in the compact case, their equations of motion, and degenerate bilocal operators are given in the appendices.

\section{Defects in BF: Complex SYK and Global Symmetries}\label{sec:compact}
In this section we will review and give some details regarding the low-energy modes of an SYK-like QM theory with global symmetries. The simplest example is that of complex SYK theory with a U(1) symmetry \cite{Davison:2016ngz,Fu:2016vas,Peng:2017spg,Bulycheva:2017uqj}, although we will also extend it to non-abelian symmetries. The low-energy system coming from a symmetry $G$ is equivalent to the QM theory of a particle on the group $G$. We will review the solution of this theory and show how turning on a chemical potential is equivalent to inserting a defect in the bulk two-dimensional description. 

\subsection{Low Energy Action}
In the IR, systems like the complex SYK model have an emergent $\sltr \times$ U(1) symmetry. In the Majorana SYK model, the low energy dynamics is governed by a reparametrization mode $f\in {\rm Diff}(S^1)$. Due to the U(1) global symmetry an additional mode appears in the complex SYK case, related to an emergent gauge transformation $e^{iq\Lambda} \in U(1)$. The field $\Lambda$ is periodic, with a size $2\pi/q$ fixed by the fundamental fermion charge $q$.
\\~\\
According to \cite{Davison:2016ngz}, the effective action of this coupled set of pseudo-Goldstone modes is given by 
\begin{equation}
\label{lecsyk}
S[f,\Lambda] = -C \int_{0}^{\beta}d\tau \left\{\tan \frac{\pi}{\beta} f(\tau), \tau \right\} -\frac{K}{2} \int_{0}^{\beta}d\tau \left(\Lambda'(\tau) - i \mu f'(\tau)\right)^2,
\end{equation}
with $\mu = -\frac{2\pi \mathcal{E}}{\beta}$.\footnote{The total chemical potential for complex SYK is given by $\mu_0 + \mu$, where $\mu_0$ is a zero temperature chemical potential \cite{Davison:2016ngz}. Just as $E_0$ or $S_0$, this part is not captured by these Schwarzian-like actions and only deviations from extremality are accounted for.} The coupling constant $K$ and $C$ correspond respectively to compressibility and heat capacity of the underlying QM system such as the complex SYK. Both of them are of order $K,C\sim N/J$ in the complex SYK system \cite{Davison:2016ngz}.  A shift of the fundamental charge from $q\to q'$ is equivalent to a shift of $K \to K (q/q')^2$ and $\mu \to (q'/q) \mu$. Therefore from now on we will fix $q=1$, making $\Lambda \sim \Lambda + 2 \pi $. The dimensionful parameters are $C$, $K$, $\mu$ and $\beta$. By a choice of units we can fix any one of these four parameters, leaving three dimensionless parameters $2 \pi C/\beta$ , $2 \pi K/\beta$ and $\mathcal{E}=- \mu\beta/2\pi$. 
\\~\\
Besides the U(1)$_{\rm global}$ shift invariance, $\Lambda \to \Lambda + \alpha$, the action \eqref{lecsyk} is left invariant by the following combination of $\sltr$ and U(1) transformations, mapping a configuration $(f,\Lambda)$ into a new one $(f_n,\Lambda_n)$ by:
\begin{align}
\tan \frac{\pi}{\beta} f  = \frac{a \tan\frac{\pi}{\beta}f_n + b}{c \tan\frac{\pi}{\beta} f_n + d}, \quad \Lambda = \Lambda_n + i \mu (f - f_n), \quad ad-bc=1.
\end{align}
To produce a well-defined path integral, we need to treat these symmetries as gauge symmetries. This means the path integral is over $({\rm Diff}(S^1) \times L\text{U(1)})/ (\sltr \times \text{U(1)}_{\rm global})$. This theory is further studied in \cite{Sachdev,Liu:2019niv} and was analyzed from a warped CFT perspective in \cite{Chaturvedi:2018uov}. In the rest of this section we will point out that it can be solved exactly by combining ingredients that have separately been solved before \cite{Mertens:2017mtv,Mertens:2018fds}.  

\subsection{Partition Function}  
 The action \eqref{lecsyk} couples the $\sltr$ with the U(1) mode. Defining $\tilde{\Lambda}(\tau) = \Lambda(\tau) - i \mu f(\tau)$, we can decouple both sectors at the expense of a twisted periodicity for $\tilde{\Lambda}$:
\begin{equation}
\tilde{\Lambda}(\beta) = \tilde{\Lambda}(0) - i \mu \beta+2 \pi m,~~m\in\mathbb{Z}.
\end{equation}
 Since the gauge mode is periodic, we should really think of this as a group element $g \equiv e^{i\tilde{\Lambda}}$ with twisted boundary condition along the thermal circle $e^{i \tilde{\Lambda}} \to e^{i \tilde{\Lambda}} e^{\mu \beta}$. From \eqref{lecsyk}, the action for $\tilde{\Lambda}$ becomes a free scalar with this boundary condition.
\\~\\
Since the two modes are decoupled we can compute the partition function of \eqref{lecsyk} as a product of each contribution separately. Since it is a free theory, the partition function of the boson $\tilde{\Lambda}$ with the periodicity above can be computed as a sum over  classical solutions with winding\footnote{We use $\theta_3(\tau,z)\equiv \sum q^{n^2/2} \eta^n$, with $q=e^{2\pi i \tau}$ and $\eta=e^{2 \pi i z}$.} \footnote{The path integral is done over the loop group $LU(1)$ modulo a global $U(1)$ thus eliminating the zero mode of the field. This introduces an ambiguity in the overall factor of the partition function that we will ignore since it can be absorbed anyways by $S_0$ in equation \eqref{pfschw} below, after the Schwarzian mode is included. }
\begin{eqnarray}\label{eq:partu1}
Z_{\text{U(1)}}&=& \int [\mathcal{D}\tilde{\Lambda}]e^{-\frac{K}{2} \int d\tau \dot{\tilde{\Lambda}}^2} =\sqrt{\frac{2\pi K}{ \beta}} e^{\frac{K\mu^2 \beta}{2}} \sum_m  e^{-\frac{2 \pi^2 K}{\beta }  m^2 }e^{-i\frac{2 \pi^2 K}{\beta  } \frac{\mu\beta}{\pi}   m  }\\
\label{theta3}
&=&\sqrt{\frac{2\pi K}{\beta}} e^{\frac{K\mu^2 \beta}{2}} ~\theta_3\left(i\frac{2 \pi K}{\beta } , -\frac{2\pi K}{\beta }  \frac{\mu\beta}{2\pi}  \right) \\
&=& ~\theta_3\left(i\frac{\beta }{2 \pi K} , i  \frac{\mu\beta}{2\pi}  \right) = \sum_n e^{- \beta \left( \frac{n^2}{2K} - \mu n \right) }. 
\label{sumcharge}
\end{eqnarray}
In the first line we compute the fixed winding $m$ partition function, which is straightforward since the theory is free, and then sum over $m$. In the second line we use the definition of the theta function, while in the third line we use the modular property of the theta function to write it as a sum over a Hilbert space. The integers $m$ and $n$ can be interpreted as winding and charge respectively, the first line coming from a direct path integral evaluation, and the last line from a Hamiltonian treatment of a non-relativistic particle on $S^1$. Using the definition of the theta function in the last line gives a Hilbert space interpretation $Z_{\text{U(1)}} = {\rm Tr}_{\text{U(1)}} ~e^{-\beta (H - \mu Q)}$, with $H=Q^2/2K$ where we sum over integer charges $Q=n$. The first term is like a capacitor potential and the second the chemical potential part of the grand canonical potential.  \\
The winding representation (in $m$) is better suited for a semiclassical large $K$ limit which gives $\log Z_{\text{U(1)}} \sim K \mu^2 \beta/2$. On the other hand, the charge representation (in $n$) directly gives the small $K$ limit $Z_{\text{U(1)}}\sim1$. 
\\~\\
The partition function of the reparametrization mode $f$ can be separately solved exactly \cite{wittenstanford, Mertens:2017mtv}. The result is
\begin{equation}
\label{pfschw}
Z_{\sltr} = e^{S_0} \left( \frac{2\pi C}{\beta} \right)^{3/2}  e^{\frac{2 \pi^2 C}{\beta}},
\end{equation}
where $S_0$ is a zero-temperature entropy which is not determined by the IR theory. The full partition function is given by combining \eqref{pfschw} and \eqref{theta3}:
\begin{eqnarray}
Z&=&Z_{\sltr}Z_{\text{U(1)}} \nonumber\\
&=&e^{S_0} \left( \frac{2\pi C}{\beta} \right)^{3/2} \left(\frac{K}{2\pi \beta q^2}\right)^{1/2} e^{\frac{2 \pi^2 C}{\beta}+\frac{K \mu^2 \beta}{2} }\theta_3\left(i\frac{2 \pi K}{\beta q^2} , -\frac{2\pi K}{\beta q^2 }  \frac{q \mu\beta}{2\pi}  \right) ,
\end{eqnarray}
where we include the dependence on fundamental charge $q$. It is an interesting problem studied in \cite{Sachdev} to find the density of states as a function of energy and charge from this expression. 
\\~\\
Even though it does not appear relevant for SYK applications, we can wonder what happens if the size of the U(1) group goes to infinity. Then winding ($m\neq 0$) contributions in \eqref{eq:partu1} become suppressed and the sum over charge becomes continuous. This gives 
\begin{equation}
Z_{\text{U(1)}\to \mathbb{R}} = \left(\frac{2\pi K}{ \beta}\right)^{1/2} e^{\frac{K\mu^2 \beta}{2}} = \int dq~ e^{-\beta ( \frac{q^2}{2K} - \mu q)}.
\end{equation}

\subsection{Correlators}
In this section we will analyze and compute the observables of the low energy theory \eqref{lecsyk}. We will follow the same procedure as in SYK: find the IR solution to the Schwinger-Dyson equations with the emergent enhanced symmetries, and finally integrate over the broken transformations with the action presented in the previous section. This generates $\sltr \times$ U(1) invariant observables that are natural to study from a purely low energy perspective, regardless of their SYK theory origin. 
\\~\\
The IR saddle solution to the complex SYK model analyzed in \cite{Davison:2016ngz} is given by
\begin{equation}
G(\tau_1,\tau_2) = e^{- \frac{2\pi \mathcal{E}}{\beta}\tau_{12}}\left(\frac{1}{\frac{\beta}{\pi} \sin \frac{\pi}{\beta}\left|\tau_{12}\right|}\right)^{2\ell}[ \Theta(\tau_{12}) +e^{-2\pi \mathcal{E}}\Theta(-\tau_{12}) ], 
\end{equation}
where $\tau_{12}=\tau_1-\tau_2$. The asymmetry under time reversal is due to the presence of a chemical potential. To simplify the discussion we will focus on $0<\tau_{12}<\beta$. Therefore we can focus on the building block of the correlator as our observable
\begin{equation}
G(\tau_1,\tau_2) = e^{- \frac{2\pi \mathcal{E}}{\beta}\tau_{12}}\left(\frac{1}{\frac{\beta}{\pi} \sin \frac{\pi}{\beta}\left|\tau_{12}\right|}\right)^{2\ell}.
\end{equation}
 Transforming this to an arbitrary solution with $f(\tau)$ and $\Lambda(\tau)$, we write\footnote{This is allowed since $\Theta(\tau_{1}-\tau_{2}) = \Theta( f(\tau_1)-f(\tau_2))$ for any $f\in {\rm Diff}(S^1)$ so it does not play any role in the path integral over Schwarzian mode.}
\begin{equation}
\label{bilocals}
G'(\tau_1,\tau_2) = e^{- \frac{2\pi \mathcal{E}}{\beta}(f(\tau_1)-f(\tau_2))}e^{i(\Lambda(\tau_1)-\Lambda(\tau_2))}\left(\frac{\sqrt{f'(\tau_1) f'(\tau_2)}}{\frac{\beta}{\pi} \sin \frac{\pi}{\beta}\left|f(\tau_1)- f(\tau_2)\right|}\right)^{2\ell},
\end{equation}
and we can compute $\langle G(\tau_1,\tau_2) \rangle = \int [\mathcal{D}f][\mathcal{D}\Lambda]~e^{-S[f,\Lambda]} G'(\tau_1,\tau_2)$ with action \eqref{lecsyk}. 
 Setting $\tilde{\Lambda}(\tau) = \Lambda(\tau) - i \mu f(\tau)$, we can decouple not only the action, but also the two point function
\begin{equation}\label{obsredl}
G'(\tau_1,\tau_2) = e^{i(\tilde{\Lambda}(\tau_1)-\tilde{\Lambda}(\tau_2))}\left(\frac{\sqrt{f'(\tau_1) f'(\tau_2)}}{\frac{\beta}{\pi} \sin \frac{\pi}{\beta}\left|f(\tau_1)- f(\tau_2)\right|}\right)^{2\ell},
\end{equation}
and the coupling between the Schwarzian and U(1) sectors is gone. We can generalize this observable to a similar one with U(1) charge $Q$, an integer in units of the fundamental charge $q$. 
\\~\\
The $\sltr$ part of the correlator was solved in \cite{Mertens:2017mtv} 
\begin{eqnarray}
\label{sch2pt}
\left\langle \left(\frac{\sqrt{f'_1 f'_2}}{\frac{\beta}{\pi} \sin \frac{\pi}{\beta}\left|f_1- f_2\right|}\right)^{2\ell} \right\rangle = \int d\mu(k_1) d\mu(k_2) e^{- \frac{k_1^2}{2C} \tau - \frac{k_2^2}{2C}(\beta-\tau)} \frac{\Gamma(\ell \pm i k_1 \pm i k_2)}{\Gamma(2\ell)},
\end{eqnarray}
where $d\mu(k)=2 k \sinh 2 \pi k$ is the Schwarzian density of states, and we define $\tau=|\tau_{12}|$. The integral over $k$ can be thought of as an integral over $\sltr$ continous series representations. 
\\~\\
Now we can focus on the new part, the U(1) factor. We will call $g(\tau)=e^{i Q \Lambda(\tau)}$ and we will compute $\langle g(\tau_1) g^{-1} (\tau_2)\rangle$. Since the theory is free, the two-point bilocal correlator can be easily computed for the zero-winding sector  
\begin{equation}
\langle g(\tau_1) g^{-1} (\tau_2)\rangle_{m=0} = \int [\mathcal{D}\tilde{\Lambda}]e^{i Q (\tilde{\Lambda}_1-\tilde{\Lambda}_2)}e^{- \frac{K}{2} \int d\tau \dot{\tilde{\Lambda}}^2} = \left(\frac{2\pi K}{\beta}\right)^{1/2}e^{\frac{K \beta \mu^2}{2}} e^{Q \mu \tau} e^{- \frac{Q^2 \tau(\beta-\tau)}{2 K \beta} },
\end{equation}
where the first term is the one-loop contribution and the rest comes from the classical solution (that includes the charge $Q$ backreaction). Since the theory is free, this is exact. Including the sum over windings gives  
\begin{eqnarray}
\langle g(\tau_1) g^{-1} (\tau_2)\rangle &=&\left(\frac{2\pi K}{\beta}\right)^{1/2}e^{- \frac{Q^2 \tau(\beta-\tau)}{2 K \beta} } e^{Q \mu  \tau} e^{\frac{K\mu^2 \beta}{2}}  \sum_m e^{- i ( 2 \pi m)Q \frac{\tau}{\beta}} e^{i (\pi i) \frac{2 \pi K}{\beta}m^2}e^{- K\frac{(i \mu\beta 2 \pi m) }{\beta}}\nonumber\\
&=&\left(\frac{2\pi K}{\beta}\right)^{1/2}e^{- \frac{Q^2 \tau(\beta-\tau)}{2 K \beta} } e^{Q \mu  \tau} e^{\frac{K\mu^2 \beta}{2}}  \theta_3 \left(i \frac{2 \pi K}{\beta}, - \frac{2 \pi K}{\beta} \frac{\mu\beta}{2\pi} - Q\frac{\tau}{\beta}\right).
\end{eqnarray}
 Including the partition function normalization \eqref{eq:partu1}, gives 
 \begin{equation}
\langle g(\tau_1) g^{-1} (\tau_2)\rangle = e^{- \frac{Q^2 \tau(\beta-\tau)}{2 K \beta} } e^{Q \mu  \tau} \frac{ \theta_3 \left(i \frac{2 \pi K}{\beta}, - \frac{2 \pi K}{\beta} \frac{\mu\beta}{2\pi} - Q\frac{\tau}{\beta}\right)}{ \theta_3 \left(i \frac{2 \pi K}{\beta}, - \frac{2 \pi K}{\beta} \frac{\mu\beta}{2\pi} \right)}.
\end{equation} 
We can see here how large $K$ fixes $m=0$, to exponential accuracy. Moreover, since in this limit $e^{- \frac{Q^2 \tau(\beta-\tau)}{2 K \beta} } \to 1$, the full correlator is given by the semiclassical answer $\langle g(\tau_1) g^{-1} (\tau_2)\rangle \approx e^{Q \mu \tau}$ as expected.
\\~\\
Using the modular property of the theta function we can write this in the charge representation as 
\begin{equation}
\langle g(\tau_1) g^{-1} (\tau_2) \rangle = \frac{1}{Z_{\text{U(1)}}} e^{-\frac{Q^2\tau}{2 K} } \theta_3 \left(i \frac{\beta}{2 \pi K}, \frac{i}{2\pi} (\mu\beta + \frac{Q \tau}{K} )\right),
\end{equation}
which adding the normalization gives 
\begin{equation}
\label{eqtoref}
\langle g(\tau_1) g^{-1} (\tau_2) \rangle =e^{-\frac{Q^2\tau}{2 K} } \frac{  \theta_3 \left(i \frac{\beta}{2 \pi K}, \frac{i}{2\pi} (\mu\beta + \frac{Q \tau}{ K} )\right)}{ \theta_3 \left(i \frac{\beta}{2 \pi K}, \frac{i}{2\pi}\mu\beta \right)}.
\end{equation}
If we send $K\to 0$ then $\langle g_1 g_2^{-1}\rangle\sim e^{-\frac{Q^2\tau}{2 K} } $. By dimensional analysis, this should be the same as taking large $\tau$, and so the two-point function goes from $e^{Q \mu \tau}$ at small times to $e^{-\frac{Q^2\tau}{2 K} }$ for long times. \\
Using the definition of the theta function, we can write \eqref{eqtoref} as a sum over charges 
\begin{equation}\label{eq:u12ptc}
\langle g(\tau_1) g^{-1}(\tau_2)\rangle =\sum_{n_1,n_2} ~e^{- \frac{n_1^2}{2K} \tau}e^{ - \frac{n_2^2}{2K} (\beta-\tau)}~e^{-\mu \beta n_2} ~\delta_{n_1,n_2+Q}.
\end{equation}
We can interpret the factor of $\delta_{n_1,n_2+Q}$ as a 3j-symbol of U(1) between representations labeled by charges $n_1$, $n_2$ and $Q$, and $e^{- \mu \beta n}$ can be interpreted as a representation $n$ character. This structure will appear in the non-abelian case as well. 
\\~\\
Again, we can take the size of U(1) to infinity, effectively becoming $\mathbb{R}$. Then the correlator is the same, but with a continuous charge 
\begin{equation}
\langle g_1 g^{-1}_2\rangle =\int dq ~e^{-\frac{q^2}{2K}\tau}e^{-\frac{(q-Q)^2}{2K}(\beta- \tau)}e^{\mu \beta (Q-q)}=\left(\frac{2\pi K}{\beta}\right)^{1/2}e^{\frac{K \beta \mu^2}{2}} e^{Q \mu \tau} e^{- \frac{Q^2 \tau(\beta-\tau)}{2 K \beta} } .
\end{equation}

\noindent To summarize, full correlation functions hence factorize in these U(1) pieces times the Schwarzian correlator determined in \cite{Mertens:2017mtv}. This then fully solves the low-energy dynamics of complex SYK, and explicitly the theory defined by \eqref{lecsyk}.
\\~\\
The generalization to an $n$-point function for the $\sltr$ part is straightforward as shown in \cite{Mertens:2017mtv}. A similar thing happens for the U(1) piece, a general correlator factorizes when written in the representation as in \eqref{eq:u12ptc}. \\
On the other hand, to compute out-of-time ordered correlators (OTOC), we need to insert 6j-symbols. For the U(1) sector, these are trivially the product of four Kronecker-delta's:\footnote{Generically, most of these delta's are also present as Clebsch-Gordan coefficients.}
\begin{equation}
\sj{Q_A}{n_1}{n_4}{Q_B}{n_3}{n_2}_{\text{U(1)}} = \delta_{n_1,n_2+Q_B}\delta_{n_2,n_3+Q_A}\delta_{n_3,n_4-Q_B}\delta_{n_4,n_1-Q_A},
\end{equation}
and we obtain for e.g. the U(1) OTOC four-point correlator:
\begin{align}&{}
\begin{tikzpicture}[scale=0.65, baseline={([yshift=0cm]current bounding box.center)}]
\draw (-1.5,-1.3) node {\small $\tau_2$};
\draw (-1.5,1.3) node {\small $\tau_3$};
\draw[thick] (-1.05,1.05) -- (-.15,.15);
\draw[thick] (.15,-.15) -- (1.05,-1.05);
\draw[thick] (-1.05,-1.05) -- (1.05,1.05);
\draw[thick] (0,0) circle (1.5);
\draw[fill,black] (-1.05,-1.05) circle (0.1);
\draw[fill,black] (1.05,-1.05) circle (0.1);
\draw[fill,black] (-1.05,1.05) circle (0.1);
\draw[fill,black] (1.05,1.05) circle (0.1);
\draw (-.8,.35) node {\footnotesize $Q_B$};
\draw (.88,.35) node {\footnotesize $Q_A$};
\draw (1.5,-1.3) node {\small $\tau_4$};
\draw (1.5,1.3) node {\small $\tau_1$};
\draw[ultra thick] (0.9,0.1)--(1.1,-0.1);
\draw[ultra thick] (1.1,0.1)--(0.9,-0.1);
\draw[ultra thick] (1.4,0)--(1.6,0);
\end{tikzpicture}
= \text{Tr} \left[g(\tau_1) g(\tau_3) g^{-1}(\tau_2)g^{-1}(\tau_4) e^{-\beta (H-\mu Q)} \right] \nonumber \\
 &= \sum_{n_1,n_2,n_3,n_4 \in \mathbb{Z}}\hspace{-0.5cm}\delta_{n_1,n_2+Q_B}\delta_{n_2,n_3+Q_A}\delta_{n_3,n_4-Q_B}\delta_{n_4,n_1-Q_A} e^{-\frac{n_1^2}{2K}\tau_{31}}e^{-\frac{n_2^2}{2K}\tau_{32}}e^{-\frac{n_3^2}{2K}\tau_{42}}e^{-\frac{n_4^2}{2K}(\beta-\tau_{41})} e^{-\mu \beta n_4}.
\label{OTOC}
\end{align}
For any diagram, one can always reduce the expression to a single sum and therefore the correlators are still given by some theta function. 

This expression \eqref{OTOC} (and all correlators above) should be multiplied by the Schwarzian four-point OTOC (written in e.g. eq. (5.3) in \cite{Mertens:2017mtv}).
\\~\\
Finally, one of the applications of the exact correlators is to extract the long time limit of correlators after quantum effects become important \cite{Bagrets:2016cdf}. For simplicity, take the zero-temperature limit $\beta^{-1}\to 0$. At short times the two-point function of two fields with dimension $\ell$ and charge $Q$, behaves as $G(\tau)\sim\tau^{-2 \ell} e^{Q \mu \tau}$. On the other hand, for times larger than the scale set by $C$ and $K$, i.e. $\tau\gg C,K$, from the analysis below equation \ref{eqtoref}, we get a long time behavior $G(\tau)\sim \tau^{-3/2} e^{-Q^2 \tau/2K}$. The universal $3/2$ power law comes from the Schwarzian sector \cite{Bagrets:2016cdf,Mertens:2017mtv}, while the exponential decay comes from a Coulomb blockade-like effect with energy gap $Q^2/2K$ \cite{Bag, Sachdev}. This behavior cannot be derived from a semiclassical analysis.   

\subsection{Holography} 
The derivation of the low energy effective action \eqref{lecsyk} studied in this section from the AdS$_2$ bulk is done in appendix C of \cite{Moitra:2018jqs}, see also \cite{Gaikwad:2018dfc,Sachdev:2019bjn}. \\
 In a holographic setting, where the global symmetry corresponds to a bulk gauge field $A_\mu$, this situtation corresponds to adding a chemical potential as the time-component of $A_\mu$ in the black hole $f$-coordinates:
\begin{equation}
A_f f' = A_t, \quad A_f = \mu.
\end{equation}
This constant chemical potential is seen as varying in the $t$-frame of the wiggly boundary curve. The choice of in which coordinate frame to fix $\mu$ is correlated with the form of the bilocals \eqref{bilocals}.\footnote{Fixing $A_t$ instead leads to the same partition function. The natural bilocal operators to consider would not have the $\sim \mathcal{E}$ piece, and the resulting computation is identical. The final formulas we obtain are also valid for these bilocals.} The bulk dual is hence 2d JT gravity and 2d compact BF with a chemical potential turned on for the latter, with boundary form:
\begin{equation}
A_t = g^{-1} g' + \mu f',
\end{equation}
in terms of the boundary group element $g = e^{i\Lambda}$ and the imposed chemical potential $\mu$.

\subsection{Non-abelian extension}
So far we have considered an IR low-energy emergent $\sltr \times$ U(1) symmetry. We can generalize this to the $ \sltr \times G$ case for a non-abelian group $G$. From the QM perspective, this case might be relevant for example if one studies a $\mathcal{N}=4$ version of SYK \cite{Anninos:2016szt}, which has a $G=$ SU(2) global R-symmetry. This will also clarify the role of the chemical potential, and more directly connect with the generalizations of the Schwarzian theory we will consider in the rest of this paper.
\\~\\
There is a straightforward non-abelian generalization of the previous model and in particular of the action \eqref{lecsyk}. One couples the particle on the group $G$ model to the Schwarzian by introducing $r$ (= rank of $\mathfrak{g}$) chemical potentials:
\begin{equation}
\label{nonab}
S = -\int_{0}^{\beta}d\tau \left[C \left\{\tan \frac{\pi}{\beta} f(\tau), \tau \right\} -\frac{K}{2} \text{Tr}\left[g^{-1} g' + i \mu f'\right]^2\right],
\end{equation}
where we introduced a set of different chemical potentials $\mu = \bm{\mu} \cdot \bm{H} = \mu^a \lambda_a$ for each element of the Cartan subalgebra $\lambda_a$.\footnote{$\bm{H}$ is the vector of Cartan generators.} A brief discussion on the classical equations of motion of the particle-on-the-group action and winding quantum numbers is given in appendix \ref{app:qmgroup}.
\\~\\
Similarly to the abelian case, we can decouple the Schwarzian mode from the gauge mode by a field redefinition. From now on, we will omit the contribution to observables from the Schwarzian mode. Therefore if we fix $K=1/2$, expressions will depend on temperature $\beta^{-1}$ and chemical potential $\mu$. In order to decouple the modes we can redefine the group element as 
\begin{equation}
\tilde{g} = g \cdot Y, ~~~~{\rm with}~~~Y' \cdot  Y^{-1}= i\mu f'(\tau),
\end{equation}
with $Y$ defined in terms of $f$ by solving this differential equation, and 
\begin{equation}\label{defM}
\tilde{g}(\beta) =  \tilde{g}(0) \cdot z , \qquad z = \mathcal{P} e^{i\mu \int_{0}^{\beta}d\tau f'(\tau)} = e^{i\mu \beta}\in T,
\end{equation}
where $T$ is the maximal torus of $G$. These are formally the same formulas as those for the coadjoint orbit action for non-constant orbits \cite{Barnich:2017jgw}. Just like the abelian case, the price to pay to decouple the fields is to introduce twisted boundary conditions for the gauge mode. We will focus in the rest of this section on the gauge mode part, leaving implicit a factorized contribution from the Schwarzian correlators.
\\~\\
To begin with, we can compute the partition function, generalizing the result in equation \eqref{eq:partu1}. The gauge piece is given by the twisted partition function for the particle on the group $G$ model:\footnote{The chemical potential Cartan matrix defined here differs from the U(1) case by an additional $i$ factor.} \footnote{Strictly speaking, the symplectic path integral is over $LG/G$, and hence there should be an overall $1/\text{vol }G$ factor. This factor can be absorbed into the zero-temperature entropy $S_0$ and is dismissed here. It will cancel out in any case for correlation functions. For non-compact groups it is important and it is precisely this factor that causes the Schwarzian partition function \eqref{pfschw} not to have an overall volume factor. Some more details on the compact case can be found in Appendix C of \cite{Mertens:2018fds}.}
\begin{equation}
Z(\beta,z) \equiv \text{Tr}~e^{-\beta H}e^{i\mu \beta} = \sum_{R} \dim R \, \chi_R (z) e^{-\beta \mathcal{C}_R},\label{twistedpog}
\end{equation}
where the sum is over all irreducible representations $R$ of the group $G$ and $\mathcal{C}_R$ denotes its Casimir. The factor $\chi_R(z) \equiv \text{Tr}_R~ z$, with $z\in T$, contains the information on the chemical potentials through equation \eqref{defM}. The element $z$ is a non-abelian generalization of the fugacity in statistical mechanics. If we set $z=\mathbf{1}$ and therefore $\chi_R(z) = \text{dim }R$, we obtain the answer for the particle on a group partition function. The role of the chemical potential is to introduce a defect-like factor of $\chi_R(z)/\text{dim }R$. \\
The abelian version of the expression above gives $Z_{\text{U(1)}}(\beta,z) = \sum_{n=-\infty}^{\infty} z^n e^{- \beta n^2}$, with representations labeled by the charge $n$ and fugacity $z=e^{i \mu \beta}$, which is expression \eqref{sumcharge} above. \\
Next, we consider the simplest non-abelian case of SU(2). One has only one Cartan generator, and the partition function is
\begin{equation}
\label{sutwo}
Z_{\text{SU(2)}}(\beta,z) = \sum_{j \in \mathbb{N}/2} (2j+1) \, \frac{\sin\left(\left(j+\frac{1}{2}\right)\mu\beta\right)}{\sin \mu \beta} e^{-\beta j(j+1)},
\end{equation}
where the sum is over the half-integer spin $j$ that labels the representation. \\
Note that these expressions are periodic as $\mu\beta \to \mu \beta +2\pi $. This indicates that it already includes non-trivial winding modes, analogous to the sum in the first line of \eqref{eq:partu1}. 
\\~\\
The observables of this theory are bilocal fields defined as a straightforward generalization of the U(1) case. We can define a bilocal observable as 
\begin{equation}
G_{R,M}(\tau_1,\tau_2) = [ \tilde{g}(\tau_1) \tilde{g}^{-1}(\tau_2) ]^R_{MM} ,
\end{equation}
where we consider the $MM$ matrix element of the group element in the representation $R$ of the group.\footnote{Off-diagonal bilocals can be considered, but vanish for the specific case of the two-point correlator studied here. See \cite{Blommaert:2018oro} for details.}
Based on symmetry arguments, we will take the natural observables from an SYK with non-abelian symmetry perspective to have the same structure as the abelian case. Namely the reparametrization mode part decouples from the $G$-symmetry part, when written in terms of $\tilde{g}$ with a fugacity-dependent monodromy $\tilde{g}(\tau+\beta) = \tilde{g}(\tau) \cdot z$. This is an obvious generalization of equation \eqref{obsredl}.
\\~\\
The theory of a particle on a group with a non-trivial monodromy can be solved exactly (see \cite{Mertens:2018fds,Blommaert:2018oro} for an approach relevant to the discussion here). The expectation value of the bilocal defined above is given by
\begin{align}
\left\langle G_{R,M} (\tau) \right\rangle &\equiv  \frac{1}{Z} \text{Tr}\left[G_{R,M} (\tau)e^{-\beta H}e^{i\mu \beta}\right] \nonumber \\
 &= \frac{1}{Z} \sum_{R_1,R_2,m_1,m_2} \text{dim R}_1  \, \chi_{R_2}(z) \, e^{-\mathcal{C}_{R_1} \tau}e^{-\mathcal{C}_{R_2} (\beta-\tau)}  \threej{R_1}{m_1}{R_2}{m_2}{R}{M}^2,
\end{align}
where we normalized the expectation value by the partition function $Z$ \eqref{twistedpog} which is a function of the group $G$, $\beta$ and $z$. This expression has the same structure and generalizes the abelian U(1) case, in the form given in equation \eqref{eq:u12ptc}. In the SU(2) case, the expectation value of a bilocal spin-$J$ field becomes
\begin{equation}
\label{2ptsu2}
\left\langle G_{J,M} \right\rangle \to \frac{1}{Z} \sum\displaylimits_{\tiny \left.\begin{array}{c}
j_1,j_2 \in \mathbb{N}/2 \\
\left|j_1-j_2\right| \leq J \leq j_1+j_2
 \end{array}\right.} \hspace{-0.5cm} (2j_1+1) \,\frac{\sin\left(\left(j_2+\frac{1}{2}\right)\mu\beta\right)}{\sin \mu \beta} \, e^{-j_1(j_1+1) \tau}e^{-j_2(j_2+1) (\beta-\tau)} \frac{1}{\text{2J+1}}.
\end{equation}

\noindent At very low temperatures $\beta \to \infty$, since $R_2$ is set to the identity, the correlator decays with Euclidean time as $\langle G_{R,M}(\tau) \rangle \sim e^{- \mathcal{C}_R \tau}$. We can interpret this as a non-Abelian version of the Coulomb-blockade with a blockade energy given by $E = \mathcal{C}_R/2 K$, after reinserting units.  
\\~\\
The main lessons from this section, besides the explicit solution of the theory, comes from the structure of the correlators. Take for example a six-point function, whose correlator can be diagrammatically represented as 
\begin{equation}
\label{striped}
\langle G_1G_2G_3\rangle =~~~~
\begin{tikzpicture}[scale=0.8, baseline={([yshift=-0.13cm]current bounding box.center)}]
\draw[thick] (0,0) circle (1.5);
\draw[thick] (1.06,1.06) arc (300:240:2.16);
\draw[thick] (-1.06,-1.06) arc (120:60:2.16);
\draw[fill,black] (-1.06,-1.06) circle (0.1);
\draw[ultra thick] (-0.2,1.3)--(0.2,0.9);
\draw[ultra thick] (-0.2,0.9)--(0.2,1.3);
\draw[ultra thick] (0,1.4)--(0,1.6);
\draw[thick] (-1.5,0)--(1.5,0);
\draw[fill,black] (1.06,-1.06) circle (0.1);
\draw[fill,black] (-1.06,1.06) circle (0.1);
\draw[fill,black] (1.06,1.06) circle (0.1);
\draw[fill,black] (1.5,0) circle (0.1);
\draw[fill,black] (-1.5,0) circle (0.1);
\end{tikzpicture}
\end{equation}
where the bulk represents a hyperbolic disc on which a BF theory lives with gauge group $G$. Boundary bilocal fields are represented by Wilson lines through the bulk. Each region in the bulk is labeled by a representation $R$, and each boundary segment has a propagation factor $e^{-\mathcal{C}_R \tau_{ij}}$, while a line ending on the boundary comes with a 3j-symbol. The tick on the outer circle denotes the location of the Boltzmann factor $e^{-\beta H}$ in the thermal trace, and is where we insert the chemical potential defect. For more details on these rules see \cite{Mertens:2017mtv} for the Schwarzian case and \cite{Mertens:2018fds,Blommaert:2018oro} for the compact case. \\
The chemical potential, through the twisted periodicity, is equivalent to inserting a defect in the bulk description, contributing a factor $D_{\mu}(R)  \equiv \frac{\chi_R(z)}{\text{dim R}}$ associated to the representation $R$ of the region where the defect lives.
\\~\\
This defect insertion was treated in appendix A of \cite{Blommaert:2018oro} and interpreted as a vertical Wilson line for the $\chi$-field in the bulk BF Lagrangian $\text{Tr}\chi F$.  Let us provide a sketch of how the bulk defect relates to the monodromy of the boundary group element. If one inserts a defect into the BF path integral with action \eqref{BFaction} of the type $\text{Tr}_\lambda\,e^{\kappa \chi}$ for some irrep $\lambda$, then integrating over $\chi$ yields the equation $F(x) = \kappa w^{-1} \lambda w \, \delta(x-y)$,\footnote{The $w$ degrees of freedom are microscopic ``color" variables that upon integrating out, give rise to Wilson lines (see e.g. \cite{Moore:1989yh}). Mathematically, they are needed to move the trace operation into the exponential by integrating in new dummy variables $w$. See also \cite{Fan:2018wya} for a particularly transparant recent exposition.} solved by pure gauge fields $A = g d g^{-1}$ where the group element $g$ has a monodromy:
\begin{equation}
g(t+\beta) = g(t) U_\alpha, \qquad U_\alpha = e^{-2\pi \kappa \bm{\lambda} \cdot \bm{H}}.
\end{equation}
This rewrites the bulk BF theory in terms of a boundary group element $g$. Substituting this back in the BF action and taking care of the boundary action, one obtains the particle-on-a-group model \cite{Mertens:2018fds}. This argument is made more elaborate in appendix \ref{app:defmono}.

\section{Defects in JT gravity: Classification and Orbits}
\label{sec:orbits}
Here we initiate a classification of the different defects in JT gravity, and how they link to the Schwarzian models. We will combine two perspectives on this problem: the gauge-theoretic BF formulation of Jackiw-Teitelboim gravity, as studied in \cite{Mertens:2018fds,Blommaert:2018oro,Blommaert:2018iqz} and the Liouville CFT perspective offered in \cite{Mertens:2017mtv}. \\
Most of the material in this section is not new, and can be viewed as a literature survey emphasizing the parts that will be relevant to us in later sections.
\\~\\
In \cite{Jensen:2016pah, Maldacena:2016upp, Engelsoy:2016xyb}, it was proven that the JT action \eqref{JTaction}, in a region with boundary curve specified by a constant (large) value of the dilaton $\Phi \sim a/\epsilon$ and asymptotically Poincar\'e boundary conditions on the metric $g_{\mu\nu}$, can be rewritten in terms of the Schwarzian action:
\begin{equation}
S = -C\int d\tau \left\{F,\tau\right\}, \quad C = \frac{a}{16\pi G_N}.
\end{equation}
This argument is local and is insensitive to possible monodromies of the field $F$ as one goes around the thermal circle. So here we want to be more careful and understand this at the same level as we did for the compact BF case above.
\\~\\
The main statement we want to make is: \\
\emph{Defects in the gauge group $G$ BF and JT models are in one-to-one correspondence with the (constant representative) coadjoint orbits of the loop groups $\widehat{LG}$ or $\widehat{\text{diff }} S^1$}.
\\~\\
For BF with compact gauge group, the link with coadjoint orbits is reviewed in appendix \ref{app:compact}. As well-known, irreducible representations can be embedded within the orbits by quantization techniques. Integrality of the Kahler form then discretizes the orbit parameter. What we illustrate there, is that for centrally-extended loop groups $\widehat{LG}$ in the large level limit $k\to+\infty$, the discrete subset of representation labels becomes dense in the set of orbits, and one finds a one-to-one mapping between conjugacy classes (monodromies) of $G$ and orbits of $\widehat{LG}$.

\subsection{From defects to monodromies}
\begin{figure}[h]
\centering
\includegraphics[width=0.65\textwidth]{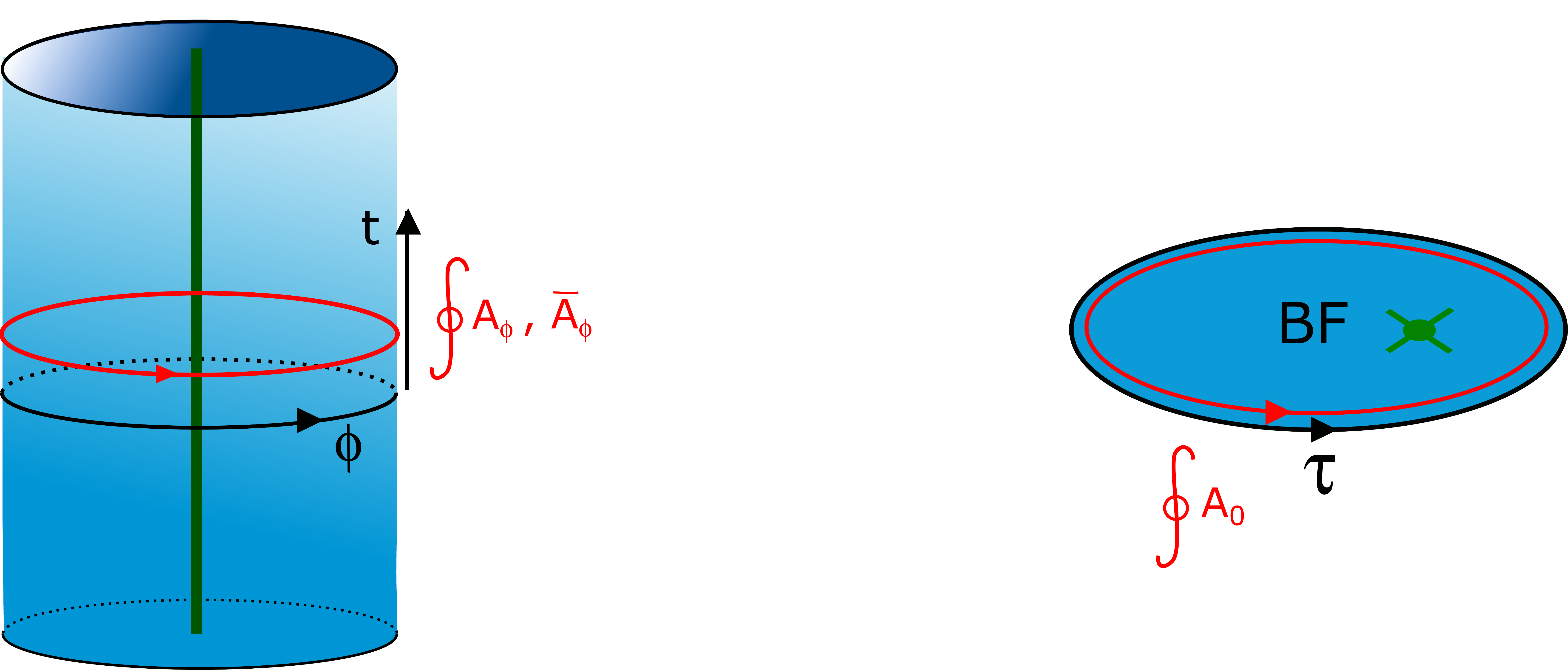}
\caption{Left: AdS$_3$ global spacetime with a defect along the time direction (green), measured by the holonomy integrals $\oint A_\phi$ and $\oint \bar{A}_\phi$. Right: Defect in 2d Jackiw-Teitelboim gravity (green), measured by the holonomy integral $\oint A_0$.}
\label{ads3}
\end{figure}

Now we study the JT gravity story, related to the non-compact centrally extended group $\widehat{{\rm Diff}}(S^1)$. Our first goal is to relate defects to boundary monodromies. \\
We first remind the reader how topological defects appear in the higher-dimensional AdS$_3$/CFT$_2$ set-up. Consider the 3d asymptotically global AdS space (Figure \ref{ads3} left), with angular coordinate $\phi$.
For any fixed value of the holonomy along the $\phi$-direction,\footnote{This is measured most naturally within the gauge theory formulation as $\mathcal{P}e^{\oint A_\phi}$ and $\mathcal{P}e^{\oint \bar{A}_\phi}$.} the gravitational path integral reduces to a theory of boundary gravitons surrounding a defect in the bulk that depends on the holonomy. Just as in the compact scenario, one can interpret these defects as vertical Wilson lines. \\
Elliptic monodromies correspond to massive point particles, with an angular deficit along their worldline, and hence a conical geometry. Hyperbolic monodromy yields macroscopic ``holes'' in the bulk, and corresponds to BTZ black holes. The parabolic case in between corresponds to the Poincar\'e vacuum.
\\~\\
In the 2d JT case, we will impose a non-trivial monodromy along the $\tau$-direction instead and insert defects into the thermal disk (Figure \ref{ads3} right). 
Analogously as in 3d, it is convenient to formulate the question on defects in terms of the possible holonomies of the gauge connection $A$. The identification of JT gravity with $\sltr$ BF gauge theory was first made in \cite{Jackiw:1992bw}. In the first-order formulation of JT gravity, one writes the gauge connection $A$ in terms of the zweibein $e^a,\,a=1,2$ and spin connection $\omega$ as:
\begin{equation}
A = e^a \lambda_a + \omega \lambda_3, \quad F = (D e)^a \lambda_a + F^3 \lambda_3,
\end{equation}
with the bulk BF action $S = \frac{K}{2\pi} \int \text{Tr} \chi F$ for which the $a=1,2$ constraints impose the geometry is torsionfree $(De)^a=0$, and the $a=3$ constraint enforces $R=-2$. Inserting a Wilson line defect of the type $\text{Tr}_\lambda e^{i \kappa \chi}$, we can write:
\begin{equation}
\text{Tr}_\lambda e^{i \kappa \chi} e^{i S_{\text{BF}}} = \int dw e^{-i \kappa \text{Tr}\big( \lambda w^{-1} \chi w\big)} e^{i \frac{K}{2\pi} \int \text{Tr} \chi F},
\end{equation}
in terms of an auxiliary group element $w$. One can perform a global $G$-transformation $\chi \to w \chi w^{-1}$, $F \to w F w^{-1}$, and we can remove the $w$-dependence completely. Since $\lambda$ is in the Cartan algebra $\sim \lambda_3$, this means it is always possible to change $G$-frame to put the defect in the curvature entirely:\footnote{This seems no longer possible when considering multiple defects, and this is a possibility we defer to future work. See also \cite{Ammon:2013hba,Raeymaekers:2014kea} for related comments in the AdS$_3$/CFT$_2$ set-up. \\
We thank K. Jensen for a discussion about this.} 
\begin{equation}
\label{tors}
F^3 = \kappa \frac{2\pi}{K}\lambda \delta(x-y) \quad \Rightarrow \quad R = -2 + \alpha \delta(x-y),
\end{equation}
which we will study further on around equation \eqref{conpath}.
\\~\\
In the case that there is a boundary, there is a particle on the group $G$ living on the boundary surface. For gravity, there are additional gravitational constraints and the $\sltr$ group element $g$ is further constrained at the holographic boundary by
\begin{equation}
\label{probl}
A_\tau  \equiv g \partial_\tau g^{-1} =  \left(\begin{array}{cc}
0 & -\frac{T(\tau)}{2} \\
1 & 0 \\
\end{array}\right) ,
\end{equation}
in terms of what turns out to be the energy $C T(\tau)$.\footnote{These conditions can be found, for example, by direct dimensional reduction from 3d AdS gravity.} Holonomies of $A$ around the time-direction are linked to monodromies of the group element $g(\tau)$ as one circles the thermal circle. The latter are classified by the well-known conformal structure of Hill's equation as we review here. Writing 
$g^{-1} =\left(\begin{array}{cc}
A(\tau) & B(\tau) \\
C(\tau) & D(\tau) \\
\end{array}\right)$ we can rewrite \eqref{probl} as
\begin{align}
\label{Hill1}
A'' + \frac{1}{2}T(\tau) A =0, &\quad B =A', \\
\label{Hill2}
C'' + \frac{1}{2} T(\tau) C = 0, &\quad D = C',
\end{align}
with $\sltr$ constraint $AC'-A'C = 1$. The most general solution $g^{-1}$ is then found as a linear combination of one specific solution, by taking one such solution and applying an arbitrary $\sltr$ matrix $S^{-1}$:
\begin{equation}
g^{-1} = S^{-1} \cdot  \left(\begin{array}{cc}
A & A' \\
C & C' \\
\end{array}\right),
\end{equation}
the $\sltr$-constraint on $S^{-1}$ coming from respecting the Wronskian condition, or one has the equivalence relation:
\begin{equation}
\label{eqrel}
g \sim g \cdot S, \qquad S \in \sltr.
\end{equation}
As well-known, depending on $T(\tau)$, the solutions of Hill's equation \eqref{Hill1} can have non-trivial monodromies:
\begin{equation}
g(\tau+\beta) = g(\tau) \cdot M,
\end{equation}
which, combined with the equivalence relation \eqref{eqrel}, leads to only dependence on conjugacy classes of monodromies:
\begin{equation}
M \sim S \cdot M \cdot S^{-1},
\end{equation}
the latter having been classified into elliptic, parabolic and hyperbolic classes. \\
Fixing $M$ to any given desired form by tuning $S$, is not in general a unique procedure: all $S$ satisfying $\left[S,M\right] = 0$ preserve $M$. As this residual symmetry transformation boils down from the global $G$ invariance \eqref{eqrel}, it is a residual gauge redundancy that should be modded out.

Under the monodromy transformation, the ratio $A/C \equiv F$, which is the uniformizing coordinate $F$, transforms under the M\"obius transformation:
\begin{equation}
F(\tau+\beta) = M \cdot F(\tau) \, \equiv\, \frac{aF(\tau) + b}{cF(\tau) +d}, \qquad M^{-1} = \left(\begin{array}{cc}
a & b \\
c & d \\
\end{array}\right).
\end{equation}
This variable, being the ratio of two independent solutions of Hill's equation with unit Wronskian, satisfies 
\begin{equation}
\{ F(\tau), \tau \} = T(\tau) ,
\end{equation}
and can be interpreted as a uniformization coordinate associated to an energy density $T(\tau)$.
\\~\\
Conjugacy classes of the monodromy matrix can be divided into hyperbolic, parabolic and elliptic. Before reviewing more details on them, we will explain a connection with Virasoro coadjoint orbits that will be central in what follows. This idea is not new and it is nicely reviewed, for example, in \cite{Witten:1987ty,Balog:1997zz}, which we summarize below. 

\subsection{From coadjoint orbits to monodromies}
The Virasoro group, which we simply denote by ${\rm Diff}(S^1)$, is given by a central extension of reparametrizations of the circle, which we will parametrize by $\tau \in [0,2\pi)$.\footnote{After talking the Schwarzian limit, this coordinate will be interpreted as Euclidean time with a size that will be rescaled to $\beta$.} Vectors in the algebra are given by infinitesimal reparametrizations $\xi(\tau) \partial_\tau$ plus the central element. Coadjoint elements are given by quadratic differentials which transform as a chiral stress-tensor of a 2d CFT. We will conveniently denote them by $T(\tau) d\tau^2$ (plus a central element which we will take to be constant and equal to $c$).
\\~\\
Associated to each $T(\tau)$ we can generate a coadjoint orbit by an arbitrary reparametrization. Classifying coadjoint orbits consist on finding the stabilizer group $H\in {\rm Diff}(S^1)$ leaving the profile $T(\tau)$ invariant, and also finding the reparametrization inequivalent profiles of $T$. This gives rise to the orbit ${\rm Diff}(S^1)/H$. Orbits with a constant representative $\partial_\tau T=0$ are usually labeled by the parameter $b_0=T$.  \\
The condition for an infinitesimal element $\xi(\tau) $ to leave the profile invariant is 
\begin{equation}
\delta_\xi T= \xi \partial_\tau T + 2 T \partial_\tau \xi - \frac{c}{12} \partial_\tau^3 \xi=0.
\end{equation}
Then the problem of classifying orbits ends up being equivalent to classifying solutions of $\delta_\xi T=0$. If $\psi_1(\tau)$ and $\psi_2(\tau)$ are independent solutions of Hill's equation $\partial_\tau^2 \psi = \frac{6}{c} T(\tau) \psi$, then $\xi(\tau)\in H$ is a linear combination of $\psi_i(\tau) \psi_j(\tau)$ for $i,j=1,2$. Therefore, the possible stabilizers $H$ and inequivalent $T(\tau)$ (the distinct coadjoint orbits) are in correspondence with conformally non-equivalent solutions of Hill's equation, which are labeled by the monodromy conjugacy class of $\psi(\tau)$. The number of generators in $H$ depend on the possible ways to combine $\psi_1$ and $\psi_2$ into a reparametrization $\xi(\tau)$ with correct periodicity. 
\\~\\
Having described the connection between monodromy and coadjoint orbits we give a list of representative of each class
\begin{itemize}
\item \textbf{Elliptic Case:} A representative of the elliptic conjugacy class of monodromies is 
\begin{equation}
M = \left(\begin{array}{cc}
\cos(\pi \theta) & \sin(\pi\theta) \\
-\sin(\pi\theta) & \cos(\pi \theta) \\
\end{array}\right) \quad \in \text{P}\sltr.
\end{equation}
This class is related to the coadjoint orbit ${\rm Diff}(S^1)/$U(1) for $\theta \neq \mathbb{Z}$. Representatives of this orbits are the constant coadjoint elements $T=b_0=-\frac{c}{24} \theta^2$. When $\theta=n\in \mathbb{Z}$ the stabilizer group gets enhanced to $\slnr$, giving rise to the orbits ${\rm Diff}(S^1)/\slnr$ with representative $T=b_0=-\frac{c}{24}n^2$. For $n=1$ this gives the vacuum orbit ${\rm Diff}(S^1)/\sltr$, corresponding to JT gravity without defects. 

\item \textbf{Hyperbolic Case:} A representative of the monodromy in this case is 
\begin{equation}
M = \left(\begin{array}{cc}
\cosh(\pi \lambda) & \sinh(\pi\lambda) \\
\sinh(\pi\lambda) & \cosh(\pi \lambda) \\
\end{array}\right) \quad \in \text{P}\sltr.
\end{equation}
This corresponds to an analytic continuation $\theta \to i \lambda$ of the orbits ${\rm Diff}(S^1)/$U(1) with representative $T=b_0=\frac{c}{24}\lambda^2$. This can be generalize to $M \to (-1)^n M$ for integer $n$, giving rise to the exceptional orbits ${\rm Diff}(S^1)/\mathcal{T}_{n,\lambda}$, characterized by a $\tau$-dependent representative $T(\tau)$. 
\item \textbf{Parabolic Case:} A representative of this case is 
\begin{equation}
\label{para}
M = \left(\begin{array}{cc}
1 & q \\
0 & 1 \\
\end{array}\right) \quad \in \text{P} \sltr,~~~q=\pm1,
\end{equation} 
which corresponds to a particular case of ${\rm Diff}(S^1)/$U(1) with $T=b_0=0$. Similarly by generalizing by $M \to (-1)^n M$ is associated to the exceptional orbits ${\rm Diff}(S^1)/\mathcal{T}_{n,{\rm sgn}(q)}$, also with a non-constant profile $T(\tau)$.
\end{itemize} 

\subsection{Twisted Schwarzian Models}
This list of monodromies and Virasoro orbits in the previous section has been known for a long time. Associated to each of the orbits, is a phase space path integral which can be interpreted as a 2D field theory
\begin{equation}
\label{virchar}
Z_H(q=e^{-\beta_{AS}}) = \int_{{\rm Diff}(S^1)/H} [\mathcal{D}f]~e^{-S[f]}, \quad h \equiv b_0 + \frac{c}{24},
\end{equation}
with an action found by Alekseev and Shatashvili \cite{Alekseev:1988ce}
\begin{equation}
\label{newgeoma}
S = \frac{c}{12\pi}\int_0^{\beta_{AS}} d\tau_{AS} \int_{0}^{2\pi} d\sigma_{AS}  \left(i\left[\frac{\dot{f}}{4 f'}\left(\frac{f'''}{f'}- 2 \left(\frac{f''}{f'}\right)^2 \right) - b_0 \dot{f}f'\right] + \left\{F \circ_H f,\sigma_{AS} \right\}\right),
\end{equation}
where primes denote derivatives with respect to the coordinate $\sigma_{AS}$ and dots with respect to $\tau_{AS}$. Compared to the previous section, we interpret $\tau \equiv \sigma_{AS}$ as the 2d spacelike coordinate. The phase space of the Alekseev-Shatashvili theory is spanned by $f(\sigma_{AS})\in {\rm Diff}(S^1)$, and the 2d time evolution is parametrized by $\tau_{AS}$. 

The precise function $F \circ_H f$ depends on the particular orbit and will be given below. The parameter $b_0$ labels the representative coadjoint element of the orbit. $h$ as a function of $b_0$ denotes the dimension of the single Virasoro representation appearing in the Hilbert space defined by quantizing the coadjoint orbit with label $b_0$ \cite{Alekseev:1990mp}.

Here we study the models that can be obtained by performing a double-scaling limit mentioned above \cite{Mertens:2017mtv}, to find the full list of twisted Schwarzian models. This limit dimensionally reduces this theory such that the $\tau_{AS}$ dependence disappears, and we are left with only the $\sigma_{AS} \equiv \tau$ dependence. This is intended to stress the fact that the spatial circle in the Alekseev-Shatashvili theory becomes the Euclidean time circle in the Schwarzian theory. 

\begin{figure}[h]
\centering
\includegraphics[width=0.5\textwidth]{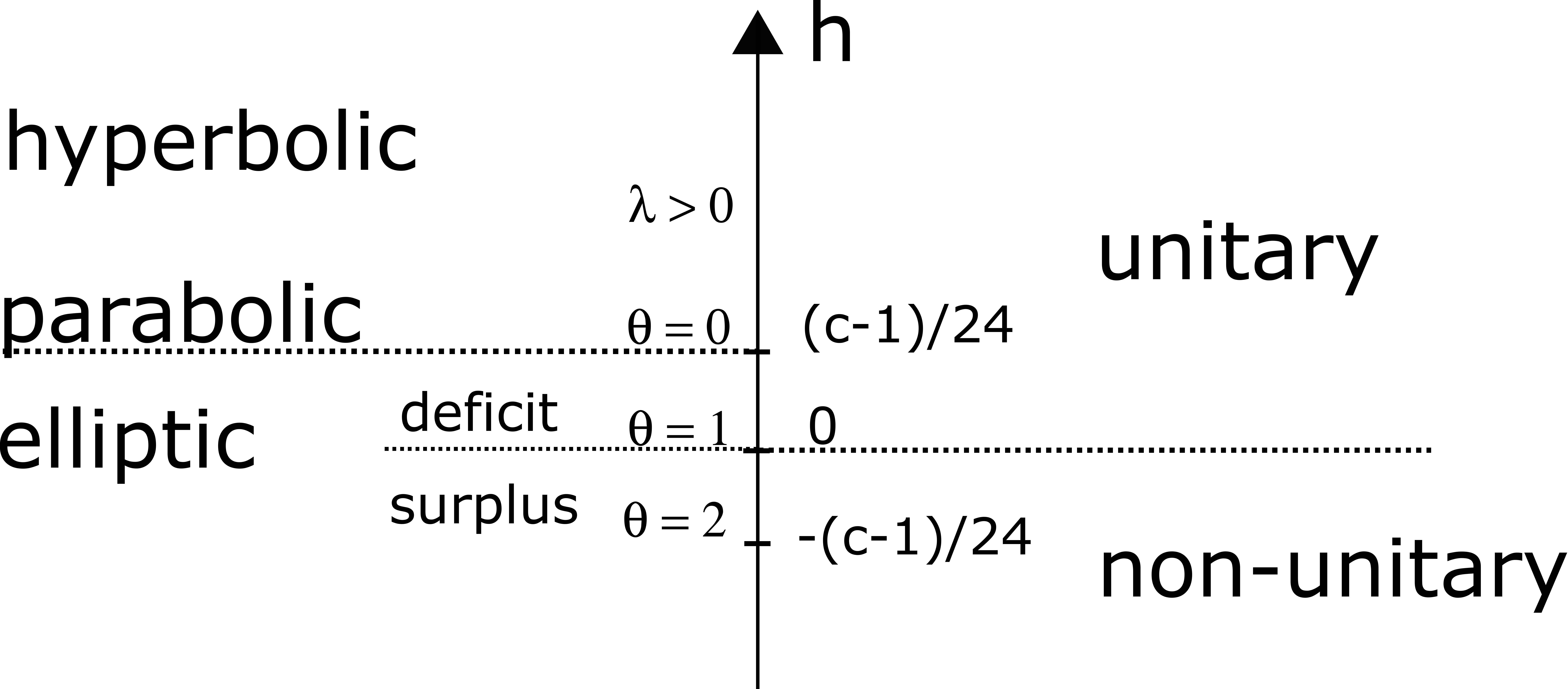}
\caption{Scheme of all Virasoro highest weight representations at large $c$. Hyperbolic, parabolic and elliptic orbits span all representations. Both unitary and non-unitary representations are required.}
\label{virRep}
\end{figure}

For a constant choice of the parameter $b_0$, these path integrals compute characters $Z_H(q) = \chi_h(q)$ of the Virasoro highest-weight representations since this is the full content of the Hilbert space. Moreover, at large $c$, the set of all Virasoro highest-weight representations is spanned by all constant $b_0 \in \mathbb{R}$ (Figure \ref{virRep}).\footnote{For $c<1$, one has to consider the minimal models series as well. But we are interested in the opposite regime.} 
The relation between the coadjoint element $b_0$ and the monodromy label $\lambda \equiv i\theta$ is given in the large $c$ limit by:
\begin{eqnarray}
h = b_0 + \frac{c}{24} &=& \frac{c}{24} + \frac{c}{24}\lambda^2 ~~~({\rm hyperbolic})\\
&=& \frac{c}{24}-\frac{c}{24}\theta^2~~~({\rm elliptic}).
\end{eqnarray}
We see that this matches the values of $b_0$ given in the previous section after removing the zero-point energy $\frac{c}{24}$. 

\begin{figure}[h]
\centering
\includegraphics[width=0.75\textwidth]{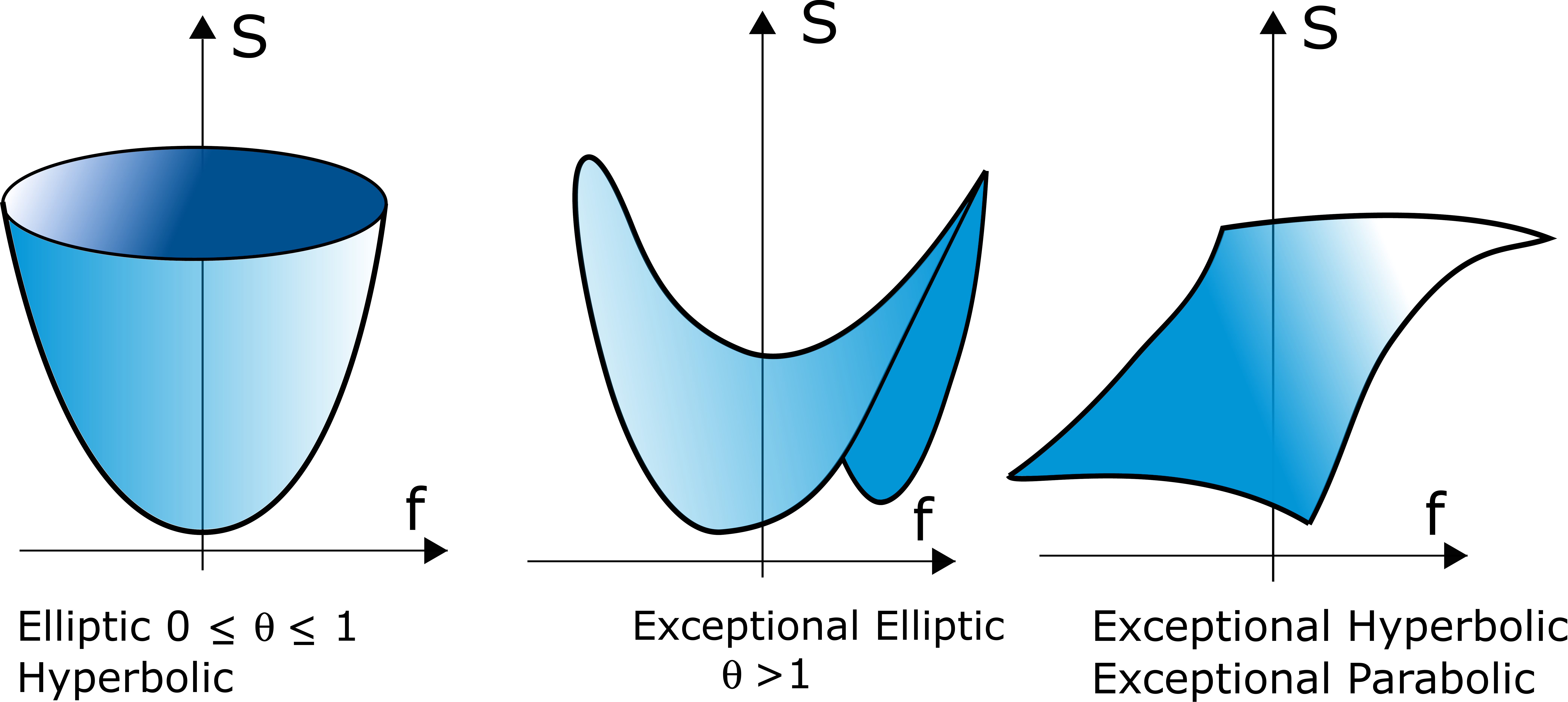}
\caption{Schematic situation of the action $S[f]$ in functional space. The elliptic ($0\leq \theta \leq 1$) and hyperbolic orbits have a unique and stable saddle solution. The exceptional elliptic ($\theta>1$), including the higher $\theta=n $ orbits, have a unique saddle solution, but it has unstable directions. The exceptional hyperbolic and parabolic orbits have no extrema at all and are unbounded in action both from above and below.}
\label{orbitsFuncSpace}
\end{figure}

In a dimensional reduction where $c\to+\infty$ and simultaneously $\beta_{AS}=\int d\tau_{AS} \to 0$ keeping the product $C\equiv \beta_{AS} \cdot c$ fixed, one finds \eqref{virchar} reduces to the Euclidean path integral:
\begin{equation}
\label{twsch}
Z_H(\beta) = \int_{{\rm Diff}(S^1)/H} [\mathcal{D}f]~e^{C \int_0^\beta d\tau \{ F \circ_H f(\tau), \tau\} },
\end{equation}
with the usual measure $[\mathcal{D}f]= \prod_{\tau_i} df(\tau_i)/f'(\tau_i)$ \cite{Alekseev:1988ce, Alekseev:1990mp}.\footnote{If we would keep the full 2D before taking semiclassical limit, we would get an effective theory for the pseudo-Goldstone mode of broken conformal transformations in 2D \cite{Turiaci:2016cvo}.} As anticipated, note that time and space have switched role here and we relabeled $\sigma_{AS} \to \tau$. We also rescaled the size of the circle from $2\pi$ to $\beta$.  These can be called twisted Schwarzian models.\footnote{Since the size of the $\tau_{AS}$ circle, $\beta_{AS}$, is going to zero, configurations with $\dot{f}(\sigma_{AS},\tau_{AS}) \neq 0 $ are highly suppressed. Restricting to $\tau_{AS}$ independent fields kills the term in brackets appearing in the action \eqref{newgeoma}, giving the simpler 1D action appearing in \eqref{twsch}.} \\
The uniformizing variable $F(\tau)$ is given in terms of $f(\tau)$ in a way that depends on the orbit. For the elliptic case it is given by $F\circ_\theta f(\tau) = \tan \frac{\pi}{\beta} \theta f(\tau)$ (this relation being still valid even if $\theta \in \mathbb{Z}$). Hyperbolic orbits have $F \circ_\lambda f(\tau) = \tanh \frac{\pi}{\beta}\lambda f(\tau)$. Parabolic orbits have $F(\tau)=f(\tau)$.  The exceptional hyperbolic orbits $\mathcal{T}_{n,\lambda}$ can be made by using the reparametrization
\begin{equation}
F \circ_{n,\lambda} f(\tau) = e^{\lambda f(\tau)}\left(\tan\frac{nf(\tau)}{2} + \frac{\lambda}{2n}\right).
\end{equation}
Finally, the exceptional parabolic orbits $\mathcal{T}_{n,\epsilon}$, where $\epsilon = \pm 1$, use
\begin{equation}
F \circ_{n,\epsilon} f(\tau) = \frac{\epsilon f(\tau)}{2\pi} - \frac{2}{n}\cot\frac{nf(\tau)}{2}.
\end{equation}
The vacuum ($n=1$) $\sltr$ orbit is ubiquitous in the entire SYK literature. The hyperbolic U(1) orbit appeared first in this context in \cite{wittenstanford} and is studied further in \cite{sss2, Blommaert:2018iqz} to glue JT disks together and obtain more general topologies. The different  Schwarzian models also appear in \cite{Anninos:2018svg} in a de Sitter context, where they are dubbed the $\gamma$-Schwarzians.
\\~\\
Next we characterize the structure of the functional space in each of these models \eqref{twsch}. The Virasoro coadjoint orbit energy functional can be written as:
\begin{equation}
E_H[f] \equiv \int_{0}^{\beta}d\tau\, \left\{F \circ_H f(\tau),\tau\right\} ,
\end{equation}
and is identified as the Schwarzian action $S[f]$ we use. It can be shown directly that for each orbit this functional is unbounded from above.  \\
Boundedness of the energy functional from below is a sufficient condition for convergence of the Euclidean Schwarzian path integral. Moreover, whenever a minimal energy configuration exists in the orbit, this is also the unique Euler-Lagrange solution of the Schwarzian system. \\
The elliptic (with $0 \leq \theta \leq 1$), the hyperbolic and the vacuum orbit ($\theta=1)$ all have the saddle as the lowest action configuration.\\
The exceptional elliptic ($\theta>1$, including $\theta=n$) orbits on the other hand are unbounded below.\footnote{This means the Schwarzian path integral is not guaranteed to be convergent. However, our derivation demonstrates the final answer is finite as well. The would-be Duistermaat-Heckman result is indeed the correct one, and the partition function is one-loop exact, but a perturbative calculation is a priori not allowed.}
Indeed, linearizing the Schwarzian with $f(\tau) = \tau+ \epsilon(\tau)$, with $\epsilon(\tau+2\pi) = \epsilon(\tau)$, one writes $\left\{\tan \theta\frac{f}{2},\tau\right\} = - \frac{1}{2}\left[\theta^2\epsilon'^2 - \epsilon''^2\right]$ (here we set $\beta=2\pi$ to simplify). Expanding the general fluctuation in modes $\epsilon(\tau) = \sum_{m\in\mathbb{Z}} \epsilon_m e^{im\tau}$ with $\epsilon_{-m}=\epsilon_m^*$, we get the action:
\begin{equation}
\frac{1}{2}\int_{0}^{2\pi}d\tau \left[\epsilon''^2 - \epsilon'^2\right] = \pi \sum_{m\in\mathbb{Z}}\left|\epsilon_m\right|^2 (m^4-\theta^2m^2),
\end{equation}
which is always positive for $0 \leq \theta \leq 1$, but can change sign for $\theta>1$. There are hence unstable directions for the winding orbits $\theta >1$. Setting $\theta \to i \lambda$ shows that the hyperbolic orbits for any real $\lambda$ are stable. \\
The exceptional orbits $\mathcal{T}_{n,\lambda}$ and $\mathcal{T}_{n,\epsilon}$ do not have elements in their orbits with constant value of $\left\{F(\tau),\tau\right\}$. But the Schwarzian equation of motion, for all orbits, is energy conservation:
\begin{equation}
\partial_\tau \left\{F(\tau),\tau\right\} = 0.
\end{equation}
This means there is no classical solution to the equation of motion in this case, and there are hence no relative extrema to be found in the action functional $S[f]$, which is unbounded from above and below. \\
A cartoon of the action functional is shown in figure \ref{orbitsFuncSpace}.

The purpose of the rest of the paper is to solve this theory for each choice of orbit, therefore solving JT gravity in the presence of defects. We will do this by exploiting the connection between this theories and a semiclassical finite temperature limit of 2d Virasoro CFT developed in \cite{Mertens:2017mtv}. We will summarize the main features in the next section. 

\subsection{Orbits and Branes}
There is a convenient way to think about the Virasoro representations in terms of 2d Liouville CFT between suitable branes. The connection between the Schwarzian theory and 2d Liouville was extensively discussed in \cite{Mertens:2017mtv}, see also \cite{Lam:2018pvp, Mertens:2018fds}. Here we will point out the most important features needed to solve the theory defined in the previous section by getting the twisted Schwarzian theories as limits of Liouville between different kinds of branes.
\\~\\
Start with the case of the Schwarzian theory with $H=\sltr$, which is connected with Liouville in the following way. Consider 2D Liouville theory with central charge $c=1+6 Q^2$, $Q=b+b^{-1}$, between two identity ZZ-branes \cite{{Zamolodchikov:2001ah}} separated by a distance $\beta/2$, at finite temperature in an annulus with modular parameter $q$. These boundary conditions are described, following Cardy, as a boundary state $|B\rangle$ expanded in Ishibashi states as $|B\rangle = \int dP~ \Psi_B(P) |\hspace{-0.3mm}|P\rangle \hspace{-0.8mm}\rangle$. According to Cardy's classification, the possible boundary states $B$ are labeled themselves by primary operators of the theory. Here $P$ denotes the Liouville momentum which labels states of dimension $\Delta = \alpha (Q-\alpha)$ with $\alpha = \frac{Q}{2} + i P$. For the identity brane the wavefunction is\footnote{The label on the identity $(1,1)$ arises from the set of degenerate Virasoro modules labeled by two integers $(m,n)$ with momentum $P=\frac{i}{2}(m/b+nb)$, which for $m=n=1$ becomes the identity.}
\begin{equation}
\Psi_{(1,1)}(P)=\frac{2 \pi i P }{\Gamma(1-2 i b P)\Gamma(1-\frac{2iP}{b})}.
\end{equation}
 This theory can be mapped to a chiral CFT on the torus, where the role of the ZZ-branes is to project the Hilbert space to the identity representation. 
\\~\\
Take a finite temperature semiclassical limit $c\to \infty$ and $ q\to 1$ with $q^c$ fixed. Then this dimensionally reduces the theory to a circle $\tau\in S^1$ of size $\beta$. In the closed channel, this semiclassical limit zooms into $P=k b$ (or dimensions very close to $\frac{c}{24}$). This 1d reduction becomes the Schwarzian theory, with the spatial 2d dimension becoming the Euclidean time in the 1d theory.\footnote{This is very similar to the reduction that turned the Alekseev-Shatashvili 2d action into the Schwarzian 1d action. The reason is simply that both 2d theories are actually the same up to a change of variables \cite{Mertens:2017mtv}.}\\

Below we will enumerate all ways to reproduce the path integrals over any coadjoint orbit by replacing one of the boundary states in the Liouville construction. 
\begin{itemize}
\item \textbf{Elliptic Case $\slnr$:} Let's begin with the $H=\sltr$ $n=1$ case. This case is constructed from the semiclassical limit of 2d Liouville between identity ZZ-branes described above. The Hilbert space of the 2d theory is the identity module of Virasoro, which in the semiclassical limit becomes the phase space ${\rm Diff}(S^1)/\sltr$. The generalization to $H=\slnr$ is obtained by simply replacing one of the identity branes by a $(1,n)$ ZZ-brane, with wavefunction 
\begin{equation}\label{eq:ZZnbs}
\Psi_{(1,n)}(P)=\Psi_{(1,1)}(P)\frac{\sinh 2\pi n P/b}{\sinh 2\pi P/b}.
\end{equation} 
This produces a Hilbert space consisting of only the $(1,n)$ module. In the semiclassical limit this becomes the phase space ${\rm Diff}(S^1)/\slnr$.\footnote{\label{foot} These degenerate representations are not unitary at large $c$ unless $n=m=1$. We will see a similar problem in the $H=\slnr$ Schwarzian in section \ref{sect:ell}.}
\item \textbf{Hyperbolic Case:} To obtain the hyperbolic $H=$ U(1) orbits we consider one identity ZZ-brane and one FZZT-brane, labeled by Liouville momentum $s$ with wavefunction 
 \begin{equation}\label{eq:FZZTbs}
\Psi_{s}(P) = \Psi_{(1,1)}(P) \frac{\cos 4 \pi s P}{2\sinh(2\pi b P)\sinh(2 \pi P/b)},
\end{equation} 
which has the Hilbert space consisting of a single module with momentum $s$ and dimension $\Delta = \frac{Q^2}{4} + s^2$. To match with the semiclassical phase space ${\rm Diff}(S^1)/$U(1) we set $s = \lambda /2 b$ in the $b\to0$ limit. Their dimension scale as $\Delta = \frac{c-1}{24}(1+\lambda^2)$ and are above the $\frac{c-1}{24}$ threshold. 
\item \textbf{Elliptic Case U(1):} To obtain the $H=$ U(1) orbits we consider one identity ZZ-brane and one FZZT-brane with parameter $s$. For the orbit to be elliptic we need to continue $s= i \theta/ 2b$, giving $\Psi_{is}(P) \sim \cosh 4 \pi |s| P$. The momentum is related to the size $2\pi/\theta$ of the U(1) group. The dimensions of these states scales like $= \frac{c-1}{24}(1-\theta^2)$ and sit below the $\frac{c-1}{24}$ threshold. Since these FZZT-branes carry an imaginary momentum label we will refer to them as $\mathfrak{i}$FZZT-branes to distinguish between both cases. 
\item \textbf{Parabolic Case:} This case is obtained by considering one identity ZZ-brane and one FZZT-brane with $s\to0$ (this limit can be taken either in the elliptic or hyperbolic direction). The Hilbert space corresponds to a single module of momentum $s=0$ and therefore sits at the threshold $\Delta = \frac{c-1}{24}$.
\end{itemize}

\noindent In section \ref{sect:ell} we will study in more detail JT gravity with elliptic defects and its corresponding elliptic deformation of the Schwarzian theory. In section \ref{sect:hyper} and \ref{sect:para} we will study the hyperbolic and parabolic cases respectively. So far we have ignored the exceptional orbits, leaving some comments for section \ref{sec:excep}.

\section{Elliptic orbits $U(1)_\theta$ and SL$^n(2,\mathbb{R})$: conical singularities}
\label{sect:ell}
In this section we will focus on orbits related to solutions of Hill's equation with elliptic monodromy. Following the notation of the previous section, setting $F(\tau) = \tan \frac{\pi}{\beta} \theta f(\tau)$, with $f(\tau+\beta) = f(\tau) + \beta$ leads to an elliptic monodromy since:
\begin{equation}
F(\tau+\beta) = \tan\left( \frac{\pi}{\beta} \theta f(\tau) + \pi \theta \right) = \frac{F(\tau) + \tan(\pi \theta)}{1- \tan(\pi \theta) F(\tau)},
\end{equation}
identifying the monodromy matrix $M$ as:
\begin{equation}
\label{elli}
M = \left(\begin{array}{cc}
\cos(\pi \theta) & \sin(\pi\theta) \\
-\sin(\pi\theta) & \cos(\pi \theta) \\
\end{array}\right) \quad \in \text{P}\sltr,
\end{equation}
indeed in the elliptic class. Setting $\theta = n \in \mathbb{Z}$ leads to no holonomy, and is the standard thermal parametrization of $F(\tau)$ with $F(\tau+\beta) = F(\tau)$. This degenerate case is often also called parabolic monodromy, although distinct from the parabolic case we study further on.
\\~\\
The full system plus the boundary condition $F(\tau+ \beta) = M \cdot F(\tau)$ is left invariant under $F(\tau) \to S \cdot F(\tau), \, \forall \tau$ for all matrices $S$ that commute with $M$. For $M$ of elliptic monodromy type \eqref{elli}, $S$ is a rotation matrix. Applying a rotation to $F(\tau)$, one finds a U(1) symmetry transformation $f \to f + {\rm const}$, rigid rotations around $S^1$. This is hence related to the ${\rm Diff}(S^1)/$U(1) orbit of the Virasoro group. The size of the U(1) stabilizer circle is related to the parameter in the monodromy matrix by $R_{\text{U(1)}}=\beta/\theta$. In the special case that $\theta \in \mathbb{Z}$, the entire $\sltr$ is a gauge redundancy leading to the SL$^n(2,\mathbb{R})$ orbits.

\subsection{AdS$_2$ perspective}
We now illustrate that this type of defect is holographically connected to a conical deficit in the bulk geometry. In terms of the conformal gauge bulk metric
\begin{equation}
ds^2 = \frac{\partial_+ X^+ \partial_- X^-}{(X^+(x^+)-X^-(x^-))^2}dx^+dx^- = e^{2\phi(x^+,x^-)}dx^+dx^-,
\end{equation}
describing a patch of AdS$_2$, one can extrapolate the boundary reparametrization into the bulk using $X^{\pm}(x) = F(x)$.\footnote{Of course, other bulk extrapolations exist, corresponding to choosing a bulk gauge, but the presence (or absence) of a conical singularity is a diff-invariant property, and we are hence free to choose the extrapolation that is most convenient for us.} 
The relation $F(t) = \tan \frac{\pi}{\beta} \theta f(t)$ then leads to the bulk frame $X^{\pm}= \tan \frac{\pi}{\beta} \theta x^{\pm}$ with orbit parameter $\theta$. One finds the bulk metric
\begin{equation}
\label{saddlem}
ds^2 = \frac{dx^+dx^-}{\sinh^2 \frac{\pi}{\beta} \theta (x^+-x^-)},
\end{equation}
with $\frac{x^++x^-}{2} \sim \frac{x^++x^-}{2} + \beta$, the thermal identification. Close to the horizon $x^+-x^- \to +\infty$, setting $r = e^{-  \frac{\pi}{\beta} \theta (x^+-x^-)}$, one obtains the metric:
\begin{equation}
ds^2 \approx r^2 d\tau^2 + \left(\frac{\beta}{2\pi \theta}\right)^2dr^2, \quad \tau \sim \tau + \beta.
\end{equation}
With the time coordinate $\tilde{\tau} = \frac{2\pi \theta}{\beta} \tau$, this is flat space with a conical identification $2\pi \theta$. The case where $\theta \in \mathbb{N}$ consists of replicated geometries. The metric \eqref{saddlem} describes a Liouville worldsheet metric with a local puncture, interpreted as a conical deficit in JT gravity. The fact that the gravitational bulk carries the Liouville metric identifies the bulk as the kinematic space of the boundary Schwarzian \cite{Mertens:2017mtv,Callebaut:2018nlq,Callebaut:2018xfu}. \\
The total energy $T(\tau)$ of this configuration is
\begin{equation}
\label{clfirst}
T(\tau) = C \left\{F,\tau\right\} = \frac{2 \pi^2 C}{\beta^2}\theta^2.
\end{equation}
We start anew with the JT path integral and insert a defect of the dilaton as follows:\footnote{This seems to define a non-diff invariant operator location $y$. However, the result will only depend on the topological sector of the location $y$ in the bulk diagram.} 
\begin{equation}
\label{conpath}
\int \left[\mathcal{D} \Phi \right] \left[\mathcal{D} g_{\mu\nu} \right]e^{\alpha \Phi(y)} e^{-\frac{1}{16\pi G}\int \Phi(R+2) + S_{\text{bdy}}}.
\end{equation}
Path-integrating $\Phi$ results in the constraint $R(x)=-2 + \alpha 16\pi G\delta(x-y)$, which is AdS$_2$ space with a conical singularity at the prescribed point $y$.\footnote{As emphasized around \eqref{tors}, one can always choose a global $G$-frame to write the defect in this way.} The thermal path integral with this insertion can be viewed as defining a conical ensemble: all off-shell configurations have to respect the presence of the conical deficit. \\
This conical defect is measured by taking an infinitesimal disk surrounding the defect and computing:
\begin{equation}
\label{GB}
\int R = \alpha 16\pi G = 4\pi (1-\theta),
\end{equation}
where the Gauss-Bonnet theorem for a conical deficit $2\pi(1-\theta)$ was used. This equation provides a direct relation between the parameter $\alpha$ and the orbit parameter $\theta$. This identifies the JT path integral with given $\Phi$-insertions, and the \emph{classical} Liouville problem of finding negatively curved metrics with given set of punctures. This problem, instead of being the semi-classical approximation, is lifted to the full quantum regime.

\subsection{Correlators}\label{sec:ellcorr}
In this section we will summarize some results for observables in the analogue of the Schwarzian theory when the integration manifold is over an elliptic coadjoint orbit of the Virasoro group.
\\~\\
 We will begin with the case of ${\rm Diff}(S^1)/$U(1) with non-integer size $\theta$. This theory can be solved by relating it to a semiclassical finite temperature limit of 2d Liouville theory between a ZZ identity brane and an $\mathfrak{i}$FZZT brane, with details explained in \cite{Mertens:2017mtv}. The size of the U(1) group $\theta$ is related to the Liouville momentum $s$ labeling the $\mathfrak{i}$FZZT brane. We will present the final results in this section and give more details on this perspective in section \ref{sec:verlindeloop}. \\
To begin, we will present the result for the exact partition function for the $\mathcal{M}_\theta \equiv {\rm Diff}(S^1)/ $U(1) elliptic coadjoint orbit
\begin{eqnarray}
\label{dosell}
Z(\beta,\theta) &\equiv&\int_{\mathcal{M}_\theta} [\mathcal{D}f]~e^{C \int_0^\beta d\tau \{ \tan{ \frac{\pi}{\beta} \theta f(\tau)}, \tau\} }\nonumber\\
&=& 2\int_{0}^{+\infty}dk \cosh(2\pi \theta k) e^{-\beta \frac{k^2}{2C}} = \left( \frac{2 \pi C}{\beta} \right)^{1/2} e^{\frac{2 \pi^2 C}{\beta} \theta^2},
\end{eqnarray}
and therefore the density of states for a U(1) orbit is $\mu_{\theta}(k) = 2\cosh (2 \pi \theta k )$. To obtain this result, start from the 2d Liouville setup between a ZZ-brane and an $\mathfrak{i}$FZZT brane. This can be done either using the wavefunctions of the boundary states, or realizing the 2d partition function is simply a Virasoro character, and taking the semiclassical limit expained in \cite{Mertens:2017mtv}. This partition function was independently computed by \cite{wittenstanford} using the Duistermaat-Heckman theorem.  
\\~\\
At special values of $\theta = n \in \mathbb{N}$, the gauge redundancy gets enhanced and one finds the $\mathcal{M}_n \equiv {\rm Diff}(S^1)/\slnr$ orbit. Even though the action is the same, the space of integration for $f(t)$ is completely different, and therefore for example the partition function is not equivalent to replacing $\theta \to n$. From the Liouville perspective, we need to replace an $\mathfrak{i}$FZZT brane by a ZZ brane with label $(1,n)$. Taking the appropriate limit gives the partition function
\begin{eqnarray}
\label{dospsl}
Z(\beta,n) &\equiv&\int_{\mathcal{M}_n} [\mathcal{D}f]~e^{C \int_0^\beta d\tau \{ \tan{ \frac{\pi}{\beta}n f(\tau)}, \tau\} }\nonumber\\
&=&  \int_{0}^{+\infty}dk^2 \sinh(2\pi n k) e^{-\beta \frac{k^2}{2C}} =n \left( \frac{2 \pi C}{\beta} \right)^{3/2} e^{\frac{2 \pi^2 C}{\beta} n^2},
\end{eqnarray}
and therefore the density of states for the $\slnr$ orbits is $\mu_n(k) = 2k \sinh(2 \pi n k)$.\footnote{Note that due to the symmetry enhancement $\mu_{\theta\to 1}(k) \neq \mu_{n=1}(k)$.} This coincides with the previously studied ${\rm Diff}(S^1)/\sltr$ orbit when $n\to1$ and generalizes this result. The power $3/2$ in the one-loop contribution comes from the number of zero-modes. When $n\neq 1$ these modes are unstable, making a perturbative formulation of the theory ill-defined \cite{Maldacena:2016upp}. This is related to the fact that this setup corresponds to a Hilbert space that is not unitary in 2d (see footnote \ref{foot}), see also \cite{Raeymaekers:2014kea}. Nevertheless, our 2d perspective still gives a good non-perturbative definition of the $\slnr$ Schwarzian as a limit of a 2d CFT. 
\\~\\
Following \cite{Mertens:2017mtv}, we can also compute expectation values of bilocal fields that are related to the Liouville field $\phi$. A correlator of Liouville primary operators can be used to compute expectation values of 
\begin{equation}\label{defbilocalell}
\mathcal{O}_\ell (\theta;\tau_1,\tau_2 ) = \left( \frac{ \dot{f}(\tau_1) \dot{f}(\tau_2)}{\frac{\beta^2}{\pi^2\theta^2} \sin^2 \frac{\pi}{\beta} \theta [f(\tau_1) -f(\tau_2) ]}\right)^{\ell}, 
\end{equation} 
in a way completely analogous to the case studied in \cite{Mertens:2017mtv}. 
 \\~\\
The single bilocal correlator in the case of the U(1) orbits is given by\footnote{This expression is normalized by the $n=1$ partition function, since we view these other models as deformations of the vacuum $\sltr$ orbit. For the elliptic orbits (and the parabolic orbit in section \ref{sect:para}), one could instead view the twisted partition functions \eqref{dosell} and \eqref{dospsl} as valid partition functions in their own right, and normalize by these instead, but such an interpretation will not be possible anymore for the generic hyperbolic orbits further on. An exception is the specific hyperbolic orbit related to global AdS$_2$ which we discuss in section \ref{sec:global}.} 
\begin{equation}
\label{correliptic2}
\langle \mathcal{O}_\ell (\theta; \tau_1,\tau_2 ) \rangle = \frac{2}{Z}\int dk_1^2 dk_2 \sinh(2\pi k_1) \cosh(2\pi \theta k_2)  \frac{\Gamma(\ell \pm i k_1 \pm i k_2)}{\Gamma(2\ell)} e^{-\tau \frac{k_1^2}{2C} - (\beta-\tau) \frac{k_2^2}{2C}}.
\end{equation}
Using the techniques of \cite{Lam:2018pvp}, the semiclassical (large $C$) regime is readily shown to be governed by a saddle at energy $M = \frac{k_2^2}{2C} = \frac{2\pi^2 C}{\beta^2}\theta^2$, in agreement with \eqref{clfirst}. On the other hand, we can take the strongly coupled limit of small $C$. This gives a universal power-law behavior, with power independent of $\ell$, as
\begin{equation}
\label{qcorrel}
\langle \mathcal{O}_\ell (\theta;\tau_1,\tau_2 ) \rangle \sim \frac{C^2}{\tau^{3/2} (\beta- \tau)^{1/2}}.
\end{equation}
The origin of these powers is in the low-energy asymptotics of the density of states. This result can in turn be used to deduce the long-time limit of the Lorenzian correlator. At finite temperature the correlator decays as $\langle \mathcal{O}_\ell (\theta; \tau_1,\tau_2 ) \rangle \sim 1/\tau^2$ while at zero temperature it decays as $\langle \mathcal{O}_\ell (\theta; \tau_1,\tau_2 ) \rangle\sim 1/\tau^{3/2}$. 
\\~\\
For the case of the orbits $\slnr$, the bilocal observable is given by the same formula in terms of $f(\tau)$ replacing $\theta \to n$, even though its expectaction value is different due to the symmetry enhancement. It is given by 
\begin{equation}
\label{correliptic1}
\langle \mathcal{O}_\ell (n; \tau_1,\tau_2 ) \rangle = \frac{1}{Z}\int dk_1^2 dk_2^2 \sinh(2\pi k_1) \sinh(2\pi n k_2) \frac{\Gamma(\ell \pm i k_1 \pm i k_2)}{\Gamma(2\ell)} e^{-\tau \frac{k_1^2}{2C} - (\beta-\tau) \frac{k_2^2}{2C}}.
\end{equation}
For these correlators we can verify the semiclassical large $C$ limit. We can also take the strongly coupled limit of small $C$. Nevertheless since for small energies $\mu_n (k) \sim 4 \pi n k^2 $, the long-time power laws are the same as for the $\sltr$ case, up to some $n$-dependent prefactors, scaling as $\sim 1/\tau^3$ at finite temperature and $\sim 1/\tau^{3/2}$ at zero temperature.
\\~\\
For the case $n=1$, these bilocal operators are the natural ones as dictated by the $\sltr$ symmetry, and a similar thing can be said about the $n \in \mathbb{N}$ case. Nevertheless, for the U(1) orbits the only constraints on observables is to preserve an $f$ shift symmetry. Therefore by computing \eqref{defbilocalell}, we are making a very specific choice of observable. From a physical perspective (either integrating out matter in the bulk or interpreting this Euclidean theory as a finite temperature QM), it is reasonable to restrict ourselves to observables satisfying the KMS condition under $\tau\to\tau+\beta$. In order to do this, we can define a deformation of the bilocal $ \mathcal{O}_\ell$ in the following way 
\begin{equation}\label{defbilocalellkms}
\tilde{\mathcal{O}}_\ell (\theta; \tau_1,\tau_2 ) = \sum_{w\in \mathbb{Z}} \left( \frac{ \dot{f}(\tau_1) \dot{f}(\tau_2)}{  \frac{\beta^2}{\pi^2 \theta^2} \sin^2 \frac{\pi}{\beta} \theta [f(\tau_1) -f(\tau_2) + w \beta]}\right)^{\ell}, 
\end{equation} 
which using that $f(\tau+\beta)=f(\tau) + \beta$ is equivalent to $\sum_{w\in \mathbb{Z}} \mathcal{O}_\ell (\theta; \tau_1+w \beta,\tau_2 )$. For elliptic orbits, this is formally divergent, but we will see these observables are well-defined for hyperbolic orbits, see section \ref{sect:hyper}. 

\subsection{Verlinde loop operators in the ZZ-ZZ system}\label{sec:verlindeloop}
In this section we will comment more on the structure of the results obtained for the deformed Schwarzian theory. The partition functions \eqref{dosell},\eqref{dospsl} and correlators \eqref{correliptic2}\eqref{correliptic1} are the direct Schwarzian generalization of the diagrammatic rule for compact groups explained in section \ref{sec:compact}, where adding a chemical potential to the particle-on-a-group problem requires the insertion $\chi_R(U_\lambda)/\text{dim }R$. 
\\~\\
More explicitly, this means that Schwarzian correlators for an orbit $H$ can be written by using the rules of the $n=1$ vacuum orbit of \cite{Mertens:2017mtv}, with the insertion of an additional factor
\begin{eqnarray}
H=\slnr ~&\to&~ D_{\slnr}(k)\equiv \frac{\sinh( 2 \pi n k)}{\sinh(2 \pi k)}, \\
H=\text{U(1)}~&\to&~D_{\text{U(1)}_\theta}(k) \equiv \frac{\cosh(2 \pi \theta k)}{k \sinh(2\pi k)},
\end{eqnarray} 
which takes into account the deformation to the different orbits. These factors are simply the Schwarzian limits of the terms needed to move between different boundary states in equations \eqref{eq:ZZnbs} and \eqref{eq:FZZTbs}. \\
Therefore, for example the two-point function for the orbit $H=\slnr$ is given by 
\begin{equation}
\langle \mathcal{O}_\ell (n,\tau_1,\tau_2) \rangle =\int d\mu(k_1) d\mu(k_2) ~D_{\slnr}(k_2) ~ \frac{\Gamma(\ell\pm ik_1 \pm ik_2)}{\Gamma(2\ell)}e^{-\tau \frac{k_1^2}{2C}}e^{-(\beta-\tau) \frac{k_2^2}{2C}},
\end{equation}
where $d\mu(k)=2k \sinh(2 \pi k)$. This can be similarly done for U(1) with the appropriate insertion $D_{U(1)}(k_2)$. We see very clearly here that $n$ or $\theta$ play the role of a chemical potential by comparing to the compact group case. \\
The same structure will arise for the parabolic and hyperbolic cases considered in the next sections. In the rest of this section we will interpret these defects from the 2d CFT perspective as inserting Verlinde loop operators. 
\\~\\
Using results from \cite{Drukker:2010jp} (based on the analysis of \cite{Dijkgraaf:1988tf}), one can show that the effect of an insertion of a Verlinde loop operator (or equivalently a topological defect operator) surrounding a boundary is to modify its boundary condition (or boundary state) \cite{LeFloch:2017lbt}. Verlinde loop operators are labeled by primary operators in the same way as Cardy boundary states. In the case of Liouville, one can start with the identity ZZ-brane. Different degenerate ZZ-branes or FZZT-branes can be obtained by the insertion of a Verlinde loop operator associated to that Virasoro representation. For example, in the FZZT case, this way one can go from \eqref{eq:ZZnbs}, with $n=1$, to \eqref{eq:FZZTbs}, by multiplying the boundary state by a ratio of modular S-matrices.

\begin{figure}[h!]
\centering
\includegraphics[width=0.5\textwidth]{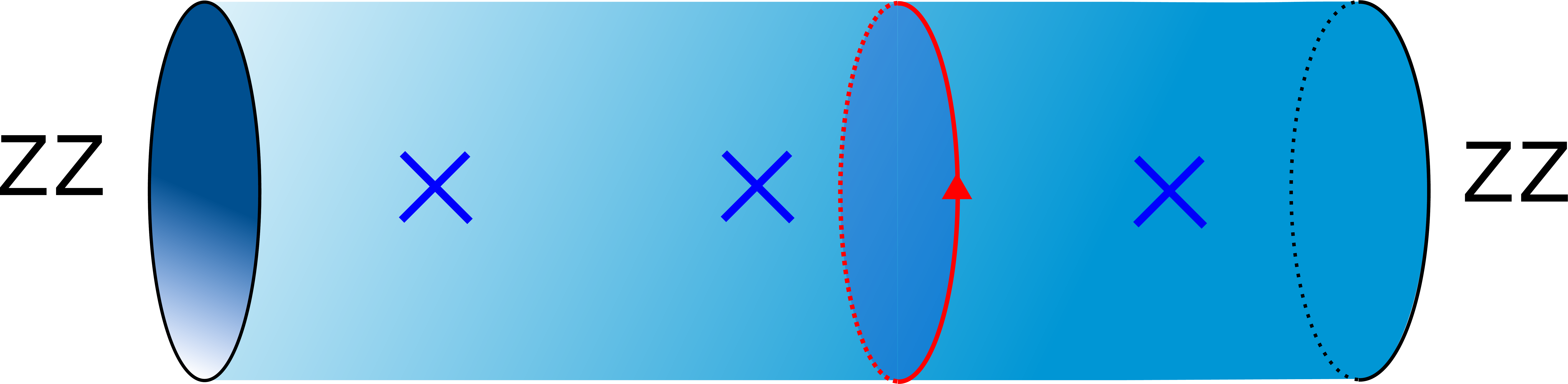}
\caption{Inserting Verlinde loop operators (red) between Liouville primary insertions (blue crosses) gives defect operators in the JT bulk.}
\label{verlindeop}
\end{figure}

\begin{figure}[h]
\centering
\includegraphics[width=0.85\textwidth]{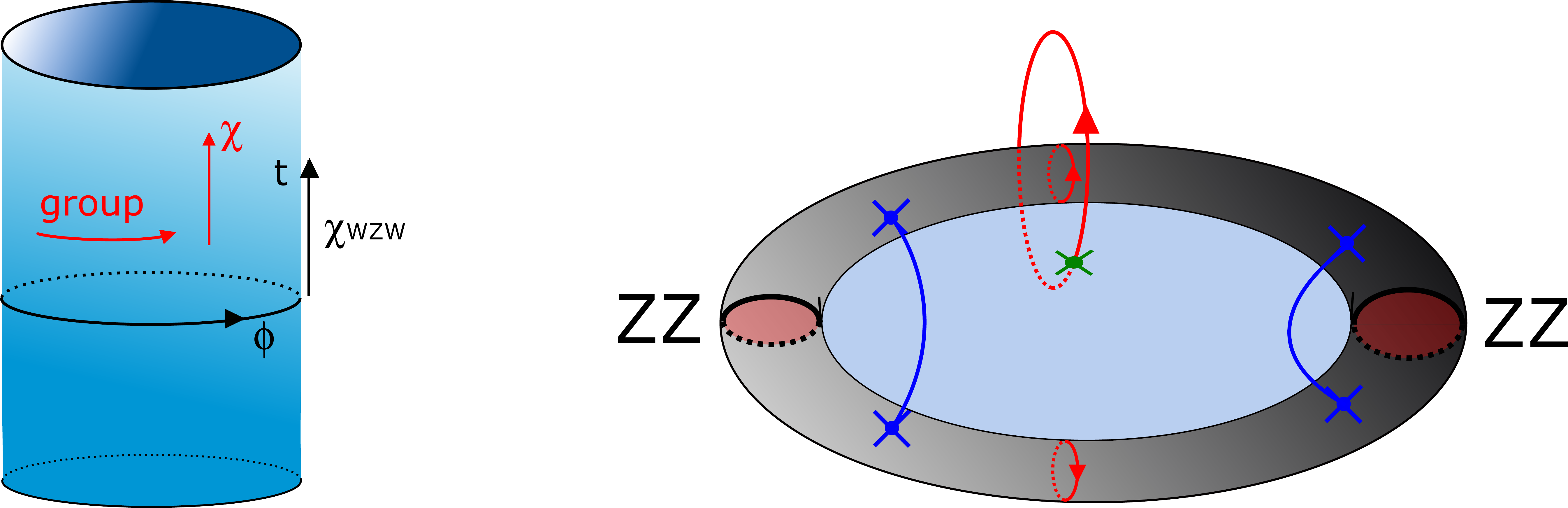}
\caption{Left: Chiral WZW model ($\chi$WZW) and its dimensional reductions along both axes. The quantum particle on the group manifold ($\partial_t = 0$) appears in the dual channel of the original chiral WZW model, which by itself reduces to the character $\chi$ of the group $G$ ($\partial_\phi = 0$). See Appendix \ref{app:compact} for the equations. Right: Liouville between ZZ-branes leads to a torus surface (grey) with mirrored operator insertions. The holographic bulk is found as the exterior of the torus. In the Schwarzian limit, the torus degenerates, the bulk becomes independent of the angular coordinate (rotating around the grey tube) and can be viewed as a disk (lightblue) bounded by the degenerate torus. Primary operator insertions (blue) lead to bilocal lines in the disk. Verlinde loop operators (red) are topologically supported. Deforming them into the bulk, they are Chern-Simons Wilson lines puncturing the BF bulk disk at an arbitrary point (green).}
\label{geometry}
\end{figure}

\begin{figure}[h]
\centering
\includegraphics[width=0.35\textwidth]{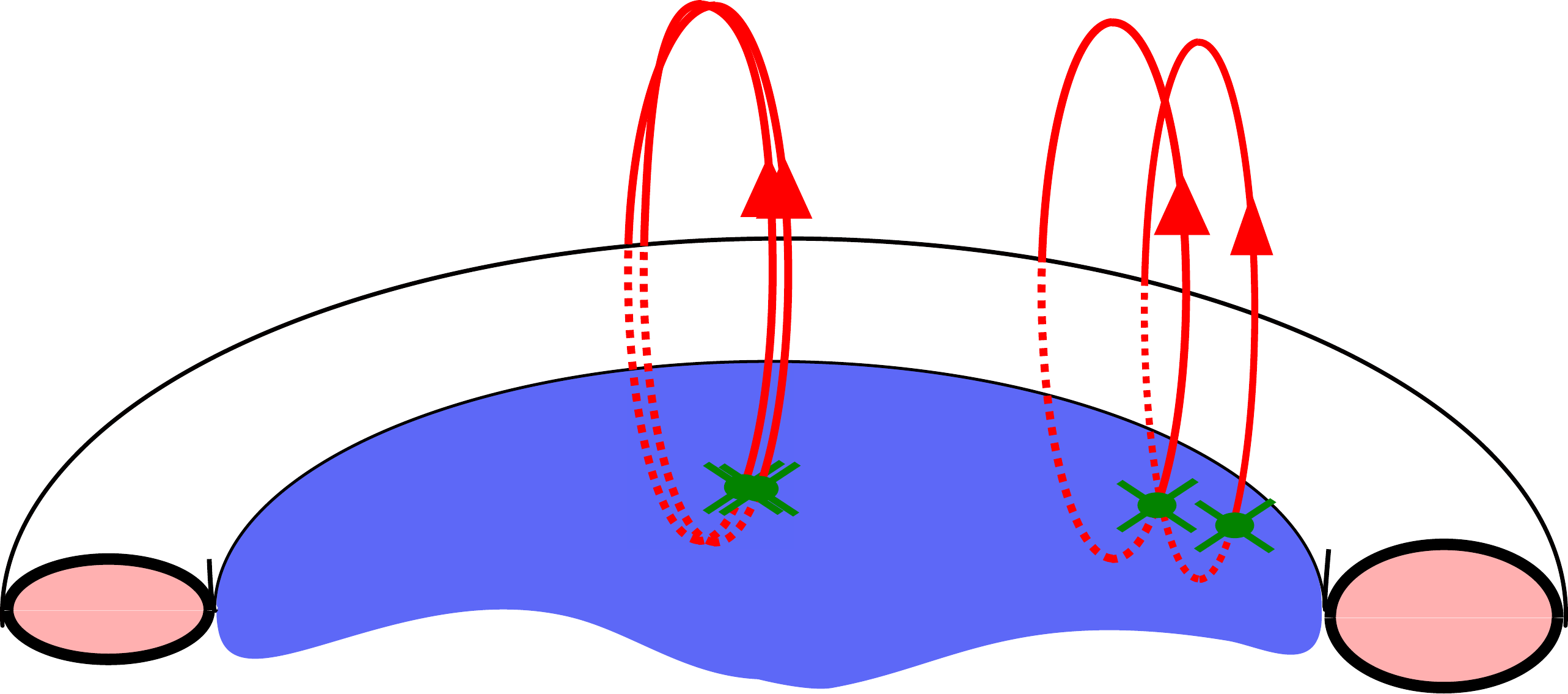}
\caption{Multiply wound Verlinde loop operators and multiple loops.}
\label{mwound}
\end{figure}

\noindent This immediately leads to a generalized situation where we can insert Verlinde loop operators in between \emph{any} Liouville operators (Figure \ref{verlindeop}), not just adjacent to the boundary ZZ-brane. Within the bulk JT theory, this corresponds to inserting a conical defect in the bulk between the bilocal lines. E.g.
\begin{equation}
\label{striped}
\begin{tikzpicture}[scale=0.75, baseline={([yshift=0cm]current bounding box.center)}]
\draw[thick] (0,0) circle (1.5);
\draw[thick] (1.06,1.06) arc (300:240:2.16);
\draw[thick] (-1.06,-1.06) arc (120:60:2.16);
\draw[fill,black] (-1.06,-1.06) circle (0.1);
\draw[ultra thick] (-0.1,0.3)--(0.1,0.1);
\draw[ultra thick] (-0.1,0.1)--(0.1,0.3);
\draw[thick] (-1.5,0)--(1.5,0);
\draw[thick] (-1.4,0.539)--(1.4,0.539);
\draw[thick] (-1.4,-0.539)--(1.4,-0.539);
\draw[fill,black] (-1.4,-0.539) circle (0.1);
\draw[fill,black] (-1.4,0.539) circle (0.1);
\draw[fill,black] (1.4,-0.539) circle (0.1);
\draw[fill,black] (1.4,0.539) circle (0.1);
\draw[fill,black] (1.06,-1.06) circle (0.1);
\draw[fill,black] (-1.06,1.06) circle (0.1);
\draw[fill,black] (1.06,1.06) circle (0.1);
\draw[fill,black] (1.5,0) circle (0.1);
\draw[fill,black] (-1.5,0) circle (0.1);
\end{tikzpicture}
\end{equation}
It is funny to realize that local Liouville punctures map to lines in JT, whereas nonlocal Verlinde operators map to local punctures in JT. As a final comment, the role of these $D$-factor insertions was also studied in \cite{LeFloch:2017lbt} in the context of the AGT dual of Liouville with boundary conditions. Then for example $D_{\slnr}(k)$ is simply an SU(2) character for a representation of spin $\frac{n-1}{2}$. This arises from the insertion of a Wilson line in the 4d gauge theory associated to that representation, as explained in \cite{Drukker:2010jp}. Similarly, the insertion $D_{\text{U(1)}_\theta}(k)$ corresponds to inserting a symmetry breaking wall with FI terms labeled by $\theta$. 
\\~\\
The full geometrical relation between the boundary ZZ-ZZ cylindrical surface and the JT bulk with defects is summarized in Figure \ref{geometry}.
As a holographic system, the JT / BF bulk lives in the \emph{exterior} of the boundary torus. The reason is that the chiral WZW system and the Virasoro coadjoint orbit action are taken to the particle-on-a-group model and the Schwarzian respectively, upon swapping the role of time $t$ and space $\sigma$ of the torus. This swapping can be viewed as a modular $S$-transform that effectively changes between the interior and the exterior of the boundary ZZ-ZZ system (mirror doubled into a torus), which in the Schwarzian double-scaling limit degenerates into a long thin tube. Verlinde loop operators are Chern-Simons Wilson lines associated to $\chi$ in the dimensionally reduced BF-action $\text{Tr} \chi F$.
\\~\\
Two elementary variants of loop operators were considered in \cite{Drukker:2009id} (Figure \ref{mwound}).
Multiply wound Verlinde loops give a single BF puncture of the form $\chi_R(M^n)$, with $n$ the number of windings. Such loops are viewed on the BF bulk as a single puncture pierced multiple times.\footnote{The angular zero-mode does not allow one to pierce the BF disk in different locations with a multiply wound Chern-Simons loop.} On the other hand, considering multiple Verlinde loops leads to integer powers $\chi_R(M)^n$. \\
Applied to the Schwarzian case at hand, the difference is in the measure factor $D(k)$. E.g. for the elliptic defect of \eqref{dosell} one has:
\begin{eqnarray}
&&D^{\rm n-wound}_{\text{U(1)}}(k) = \frac{\cosh(2\pi n\theta k)}{k\sinh( 2\pi k)},  \quad n \text{  wounds}, \\
&&D^{\rm n-loops}_{\text{U(1)}}(k)=\Big(\frac{\cosh(2\pi \theta k)}{k \sinh(2\pi k)}\Big)^n, \quad n \text{ separate loops}.
\end{eqnarray}
Moreover one can also study combinations of Verlinde loop insertions, where one inserts loops of different types, resulting in inserting factors of $D_n(k)D_\theta(k)$, and each with different windings etc. 

\subsection{Application: Horizon area operator}
\label{sect:areaop}
In the context of 2d dilaton-gravity, the role of the black hole horizon area is played by the horizon value of the dilaton $\Phi_h$. This can be understood either from the dimensional reduction of a near extremal black hole in higher dimensions, or from the fact that the effective Newton constant is $G_{\text{eff}} = G/\Phi_h$. For a thermal state dual to a black hole, the entropy is, to leading order in small $G_N$, given by
\begin{equation}
S_{\text{BH}} = \frac{\Phi_h}{4G_N}.
\end{equation}
Besides the thermofield double, this relation between entropy and minima of the dilaton was also verified in the semiclassical limit for non-thermal states in \cite{Goel:2018ubv}, for the case of JT gravity. 
\\~\\
Since JT gravity is solvable, it would be nice to embed this semi-classical result within a fully quantum-mechanical framework. As a first approach to this problem, in this section we will define a non-perturbative operator $\hat{A}$, with the property that its expectation value is given by the semi-classical minimum of the dilaton.

We will define the horizon area operator $\hat{A} \equiv \hat{\Phi}_h$ operationally. We will begin by computing its expectation value, but this can be extended to a more general case. In order to measure $\langle \hat{A} \rangle$ we will insert a factor of $e^{\alpha \Phi(y)}$ in the gravitational path integral as in \eqref{conpath}, which creates a conical deficit at the point where we want to measure the transverse area.\footnote{Since the theory is gravitational, a local diff-invariant bulk operator that measures the value of the dilaton locally, would have to be of the form $\Phi(g(y))$, which is not what we study here. Our result only depends on the topological sector of the label $y$ within the diagram and as such this is a \emph{global} observable.}  Then we will differentiate the partition function $Z_{\text{U(1)}}(\beta,\alpha)$ \eqref{conpath} with respect to $\alpha$ and set $\alpha = 0$ in the end. This leads to
\begin{equation}
\langle \hat{A}\rangle = \partial_\alpha \log Z_{\text{U(1)}}(\beta, \alpha) \Big|_{\alpha \to 0} = 4 G_N \partial_\theta \log Z_{\text{U(1)}} (\beta, \theta) \Big|_{\theta \to 1},
\end{equation}
where $\theta$ corresponds to the size of the U(1) group. The horizon expectation value of the dilaton is related to how the partition function changes when changing the conical deficit; this should be related to the entropy.

Now we can make two choices in order to define our operator $\hat{A}$. We can take the derivative by analytically continuing the integer $n$ orbit. This is not a reasonable prescription since the computation does not preserve the full $\sltr$ global symmetry, and the function $Z(\beta,\theta)$ is discontinuous for $\theta=n\in\mathbb{N}$. Since only the U(1) elliptic subgroup ($ f \to  f + {\rm const}$) is preserved by the insertion, we will take the derivative with respect to the ${\rm Diff}(S^1)/$U(1) parameter $\theta$.\footnote{The relation between the $n\in \mathbb{N}$ orbits and the generic elliptic orbits, $\frac{\partial}{\partial \theta} \cosh(2\pi \theta k) = 2\pi k \sinh(2\pi \theta k)$, is a direct consequence of the 2d boundary state equality:
\begin{equation}
\left|\text{ZZ};m,n\right\rangle = \left|\text{FZZT};s(m,n)\right\rangle - \left|\text{FZZT};s(m,-n)\right\rangle,
\end{equation}
where $s = i \left(\frac{m}{b} + nb\right)$. Indeed, this becomes the definition of a derivative:
\begin{equation}
\lim\limits_{b\to 0 } \left[\cosh(2\pi (m k + n b^2 k) )  - \cosh(2\pi (m k - n b^2 k) ) \right] \, = \, 2\pi b^2 n \, k \sinh(2 \pi m k).
\end{equation}
}
\\~\\
Using \eqref{GB}, the expectation value of our operator $\hat{A}$ is 
\begin{equation}
\label{dilavev}
\frac{1}{4G_N} \langle \hat{A} \rangle = \frac{1}{Z_{U(1)}(\beta)}\frac{\partial}{\partial  \theta}\left.\int dk \cosh(2\pi \theta k) e^{-\beta \frac{k^2}{2C}}\right|_{\theta=1} = \frac{4 \pi^2 C}{\beta},
\end{equation}
to be compared to the thermodynamic entropy $S(\beta) = \frac{4\pi^2 C}{\beta} +  \frac{3}{2} + \frac{3}{2}\ln\frac{2 \pi C}{\beta}$, where the first term is the tree-level part, and the remainder is the one-loop piece. Within the JLMS framework \cite{Jafferis:2015del}, the first term should be written as $\langle \hat{A}\rangle_{\beta}/4G$ and the above formula \eqref{dilavev} realizes this construction explicitly in 2d dilaton gravity.\footnote{Such a formula was written down suggestively in \cite{Lin:2018xkj} but without an independent computation of $\left\langle \Phi_h\right\rangle$.} We have in effect written down a suitable quantum operator for the horizon area $\hat{A}$. 
\\~\\
We argued above that in order to compute $\langle \hat{\Phi}_h\rangle$ we should continue the U(1) orbits for $\theta \to 1$ and not the SL${}^{n}(2,\mathbb{R})$ orbits. The difference between both calculations is subleading in the large $C$ limit and we take our prescription as a definition of the $\hat{\Phi}_h$ operator beyond the semiclassical limit. 
\\~\\
A similar calculation can be done for the supersymmetric $\mathcal{N}=1$ generalization of JT gravity, using the results from $\mathcal{N}=1$ Schwarzian theory. Interestingly, in this case the expectation value gives the same answer \eqref{dilavev} for either continuous or discrete orbits, for arbitrary $C$, indeed providing evidence that our definition is the correct one. 
\\~\\
We can distill the energy eigenstate expectation value by inverse Laplace transforming:\footnote{Note that we divide by the $n=1$ $\sltr$ partition function here. The previous computation can be viewed as a trick to get to $\langle \hat{\Phi}_h\rangle_\beta$, but once we have it, it is interesting to find the correct Fourier transformed defect interpretation in any diagram.}
\begin{equation}
\langle \hat{\Phi}_h\rangle_\beta \equiv \frac{1}{Z_{\sltr}}\int dM \sinh(2\pi \sqrt{M}) \langle \hat{\Phi}_h\rangle_M e^{-\beta \frac{M}{2C}},
\end{equation}
from which we obtain
\begin{equation}
\langle \hat{\Phi}_h\rangle_M = 2\pi \sqrt{M} \coth 2\pi \sqrt{M} -1,
\end{equation}
semiclassically becoming indeed $\Phi_h = 2 \pi \sqrt{M}$. This can be used for any correlation function of the type:
\begin{equation}
\left\langle \hat{\Phi}_h \mathcal{O}_\ell (\tau_1,\tau_2) \hdots \right\rangle_\beta = \int dM \sinh(2\pi \sqrt{M}) \langle \hat{\Phi}_h \rangle_M \left\langle \mathcal{O}_\ell (\tau_1,\tau_2) \hdots \right\rangle_M e^{-\beta \frac{M}{2C}},
\end{equation}
upon inserting a complete set of energy eigenstates $\left|M\right\rangle$ and using that the bilocal is diagonalized in this basis \cite{Mertens:2017mtv}. Hence, in analogy with the case studied in previous sections, the horizon area operator $\hat{\Phi}_h$ can be interpreted as a defect implemented by the insertion:
\begin{equation}
\label{areadef}
D_A(k) = 2\pi k \coth 2\pi k -1,
\end{equation}
in the region of interest in the diagram, readily generalizing this to situations with the defect between both ends of a bilocal. This kind of expression, computing the minimal value of the dilaton in any region we choose, was constructed semiclassically in \cite{Goel:2018ubv} and interpreted in terms of the interior of the black hole. Since the defect \eqref{areadef} does not alter the saddle point of the momentum integrals, the above expression indeed reduces to the discussion of \cite{Goel:2018ubv} in the heavy semi-classical limit $\ell \sim C \to +\infty$.\footnote{For the $\mathcal{N}=1$ super-Schwarzian model, the analogous defect expression is $D_A(k) = 2\pi k \tanh 2 \pi k$.}
\\~\\
With corrections to the Bekenstein-Hawking entropy in mind, it would be interesting to understand $\left\langle \log A\right\rangle$ and negative powers of $\hat{A}$. We deduce these expectation values in appendix \ref{app: f(A)} and postpone an interpretation to possible future work.

\section{Hyperbolic orbits $U(1)_\lambda$: multiboundary geometries and gluing}
\label{sect:hyper}
Setting $F(\tau) = \tanh \frac{\pi}{\beta} \lambda f(\tau)$, with $f(\tau+\beta) = f(\tau) + \beta$ leads to hyperbolic monodromy:
\begin{equation}
F(\tau+\beta) = \tanh\left( \frac{\pi}{\beta} \lambda f(\tau) + \pi \lambda \right) = \frac{F(\tau) + \tanh(\pi \lambda)}{1+ \tanh(\pi \theta) F(\tau)},
\end{equation}
with hyperbolic $M$:
\begin{equation}
\label{hypmo}
M = \left(\begin{array}{cc}
\cosh(\pi \lambda) & \sinh(\pi\lambda) \\
\sinh(\pi\lambda) & \cosh(\pi \lambda) \\
\end{array}\right) \quad \in \text{P}\sltr.
\end{equation}
In a way similar to the previous section, we can completely solve the deformation of the Schwarzian theory associated with the hyperbolic U(1) orbits. Since the analysis is very similar, we will only point out here that most formulas can be obtained from the elliptic U(1) by the analytic continuation $\theta \to i \lambda$. E.g. the twisted partition function is written as
\begin{eqnarray}
\label{doshyp}
Z(\beta,\lambda) &\equiv&\int_{\mathcal{M}_\lambda} [\mathcal{D}f]~e^{C \int_0^\beta d\tau \{ \tanh{ \frac{\pi}{\beta} \lambda f(\tau)}, \tau\} }\nonumber\\
&=& 2\int_{0}^{+\infty}dk \cos(2\pi \lambda k) e^{-\beta \frac{k^2}{2C}} = \left( \frac{2 \pi C}{\beta} \right)^{1/2} e^{-\frac{2 \pi^2 C}{\beta} \lambda^2}.
\end{eqnarray}
Due to the non-positive ``density of states'' $\rho(k) \sim \cos(2\pi \lambda k)$, this system cannot be viewed as a thermal system on its own, and should instead be interpreted as a system with density of states $\rho(k) =  2 k \sinh 2 \pi k$ in the presence of a defect.

In a similar way as we did for the elliptic case, we can compute correlators. The natural observable from the Liouville perspective is 
\begin{equation}\label{defbilocalh}
\mathcal{O}_\ell (\lambda; \tau_1,\tau_2 ) =\left( \frac{ \dot{f}(\tau_1) \dot{f}(\tau_2)}{ \frac{\beta^2}{\pi^2 \lambda^2}\sinh^2 \frac{\pi}{\beta} \lambda [f(\tau_1) -f(\tau_2) ]}\right)^{\ell}, 
\end{equation} 
The result for expectation values $\langle \mathcal{O}_\ell \rangle$ of these bilocal observables is given by the $n=1$ expressions with an insertion of the hyperbolic factor $D_{U(1)_\lambda}(k)= \frac{\cos 2 \pi \lambda k}{k \sinh 2 \pi k}$. As mentioned in the end of section \ref{sec:ellcorr}, this observable cannot arise from free matter in the bulk since it does not satisfy the KMS condition. Therefore we define an operator satisfying the KMS condition as 
\begin{equation}\label{defbilocalh2}
\tilde{\mathcal{O}}_\ell (\lambda; \tau_1,\tau_2 ) = \sum_{w\in \mathbb{Z}} \left( \frac{ \dot{f}(\tau_1) \dot{f}(\tau_2)}{ \frac{\beta^2}{\pi^2 \lambda^2} \sinh^2 \frac{\pi}{\beta} \lambda [f(\tau_1) -f(\tau_2) + w \beta]}\right)^{\ell}.
\end{equation} 
As opposed to the elliptic case, now this observable is convergent. This correlator respects both the global U(1) symmetry $f\to f+{\rm const}$, and the KMS condition associated to a finite temperature correlator. As expected we can see from this expression how the effect of the winding modes dissappear in the zero temperature limit. \\
For a fixed $w \neq 0$, this correlator describes a bilocal line encircling the defect in the bulk $w$ times. As such, following the rules of \cite{Mertens:2017mtv}, the line self-intersects and is associated to a 6j-symbol. It is possible to write down the expression, but we have yet to find an enlightening way of writing or summing this expression, which we leave to future work.
\\~\\
In the rest of this section we will point out interesting aspects mostly related to the role of these hyperbolic defects in the context of AdS$_2$ holography.

\subsection{AdS$_2$ perspective}
\begin{figure}[h]
\centering
\includegraphics[width=0.85\textwidth]{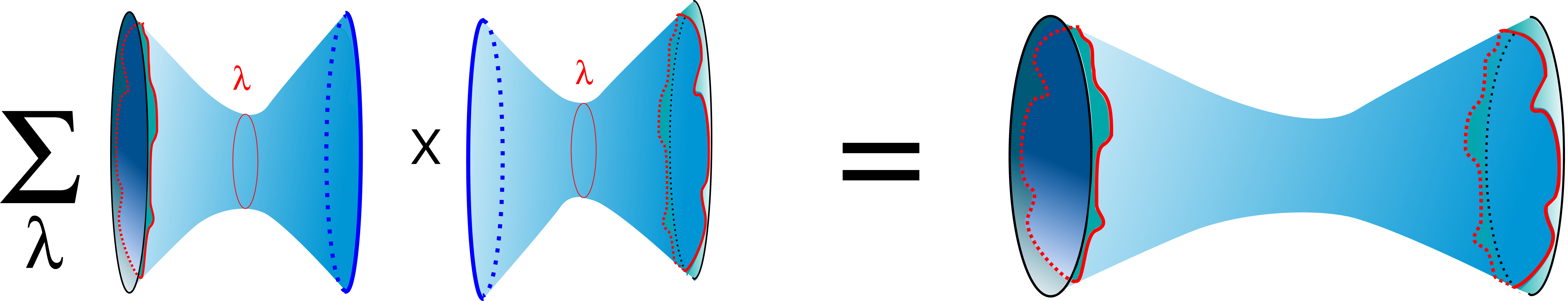}
\caption{Gluing disks with hyperbolic defect operators. Summing over orbits leads to a glued 2-boundary geometry. A single disk with a fixed hyperbolic defect only has a single fluctuating (twisted) Schwarzian boundary.}
\label{hyperbolicorbit}
\end{figure}

The hyperbolic defect geometry is wormhole-like with a minimal neck length.  Indeed, by the same argument as in \eqref{saddlem}, the bulk geometry is of the form:
\begin{equation}
\label{globalgeom}
ds^2 = 4 \left(\frac{\pi \lambda}{\beta}\right)^2\frac{d\tau^2 + dz^2}{\sin^2\frac{2\pi}{\beta} \lambda z},
\end{equation}
with boundaries at $z=0, \frac{\pi}{\lambda} \frac{\beta}{2\pi}$. The hyperbolic monodromy defect is a two-boundary geometry (Figure \ref{hyperbolicorbit}).
Within JT gravity, these can be used to glue two separate Euclidean manifolds together. One inserts matching local operators and then sums over all such hyperbolic operators.\footnote{Probably ``local'' is a misnomer, as they can be viewed as the Liouville hyperbolic insertions.} This is also the procedure to create a tube / wormhole in the effective action \cite{Coleman:1988cy,ArkaniHamed:2007js}. This logic was used to glue two disks to create an annulus in \cite{Blommaert:2018iqz} and to get generic multi-boundary systems \cite{sss2}. This again identifies the bulk JT model with 2d Liouville CFT, where cutting and gluing is achieved by inserting complete sets of precisely these hyperbolic operators, forming the Liouville spectrum.
\\~\\
The circumference of the neck of the wormhole computed halfway at the value of $z$ where the scale factor in \eqref{globalgeom} is minimal, is then given by
\begin{equation}
\ell = 2 \left(\frac{\pi \lambda}{\beta}\right) \beta = 2 \pi \lambda.
\end{equation}
We will demonstrate that this can be measured quantum-mechanically by applying a Wilson loop in the $\sltr$ BF description.
 
\subsection{Application: Length Operator}
\begin{figure}[h]
\centering
\includegraphics[width=0.35\textwidth]{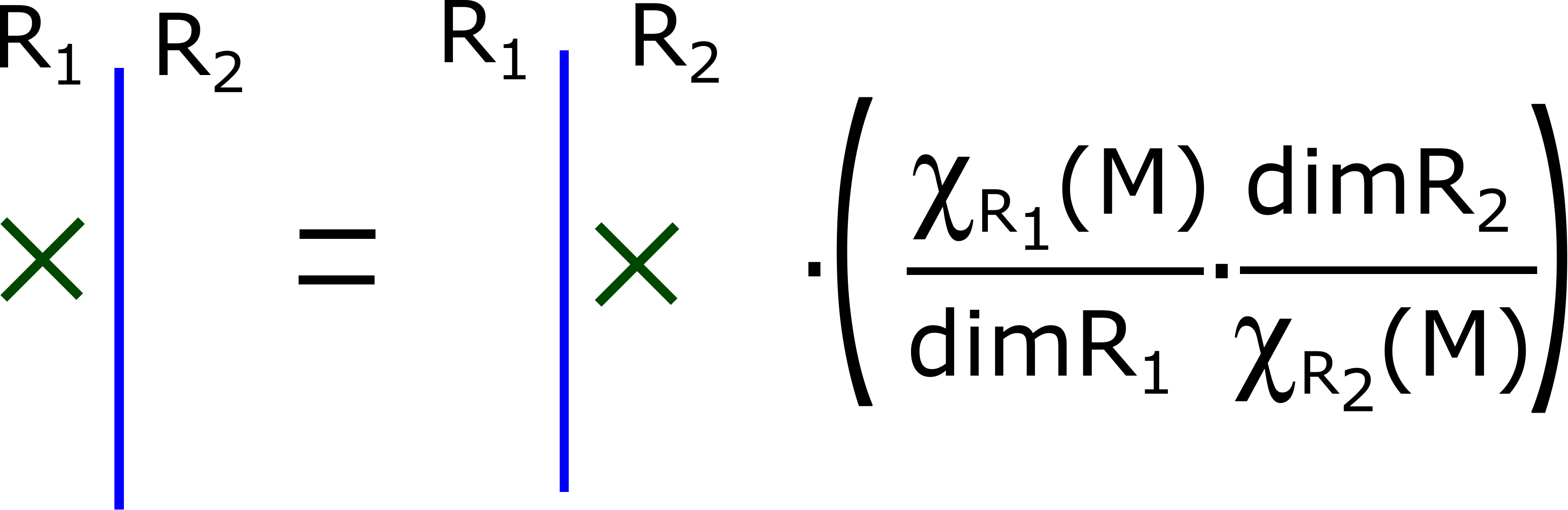}
\caption{Moving the Wilson lines through a puncture (cross) of holonomy $M$ produces an additional factor in the representation basis.}
\label{cross}
\end{figure}

The rule for including a defect in the diagram, can be used to obtain a rule for moving a defect through a Wilson line, i.e. to another sector of the diagram. Let us first write the expressions for a compact group $G$ BF model. The correcting factor to move a line through a defect of holonomy $M$ is:
\begin{equation}
\frac{\chi_{R_1}(M)}{\text{dim }R_1} \frac{\text{dim }R_2}{\chi_{R_2}(M)},
\end{equation}
where the representations $R_1$ and $R_2$ are as in Figure \ref{cross}.
The above formula can be used to find the expression for the diagram
\begin{figure}[h]
\centering
\includegraphics[width=0.6\textwidth]{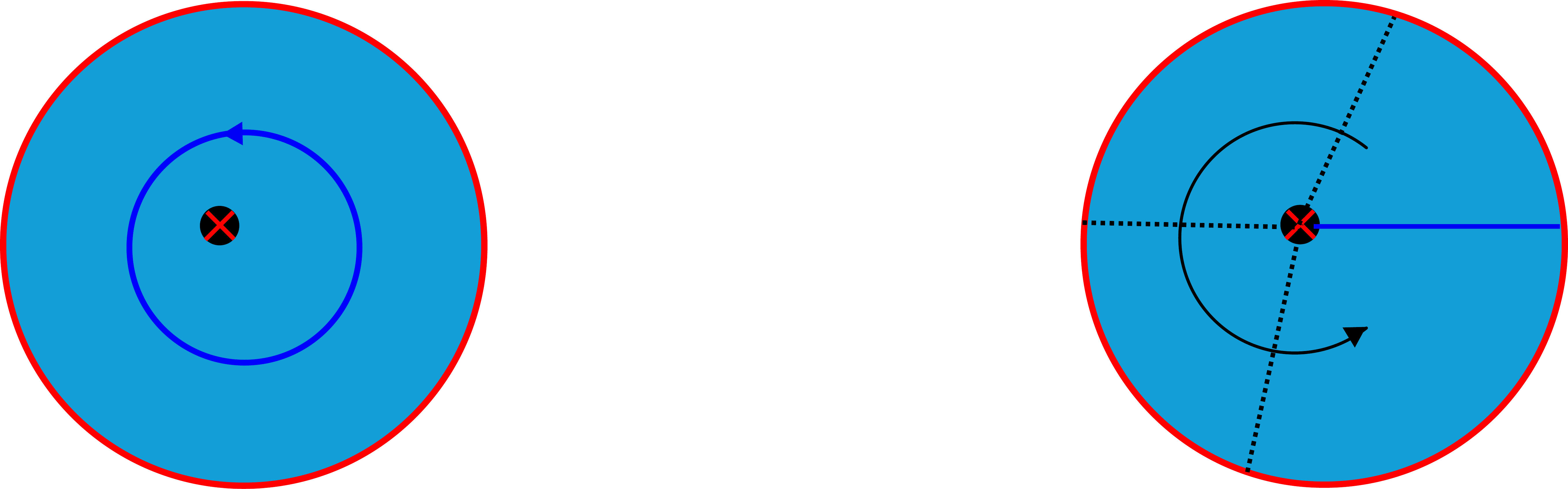}
\caption{Left: A puncture surrounded by a bulk Wilson loop. Right: Inserting a Wilson line attached to the defect in the interior and the outer boundary, is readily evaluated in angular slicing (dotted lines).}
\label{puncturediagram}
\end{figure}
by moving the puncture through its encircling Wilson loop. E.g. for the diagram of Figure \ref{puncturediagram} left, one obtains:
\begin{align}
\frac{1}{Z}\sum_{R_1,R_2} \text{dim }R_2 \chi_{R_1}(M) \frac{N_{RR_1R_2}}{\text{dim }R} e^{-\beta \mathcal{C}_{R_2}}
&=\frac{1}{Z}\frac{\chi_R(M)}{\text{dim }R}\sum_{R_2} \text{dim }R_2 \chi_{R_2}(M) e^{-\beta \mathcal{C}_{R_2}},
\end{align}
where the Verlinde formula was used to get to the second line. The second line reflects the fact that the Wilson line measures the holonomy around the puncture. One can write more generally:
\begin{equation}
\label{loopw}
\left\langle \mathcal{W}_R \hdots \right\rangle = \frac{\chi_R(M)}{\text{dim }R} \left\langle \hdots\right\rangle.
\end{equation}

\noindent For $\sltr$, it is known that the spin 1/2 Wilson loops
\begin{equation}
\mathcal{W}_{1/2}(\gamma) = \frac{1}{2}\text{Tr}_{1/2}\mathcal{P} e^{\oint_\gamma A},
\end{equation}
in the (non-unitary) 2-dimensional representation of $\sltr$ are given by the geodesic length operators in Teichm\"uller theory as
\begin{equation}
L(\gamma) \equiv \mathcal{W}_{1/2}(\gamma) = \cosh \frac{1}{2} \ell(\gamma),
\end{equation}
in terms of the length $\ell(\gamma)$ along the geodesic curve in the homotopy class of $\gamma$.
\\~\\
First we prove the analogue of \eqref{loopw} for this case. One writes:
\begin{align}
\frac{1}{Z} \int dk_1 dk_2 \cos(2\pi \lambda k_1) N_{k_1k_2\frac{1}{2}} e^{-\beta k_2^2}
& = \frac{\sinh(2\pi \lambda)}{\sinh(\pi \lambda)} \frac{1}{Z} \int dk_2 \cos(2\pi \lambda k_2) e^{-\beta k_2^2}.
\end{align}
To find this, one uses the formulas
\begin{align}
\int dk_1 \cos(2\pi \lambda k_1) N_{k_1k_2\frac{1}{2}} &= \cos(2\pi \lambda k_2 + \pi i \lambda) +  \cos(2\pi \lambda k_2 - \pi i \lambda) \nonumber \\
&= \frac{\lambda \sinh (2\pi \lambda)}{\lambda \sinh(\pi \lambda)} \cos(2\pi \lambda k_2), 
\end{align}
where in the first line we used the fusion rule:
\begin{equation}
N_{k_1k_2\frac{1}{2}} = \delta(k_1-k_2 \pm i/2).
\end{equation}
This should be compared to the 2d Verlinde formula:
\begin{equation}
\sum_{k_1} N_{k_1k_2 J} S_{k_1}^\lambda = \frac{S_{J}^\lambda}{S_{0}^\lambda}  S_{k_2}^\lambda,
\end{equation}
and is an irrational $\sltr$ extension of it, explored in \cite{Jego:2006ta}. Note that we should use the fusion coefficient of the $\sltr$ representations (used for $H_3^+$ in \cite{Jego:2006ta}) as the fusion happens deep in the bulk where the gravitational coset constraints are not felt.
\\~\\
Hence formula \eqref{loopw} applies in this case as well, and using the general formula for the character of the spin 1/2 representation:
\begin{equation}
\chi_R(M) = \frac{\lambda^{2j+1}-\lambda^{-2j-1}}{\lambda-\lambda^{-1}}, \qquad \lambda = e^{i\pi \theta}, e^{\pi \lambda}, 1, \quad j=1/2,
\end{equation}
one finds for the three classes of monodromies:
\begin{align}
\text{elliptic: } &\quad \frac{\chi_R(M)}{\text{dim }R} = \frac{\sin(2\pi \theta)}{2\sin(\pi \theta)} = \cos(\pi \theta), \\
\text{parabolic: } &\quad \frac{\chi_R(M)}{\text{dim }R}  = 1, \\
\text{hyperbolic: } &\quad \frac{\chi_R(M)}{\text{dim }R} = \frac{\sinh(2\pi \lambda)}{2\sinh(\pi \lambda)} = \cosh(\pi \lambda).
\end{align}
In the first two cases, the actual length $\ell(\gamma)$ one finds is either imaginary or zero,\footnote{The latter happening for the special orbits $\theta \in \mathbb{N}$.} signaling no minimal length curve exists, and one should take $\ell(\gamma) \equiv 0$ in this case. The hyperbolic case is non-trivial and leads to a non-zero length $\ell(\gamma) = 2 \pi \lambda$. Hence there is a shallow neck of circumference $2 \pi \lambda$ in the hyperbolic orbit, and applying such a defect creates a (Euclidean) wormhole geometry.
\\~\\
Two asymptotic Schwarzian disks can be glued together by integrating over this neck circumference \cite{Blommaert:2018iqz,sss2}. If one only inserts one fixed hyperbolic defect in a single Schwarzian disk, the other holographic boundary is generated, but it has no dynamics (i.e. no fluctuating boundary) and no gravitational boundary conditions (i.e. it is a $\sltr$ boundary model); the entire theory can be described as a single Schwarzian theory with a hyperbolic defect. This is the situation we study here and is explored more below. \\
If we study 2d JT with two boundaries on an annulus, with the additional constraint that $L(\gamma)$ is fixed to some value, then one can study the two-boundary theory without the summation over $\lambda$. Note that keeping $L(\gamma)$ fixed is indeed a diff-invariant statement.
\\~\\
We encountered the finite-dimensional representations of $\sltr$ to measure the neck length of the Euclidean wormhole. This is the first time these representations appear in the JT context.\footnote{The continuous representations constitute the Hilbert space of the theory, whereas the infinite-dimensional discrete representations are used to describe the bilocal operator insertions.} One can ask whether these representations can be used as bilocal operator insertions. They can be obtained from the usual bilocals by setting $\ell \in -\mathbb{N}/2$. Due to the $\Gamma(2\ell)$ factor in the denominator of \eqref{correliptic1}, these require a careful separate treatment, which we perform in Appendix \ref{app:deg}. Whereas we do not know of any application of these, their treatment presents an easy test-case where a closed analytic expression can be obtained for the two-point correlator.

\subsection{Application: Global AdS$_2$ and Correlators}
\label{sec:global}
\begin{figure}[h]
\centering
\includegraphics[width=0.35\textwidth]{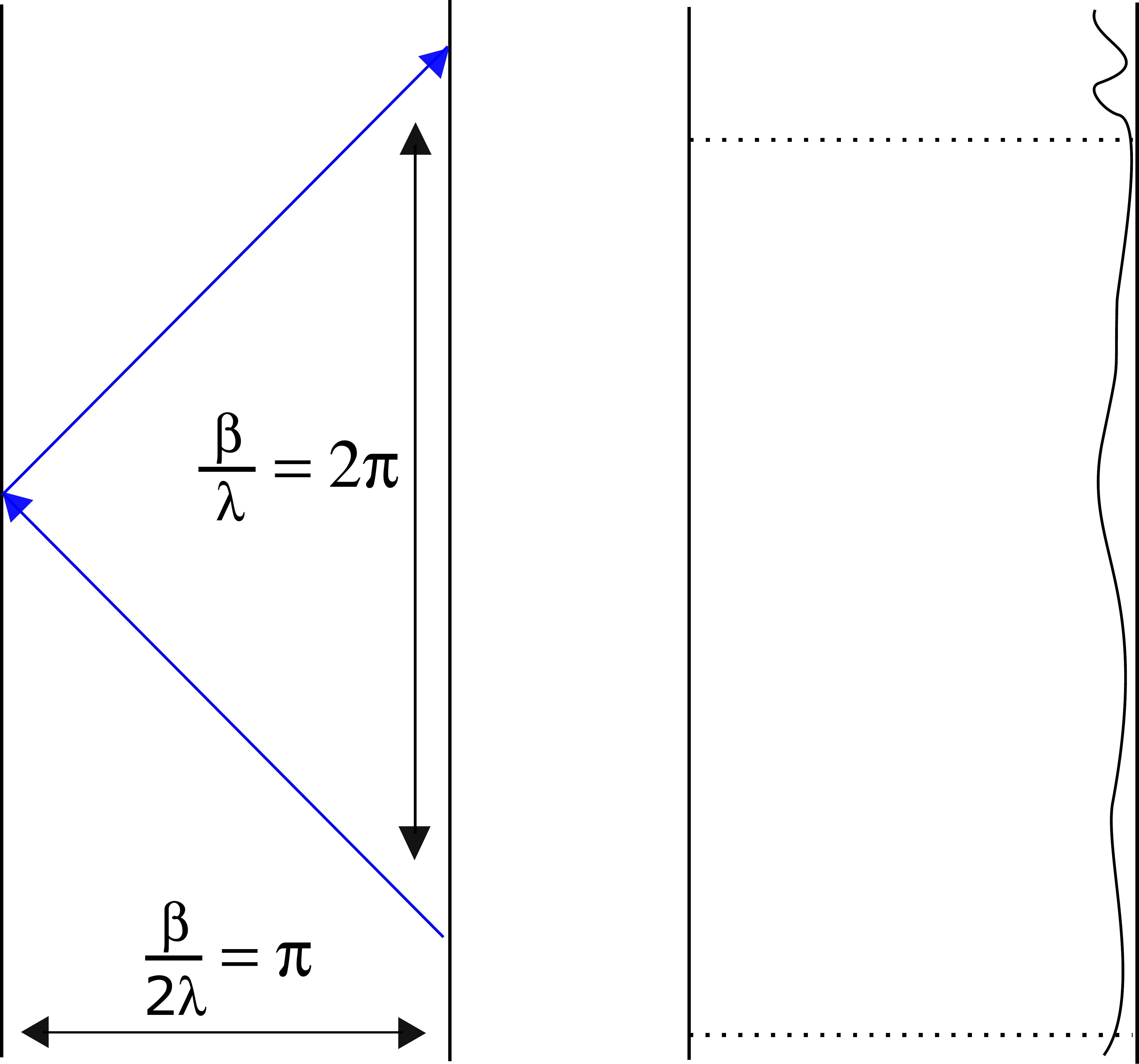}
\caption{Left: Global AdS$_2$ geometry with $\lambda = \frac{\beta}{2\pi}$. Semi-classical periodicity occurs due to reflections on the second boundary. Right: Jackiw-Teitelboim structure of the wiggly boundary curve.  Dashed lines denote the semi-classical periodicity present in correlation functions. One has no real-time periodicity condition: $f(t+2\pi) \neq f(t)+2\pi$. Hence the full bilocal correlator does not exhibit periodicity. }
\label{globalads}
\end{figure}

For any fixed $\lambda$, the geometry can be interpreted as a two-sided global AdS space. The global AdS$_2$ system, including its doubled SYK interpretation, were studied in \cite{Maldacena:2018lmt}. \\
Taking $\lambda \equiv \lambda_\beta = \beta/2\pi$, we obtain the standard normalization of global AdS$_2$ and the geometrical identification is not influenced by the temperature $\beta^{-1}$ (Figure \ref{globalads}).
The Schwarzian in Euclidean space is here a $\tanh\frac{\pi}{\beta} \lambda f = \tanh \frac{f}{2}$ orbit:
\begin{equation}
\label{globalsch}
S= -C \int_{0}^{\beta}d\tau \left\{\tanh\frac{f}{2},\tau\right\}, \quad f(\tau+\beta) = f(\tau) + \beta.
\end{equation}
This configuration occurs in JT gravity in a strip by imposing the left boundary to have no boundary dynamics: $\Phi_{LR} \sim \frac{a_{LR}}{\epsilon}$ with $a_L=0$. Only a single Schwarzian wiggly boundary curve is present whose dynamics is described by the action \eqref{globalsch}. The left boundary has no Schwarzian action, and no 1d conformal symmetry breaking. \\
\\~\\
The partition function of this model is:
\begin{equation}
\label{pfglobal}
Z(\beta,\lambda_\beta) = \int_{0}^{+\infty}dk \cos(\beta k) e^{-\beta \frac{k^2}{2C}} = \frac{e^{-\frac{\beta C}{2}}}{2}\sqrt{\frac{2 \pi C}{\beta}} = \int_{0}^{+\infty}dk ~e^{-\beta (\frac{k^2}{2C} + \frac{C}{2})},
\end{equation}
and is one-loop exact. This leads to the density of states $\rho(E) = \frac{\Theta(E-\frac{C}{2})}{2\sqrt{E-\frac{C}{2}}}$. This offset $E_{\text{gap}} = C/2$ can be interpreted as the energy required to cancel the Casimir energy of global AdS$_2$. Indeed, global AdS$_2$ has negative energy $E=-C/2$. Excitations in this theory are hence gapped and only arise when the energy becomes at least that of the Poincar\'e energy.\footnote{Of course, we could just redefine the energy of the vacuum and not deal with this shift, but when embedding global AdS into the set of models created by applying defects, this is the natural normalization.}
\footnote{Note that the modular $S$-matrix $S_0^k$ is \emph{not} the density of states in this case, as it depends on the temperature $\beta^{-1}$ to define a sensible geometric theory.}
\\~\\
The saddle of the action \eqref{globalsch}, with solution $f(\tau) = \tau$, leads to the free energy $F$, energy $E$ and entropy $S$ given by:
\begin{equation}
\beta F = -\log Z(\beta) = \frac{\beta C}{2}, \quad E =  \frac{C}{2}, \quad S = \beta E - \beta F = 0,
\end{equation}
whereas the exact result \eqref{pfglobal} gives:
\begin{equation}
\beta F = -\log Z(\beta) = \text{const} + \frac{\beta C}{2} + \frac{1}{2}\log \frac{\beta}{2 \pi C}, \quad E = \frac{1}{2\beta} + \frac{C}{2}, \quad S = \text{const} + \frac{1}{2} \log 2E.
\end{equation}
The entire entropy is hence one-loop. A similar conclusion will be reached for the parabolic $\theta = 0$ orbit in section \ref{sect:para} below.
\\~\\
Following the methods developed in previous sections, we can compute expectation values of the Euclidean bilocal correlator
\begin{equation}
\label{bilopi}
\mathcal{O}_\ell (\lambda_\beta; \tau_1,\tau_2 ) = \left(\frac{\dot{f}(\tau_1) \dot{f}(\tau_2)}{4\sinh^2 \frac{1}{2}(f(\tau_1)-f(\tau_2))}\right)^{\ell},
\end{equation}
which in the twisted Schwarzian path integral leads to
\begin{equation}
\label{globcorr}
\langle \mathcal{O}_\ell (\lambda_\beta; \tau_1,\tau_2 ) \rangle_\beta = \frac{1}{Z} \int dk_1^2 dk_2 \sinh(2\pi k_1) \cos( \beta k_2) \frac{\Gamma(\ell \pm i k_1 \pm i k_2)}{\Gamma(2\ell)} e^{-\tau \frac{k_1^2}{2C} - (\beta-\tau) \frac{k_2^2}{2C}}.
\end{equation}
Writing $\cos(\beta k_2) = \frac{1}{2}(e^{ik_2 \beta} + e^{-i k_2 \beta})$, and shifting the integration contour $k_2 \to k_2 \pm i C$, one can write:
\begin{align}
\langle \mathcal{O}_\ell (\lambda_\beta; \tau_1,\tau_2 ) \rangle_\beta = \frac{1}{Z} \int &dk_1^2 dk_2 \sinh(2\pi k_1)e^{-\tau \frac{k_1^2}{2C} - (\beta-\tau) (\frac{k_2^2}{2C}+\frac{C}{2})}e^{-\tau C} \\ \nonumber
&\times \frac{\Gamma(\ell + \frac{1}{2} \pm i k_1 + i k_2)\Gamma(\ell - \frac{1}{2} \pm i k_1 - i k_2)e^{-ik_2 \tau} \,+ \, (k_2 \to -k_2)}{2\Gamma(2\ell)},
\end{align}
which now has a standard thermodynamic interpretation, with energy $\frac{k_2^2}{2C}+\frac{C}{2}$,\footnote{Note that this expression is not in the form of $e^{-E_1 \tau}e^{-E_2(\beta-\tau)}$ for some independent energies $E_1$ and $E_2$, which has to be the case if the bilocal operator is obtained from two local operators. This is in accord with the fact that any thermal two-point function has to satisfy a KMS condition, which this operator does not.} from which the fixed-energy (microcanonical) correlator can be deduced. For example, this form is useful to get the zero-temperature correlator. Taking for $Z$ the global partition function \eqref{pfglobal}, the zero-temperature correlator is given by:
\begin{equation}
\label{zerotglobal}
\langle \mathcal{O}_\ell (\lambda_\beta; \tau_1,\tau_2 ) \rangle_{\beta\to +\infty} = \int dk_1^2 \sinh(2\pi k_1) \frac{\Gamma(\ell + \frac{1}{2} \pm i k_1)\Gamma(\ell - \frac{1}{2} \pm i k_1) }{\Gamma(2\ell)} e^{-\tau (\frac{k_1^2}{2C} + \frac{C}{2})}.
\end{equation}
This correlator is also the zero-temperature version of the KMS-satisfying bilocal \eqref{defbilocalh2}, since winding is suppressed as $\beta \to+\infty$ when $\lambda = \beta/2\pi$.
One checks numerically that the bilocal correlator \eqref{globcorr} (and \eqref{zerotglobal}) in Euclidean time in the semi-classical large $C$ regime is $1/\sinh(\frac{1}{2} \tau)^{2\ell}$. The real-time semi-classical correlator is then:
\begin{equation}
\langle \mathcal{O}_\ell (\lambda_\beta; t_1,t_2 ) \rangle_\beta \, \to \, \frac{1}{\left(\sin \frac{t}{2}\right)^{2\ell}},
\end{equation}
and has periodicity $2\pi$ in real time due to reflections off of the second boundary. The full real-time correlator is not periodic though (and does not have to be), as can be checked numerically, and as shown in Figure \ref{globalads} right.
\\~\\
Let us next characterize the second asymptotic boundary that this model generates. We demonstrate that this boundary carries the full $\sltr$ particle on it, and hence no gravitational boundary constraints are imposed. This of course makes sense as the hyperbolic defect itself is an $\sltr$-construct and does not know itself about gravity. Secondly, integrating over the hyperbolic label allows us to glue geometries together, which requires the unconstrained $\sltr$ model. \\
Consider therefore a Wilson line $\mathcal{W}^{\bar{R}}_{mn}$ attached on one side to the interior defect $M$ (Figure \ref{puncturediagram} right). Using the techniques of \cite{Blommaert:2018oro,Blommaert:2018iqz}, this diagram can be evaluated by an angular slicing to be:
\begin{align}
\sum_{R,a,b} &\text{dim }R \int dg R_{ab}(g) R_{ba}(Mg^{-1}) \bar{R}_{mn}(g)  e^{-\beta \mathcal{C}_R} \nonumber \\
 &=\sum_{R,a,b,c} \text{dim }R R_{bc}(M) \threej{R}{a}{\bar{R}}{m}{R}{a}\threej{R}{b}{\bar{R}}{n}{R}{c} e^{-\beta \mathcal{C}_R}.
\end{align}
Within JT gravity where the outer boundary is gravitational, we require the index $a=\mathfrak{i}$ which is the sole state surviving the coset \cite{Blommaert:2018oro,Blommaert:2018iqz}. Additionally setting $m=n=0$ to consider the standard bilocal operator, we reduce to $b=c$, and we get:
\begin{equation}
\label{biloc}
\int dk^2 \sinh(2\pi k)  \int ds R^k_{ss}(M) \left(\frac{\Gamma(\ell \pm 2ik) \Gamma(\ell)^2}{\Gamma(2\ell)}\right)^{1/2} \threej{k}{s}{\ell}{0}{k}{s}e^{-\beta \frac{k^2}{2C}},
\end{equation}
in terms of the hyperbolic basis label $s$ of $\sltr$ representations, and with hyperbolic monodromy matrix:
$M = \left(\begin{array}{cc}
e^{2\pi \lambda_\beta} & 0 \\
0 & e^{-2\pi \lambda_\beta} \\
\end{array}\right)$.
As a check, setting $\ell = 0$, the 3j-symbol becomes $\frac{1}{\sqrt{k \sinh 2\pi k}}$, $ \int ds R^k_{ss}(M) = \chi_k (M) = 2\cos( 2\pi \lambda_\beta k)$, and we return to the twisted Schwarzian partition function:\footnote{Strictly speaking, there should be a factor of $1/V_T$ present in \eqref{biloc}, where $V_T = \delta(k-k)$ is the volume of the set of conjugacy class elements, Fourier dual to a delta-function on the set of irreps. The limit $\ell \to 0$ gives an additional factor of $\delta(k-k)$ in the numerator from the 3j-symbols, cancelling this one and indeed recovering \eqref{twpf}.} \footnote{Instead setting $M = \mathbf{1}$, we find the amplitude for a Wilson line dangling into the bulk without ending on any special (i.e. defect) location. This evaluates to $\sqrt{k \sinh(2\pi k)} \delta_{\ell 0}$, using properties of 3j-symbols, so upon dividing by Z we end up with:
\begin{equation}
\frac{1}{Z}\int dk^2 \sinh(2\pi k) e^{-\beta \frac{k^2}{2C}} \delta_{\ell,0} =  \delta_{\ell,0},
\end{equation}
and this situation cannot occur for any non-trivial Wilson line.
 }
\begin{equation}
\label{twpf}
2\int dk \cos( 2\pi \lambda k) e^{-\beta \frac{k^2}{2C}}.
\end{equation}

\noindent The defect in the interior, which geometrically corresponds to the second asymptotic boundary, has no gravitational constraints and hence does not represent a holographic boundary; this correlator is not modeling the wormhole-crossing correlators of global AdS$_2$.
\\~\\
A different interesting bilocal operator to consider within the global AdS$_2$ model \eqref{bilopi} is
\begin{equation}
\label{bilocalworm}
\mathcal{O}_{\lambda,LR}^{\ell}(\tau_1,\tau_2) = \left(\frac{\dot{f}_1 \dot{f}_2}{\cosh(\frac{1}{2}\left(f_1 - f_2\right))^2}\right)^{\ell},
\end{equation}
obtained by taking $f_1 \to f_1 + i\pi$, i.e. rotating the \emph{global} time of one leg of the correlator.\footnote{It is quite subtle to convince oneself that this is the correct reparametrization to describe the two-sided global AdS geometry. For the TFD on the other hand, one takes $\tau \to \tau + \pi$ instead of $f$ (i.e. $\sinh (\frac{1}{2}f(\tau_1+\pi)-f(\tau_2))$). This is because by definition, thermal correlators are obtained in this way. For global AdS, it is different because the coordinate of the second boundary is at $f(t+z) = f(t-z) + \pi$ by \eqref{globalgeom}. Hence we expect a singularity at fixed location in $f$, instead of in $\tau$.} The semi-classical result is of the form $\left(\frac{1}{\cosh(\frac{1}{2}\left(t_1 - t_2\right))^2}\right)^{\ell}$ and represents a left-right boundary two-point function. We leave an exact quantum treatment of this bilocal correlator for future work.

\section{Parabolic orbit $U(1)_0$: thermal AdS$_2$}
\label{sect:para}
\begin{figure}[h]
\centering
\includegraphics[width=0.35\textwidth]{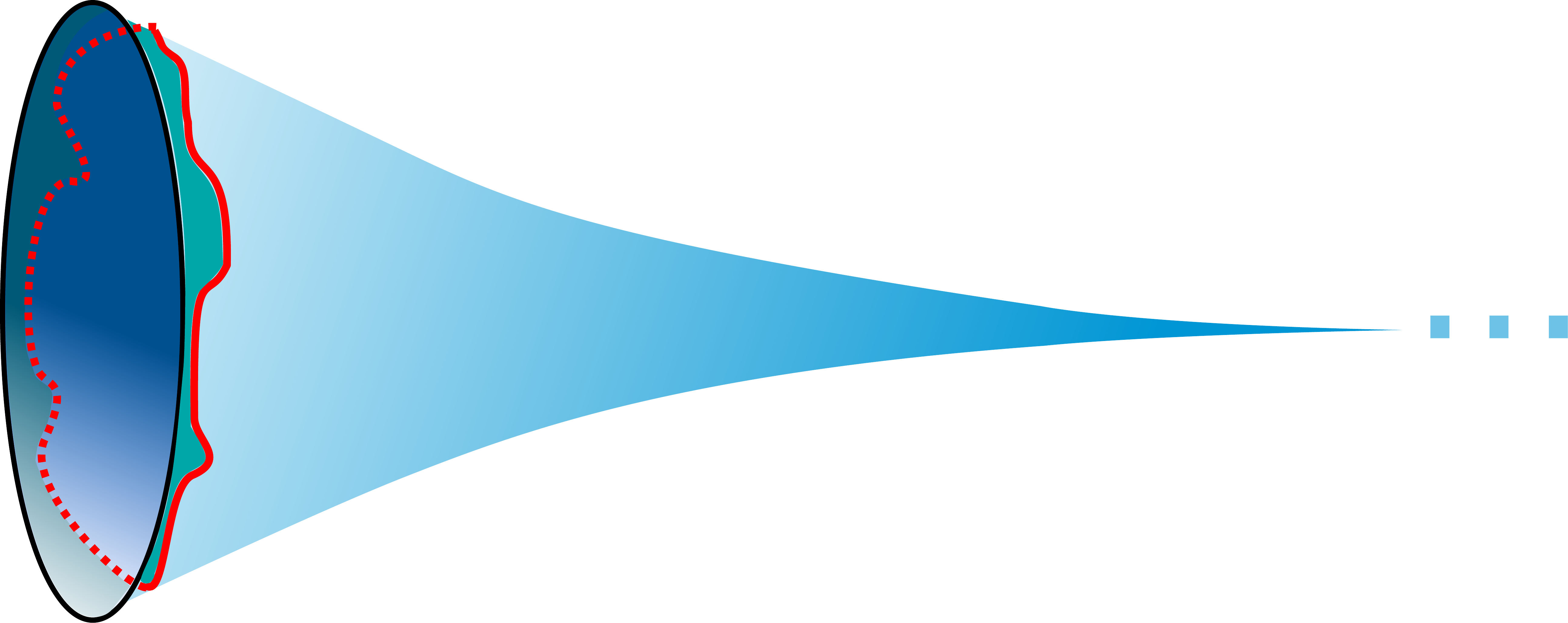}
\caption{Parabolic orbit in the Schwarzian limit leads to the thermal AdS$_2$ geometry with fluctuating boundary curve.}
\label{thermalAdS}
\end{figure}

The parabolic orbit is the limit of $\theta \to 0$ of either the elliptic or the hyperbolic orbits. It leads to thermal AdS$_2$, with
\begin{equation}
F(\tau+\beta) = F(\tau) + \beta,
\end{equation}
left invariant only under the one-parameter parabolic subgroup of $\sltr$:
\begin{equation}
\label{para}
M = \left(\begin{array}{cc}
1 & b \\
0 & 1 \\
\end{array}\right) \quad \in \text{P}\sltr,
\end{equation} 
and with a flat density of states: $d\mu(k) = dk$ (Figure \ref{thermalAdS}).

The partition function is
\begin{equation}
Z(\beta,0) = \int_0^{+\infty} dk \, e^{-\beta \frac{k^2}{2C}} = \frac{1}{2}\sqrt{\frac{2\pi C}{\beta}}.
\end{equation}
The energy average is $\left\langle E\right\rangle = -\frac{\partial \log Z}{ \partial \beta} = \frac{1}{2\beta}$, whereas the saddle itself (vacuum AdS) has zero energy.\footnote{This is contingent upon subtracting the extremal energy $E_0$ of the system.} The bilocal correlator in the parabolic orbit is:\footnote{Again one can in principle choose here to normalize with the $n=1$ partition function (as we have done here), or with $Z(\beta,0)$ instead. This depends on whether we view this system as the vacuum $n=1$ system with a defect, or as an independent model in its own right.}
\begin{equation}
\langle \mathcal{O}_\ell (0; t_1,t_2 ) \rangle = \frac{1}{Z} \int dk_1^2 dk_2 \sinh(2\pi k_1) \frac{\Gamma(\ell \pm i k_1 \pm i k_2)}{\Gamma(2\ell)} e^{-\tau \frac{k_1^2}{2C} - (\beta-\tau) \frac{k_2^2}{2C}}.
\end{equation}
The semi-classical large $C$-limit is readily checked to behave as $\sim 1/\tau^{2\ell}$.\footnote{In case the sinh is in the $k_2$-integral, one obtains instead $\sim 1/(\beta-\tau)^{2\ell}$.} The Lorentzian continuation $\tau \to i t$ gives the same semi-classical behavior over $\mathbb{R}$. At zero temperature, this correlator is the same as that for the $n=1$ elliptic orbit: the vacuum of both theories is identical, but their excited state spectrum differs.  The flat density of states indicates that this configuration has no Bekenstein-Hawking rise of states, and hence no black hole degeneracy in the bulk. \\
A more interesting thermal bilocal, satisfying KMS, is the following, obtained as the $\lambda \to 0$ of the hyperbolic correlator \eqref{defbilocalh2}: 
\begin{equation}
\label{tadscorr}
\sum_{w \in \mathbb{Z}} \left(\frac{\dot{F}(\tau_1 + w\beta)  \dot{F}(\tau_2) }{(F(\tau_1 + w\beta) - F(\tau_2) )^2}\right)^{\ell} = \sum_{w \in \mathbb{Z}} \left(\frac{\dot{F}(\tau_1) \dot{F}(\tau_2)}{(F(\tau_1) - F(\tau_2) +w\beta)^2}\right)^{\ell},
\end{equation}
an operator that is U(1) invariant $F_i \to F_i + \text{const}$, but not $\sltr$ invariant. As such, it is a new possibility for this orbit that was not available in the standard $\sltr$ theory. This is the conformal reparametrization with function $f$, of the standard CFT expression:
\begin{equation}
\sum_{w \in \mathbb{Z}} \left(\frac{1}{(\tau_1 - \tau_2 + w \beta)^2}\right)^{\ell},
\end{equation}
of the two-point correlator on a periodically identified circle. A single term in the sum \eqref{tadscorr} can be interpreted as a boundary-anchored Wilson line encircling the bulk defect $w$ times. We leave a computation of this bilocal correlator to future work.
\\~\\
While we are unaware of an application of this orbit directly as a low-energy description of an SYK computation, it is interesting to contemplate a Hawking-Page transition by combining by hand this orbit with the $n=1$ elliptic orbit. Including only the tree-level contribution, the $n=1$ orbit is always the most dominant contribution.\footnote{Indeed, this directly descends from the Hawking-Page story in AdS$_3$/CFT$_2$ where in the double-scaling limit one is forced on the saddle that corresponds to the $n=1$ orbit in the JT system.} However, including the one-loop piece, there is a Hawking-Page (HP) transition and the parabolic orbit (thermal AdS) wins at low temperatures. The transition happens at $\beta_{\text{HP}} \sim C $. This is a very low temperature of order $1/C$ corresponding to a strongly coupled theory.

This version of the Hawking-Page transition arises from a combination of classical and quantum effects. This should not alarm the reader since the partition function for both thermal AdS and the black hole can be computed exactly within the Schwarzian theory.

\section{Exceptional Orbits $\mathcal{T}_{n,\lambda}$}\label{sec:excep}
Up to now we have ignored the exceptional (sometimes also called special) orbits $\mathcal{T}_{n,\lambda}$ and $\mathcal{T}_{n,\pm}$ (see for example \cite{Witten:1987ty} for their definition). One reason to do this is the fact that their representative energy is not $\tau$-independent. 

To motivate a proposal for their interpretation, we can exploit the 2d Liouville perspective to analyze whether we have exhausted all 2d setups with a well-defined semiclassical limit. We have fully analyzed the case of identity ZZ-branes and FZZT-branes. Nevertheless, for degenerate ZZ-branes we focused so far only on the $(1,n)$ case. The reason for this being that when we take the large $c$, small $b$ limit, the ZZ-brane associated to $(m,n)$ reduces to the $(1,n)$ case. This happens due to the fact that the wavefunction is 
\begin{equation}
\Psi_{(m,n)}( P) = \Psi_{(1,1)}(P) \frac{\sinh 2 \pi n P/b}{\sinh 2 \pi P/b} \frac{\sinh 2 \pi m b P}{\sinh 2 \pi b P},
\end{equation}
while the Schwarzian limit picks momenta that scale with $b$ as $ P \sim b$.  

In this section we will comment on the only way left to generate a 1d limit different from the ones previously studied. Take Liouville between an identity ZZ-brane and an $(m,n)$ ZZ-brane with fixed integer $n$ and an integer $m$ that scales with the central charge as $m=\lambda/b^2$, giving a new continuous parameter $\lambda$ in the large $c$ limit. The partition function of the 1d theory we generate in this way gives 
\begin{eqnarray}
Z_{n,\lambda} &=& \int_0^{\infty} dk \sinh(2 \pi n k)\sinh(2 \pi \lambda k) e^{-\beta \frac{k^2}{2C}} \nonumber \\
&=& \Big( \frac{ 2 \pi C}{\beta} \Big)^{1/2}  \sinh\Big( \frac{4 \pi^2 C}{\beta} n \lambda \Big) e^{\frac{2 \pi^2 C}{\beta}(n^2 + \lambda^2)}.
\end{eqnarray}
We can also compute correlators between bilocal operators arising from the 1d reduction of a local Liouville primary operator. The results and correlators for this setup should by now be straightforward, with the result being equal to the Schwarzian $n=1$ theory, but with the insertion of 
\begin{equation}
D_{n,\lambda}(k) = \frac{\sinh(2\pi nk)}{\sinh(2\pi k)}\frac{\sinh(2\pi \lambda k)}{k}.
\end{equation}
Even though this theory has a well-defined 1d limit, the representation we are studying from the 2d perspective has a dimension scaling with $c$ in a funny way $\Delta= -\lambda^2/4 b^6$, or equivalently in terms of the 2d central charge $\Delta \sim -c^3 \lambda^2$.

Since this 1d theory is labeled by a continuous and a discrete parameter, it is tempting to associate it with the orbit $\mathcal{T}_{n,\lambda}$, especially since these orbits are known to be obtained as deformations of the exceptional elliptic orbits \cite{Witten:1987ty}, and we obtained the latter indeed from $(1,n)$ ZZ-branes. We leave a more detailed analysis of this question for future work. In particular, it would be interesting to identify the 1d action which reproduces this partition function and whether its saddle-point is consistent with the profile $T(\tau)$ associated to this monodromy class (see section 3.5.2 of reference \cite{Balog:1997zz} for example). The results of \cite{Gorsky:1993ga} might be useful for this. Equivalently, it would be interesting to identify a defect in JT gravity which is holographically related to this theory. 

\section{Some Conclusions}
\label{sec:concl}
\begin{figure}[h]
\centering
\includegraphics[width=0.55\textwidth]{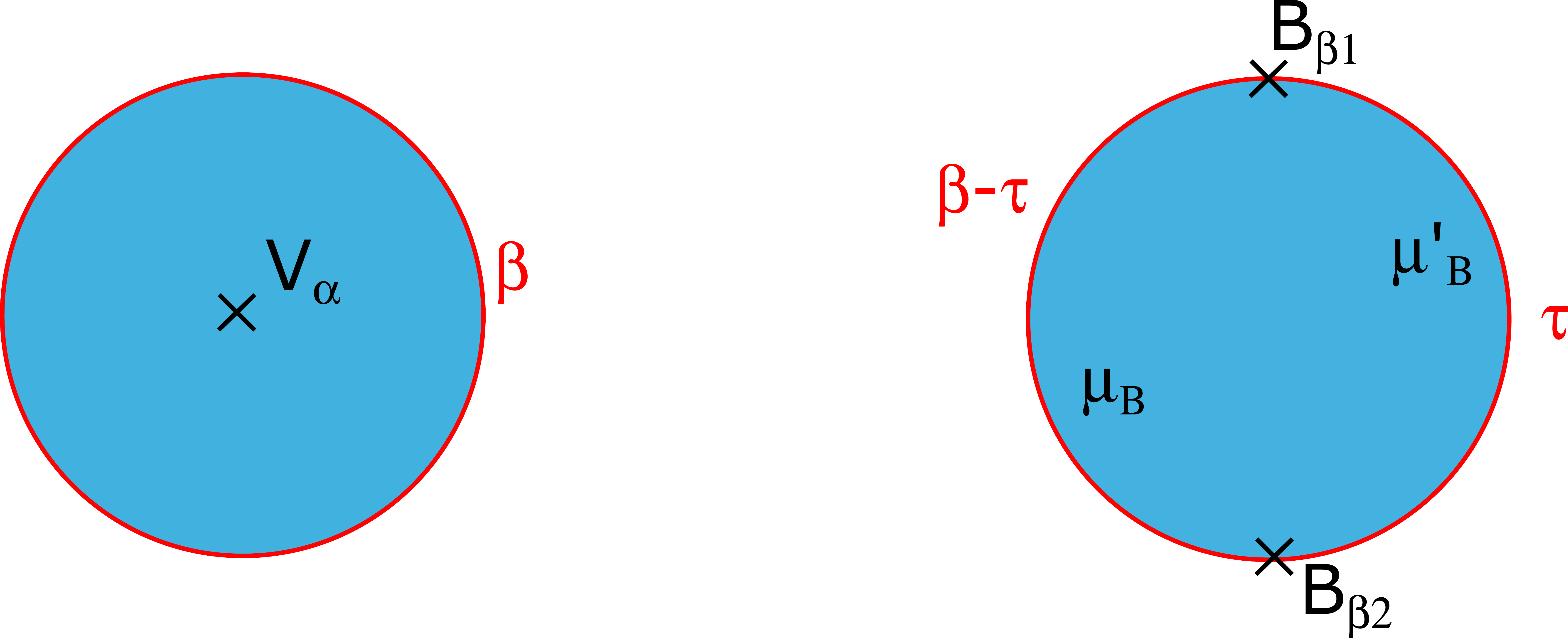}
\caption{Left: Liouville bulk one-point function with $V_\alpha = e^{2\alpha \phi}$. Right: Liouville boundary two-point function with $B_{\beta_i} =e^{\beta_i \phi}$.}
\label{minimal}
\end{figure}

The common theme in this work is the analysis of punctures, or defects, within the BF or JT gravity two-dimensional bulk. Studying this problem relates several a priori different physical situations. Let us summarize the salient points.
\\~\\
The low energy theory that appears in complex SYK (a Schwarzian and U(1) mode) is exactly solvable, also for its correlation functions. Diagrammatically, the chemical potential corresponds to inserting a defect into the BF bulk dual to the compact U(1) sector. We wrote down the generalization to non-abelian compact sectors, relevant for SYK-like models with more complicated internal symmetries (for example $\mathcal{N}=4$ SYK), which includes a defect character describing the non-abelian fugacity.
\\~\\
Within JT gravity, we classified the possible defects with Virasoro coadjoint orbits in the following categories: 
\begin{itemize}
\item Elliptic defects. These represent conical singularities and correspond to inserting dilaton punctures in JT. We illustrated how one defines the horizon area operator from these.
\item Hyperbolic defects. These represent macroscopic holes in the JT geometry. They are required to glue disk segments to each other. We defined a geodesic length operator to measure the neck length in this case. The geometry with a single such defect can be viewed as a global AdS$_2$ space which we studied.
\item Finally, the parabolic defect corresponds to thermal AdS$_2$.
\end{itemize}
Within the 2d Liouville CFT perspective on JT gravity, we pointed out that these defects can be created by inserting Verlinde loop operators between a pair of ZZ-branes, completing the set of all (interesting) operator insertions one can study in Liouville CFT.
\\~\\
Our methods have yielded a host of new Schwarzian-type models and their correlators that deserve further study in their own right. In particular, it would be interesting to learn of a direct SYK embedding of any of these other models.
\\~\\
Throughout this construction, we have defined several geometric observables, the horizon area operator $\hat{\Phi}_h$ and the length operator $L(\gamma)$, and have considered how these are implemented in correlation functions. We emphasized that they are defined in a diff-invariant way, and this is as far as one can get without explicitly defining dressed operators. Local diff-invariant bulk observables and their correlation functions were considered in \cite{Blommaert:2019hjr,Mertens:2019bvy}.
\\~\\
Let us end with some points that deserve further study. \\
Within the global AdS$_2$ and thermal AdS$_2$, we identified KMS-satisfying bilocal operators to consider, \eqref{defbilocalh2} and \eqref{tadscorr} respectively. Also the wormhole-crossing operator \eqref{bilocalworm} is structurally of the same type, where say $f_1$ is shifted compared to $f_2$. It would be interesting to better understand their Schwarzian correlation functions. \\
We studied Schwarzian bilocal correlators of the finite-dimensional $\sltr$ representations in appendix \ref{app:deg}. Although we studied these as a toy model, it would be interesting to learn of any direct physical relevance. \\
In section \ref{sec:excep}, we pointed out a suggestive approach to exceptional orbits $\mathcal{T}_{n, \lambda}$. This deserves further study. Exceptional orbits in general have not received a lot of attention in the literature. Their status (in all contexts) is still unclear and we hope our two-dimensional Liouville perspective can shed some light on their interpretation.
\\~\\
We end with a relation to the recent work \cite{sss2}. Among other things, a new way was found in that work to make contact with JT gravity as the semi-classical limit of the $(2,p)$ minimal string model, which contains a Liouville piece as the most interesting factor. Within this set-up, the Liouville theory is taken on the hyperbolic disk with a fixed-length boundary containing the Schwarzian boundary dynamics. The bulk JT model is hence directly identified as the Liouville worldsheet in the semi-classical $p\to+\infty$ regime. \\
During the course of our work, we found that determining the bulk JT geometry is identical to solving the classical Liouville uniformization problem, including microscopic punctures (elliptic defects). This geometry can be cut-and-glued along tubes along which the macroscopic Liouville primaries (hyperbolic defects) propagate.\footnote{To avoid confusion, we mention that Liouville CFT appeared in two places; firstly and studied throughout the paper, it appears between branes as a boundary model which has a Schwarzian double-scaling limit (see Figure \ref{geometry} Right). Here on the other hand, the semi-classical Liouville worldsheet is directly identified with the 2d JT bulk.} This resonates with the identification made in \cite{sss2}. \\
Let us be more explicit about this identification. We will illustrate that the Liouville bulk one-point function and boundary two-point functions indeed agree with what we already know (Figure \ref{minimal}).
The Liouville one-point function on the hyperbolic disk with boundary length $\beta$ is given by \cite{Fateev:2000ik}:
\begin{equation}
\left\langle V_\alpha \right\rangle_{\beta}\, \sim \, \mu^{(Q-2\alpha)/(2b)}K_{(Q-2\alpha)/b}(\kappa \beta), \quad \kappa^2 = \frac{\mu}{\sin \pi b^2}, \quad Q = b+ 1/b.
\end{equation}
Setting $2\alpha = Q - \theta/b$, for a heavy local Liouville operator, and using
\begin{equation}
e^{s}K_{\theta/b^2}(\kappa \beta) = \int_{0}^{+\infty}dt e^{-\kappa \beta t} \frac{\cosh\left(\frac{\theta}{b^2}\text{arccosh}(1+t)\right) }{\sqrt{t(2+t)}},
\end{equation}
and taking a double-scaling limit where $t = \frac{E}{\kappa} \to 0$ and $b \to 0$, one finds agreement with the elliptic U(1)$_\theta$ density of states \eqref{dosell}.\footnote{The exceptional elliptic case \eqref{dospsl} is found by considering the degenerate operator $V_\alpha$, $2\alpha = Q - n/b$, which requires a slightly more subtle discussion.} This is a direct extension of the computation of \cite{sss2}. \\
Setting $\theta \to - i \lambda$, we can link this to Teichm\"uller theory by $\alpha = \frac{Q}{2} + i \frac{\ell}{4\pi b}$ (see e.g. \cite{Drukker:2009id}), from which we read off $\ell = 2\pi \lambda$ as we found before. This formula links the identification of Teichm\"uller geodesic length $\ell$ and Liouville intermediate primary $\alpha$ in a tube, to the hyperbolic label $\lambda$ we indeed use for gluing here as well. \\
Next, consider the boundary two-point function (Figure \ref{minimal} right). As above, we parametrize $\beta_1 = \beta_2 = Q - b \ell$ in terms of what turns out to be the bilocal parameter $\ell$. The Liouville amplitude is proportional to \cite{Fateev:2000ik}:
\begin{equation}
\left\langle B_{\beta_1}B_{\beta_1}\right\rangle \, \sim \,\frac{S_b(Q-\beta_1  \pm i s/2 \pm i s'/2)}{\Gamma_b(Q-2\beta_1)\Gamma_b^{-1}(2\beta_1-Q)} \frac{1}{\left|x\right|^{2\Delta_{\beta_1}}}.
\end{equation}
Using $s= 2bk$ where $k$ is the Schwarzian momentum, and $S_b(bx) \to (2\pi b^2)^{x-1/2} \Gamma(x)$, one indeed finds the correct Schwarzian vertex functions of \eqref{sch2pt}. This amplitude is computed for fixed FZZT cosmological constants $\mu_B$ and $\mu_B'$, found from $\cosh^2 \pi b s = \frac{\mu_B^2}{\mu^2}\sin \pi b^2$, and should be transformed to the length basis, using the propagator distilled from the partition function. It seems clear that this approach will then indeed yield the full Schwarzian bilocal correlator. It would be interesting to study this procedure in more detail, which we leave for future study.

\section*{Acknowledgements}
We thank A. Blommaert, Y. Fan, L. Iliesiu, K. Jensen and H. Verlinde for discussions. TM gratefully acknowledges financial support from Research Foundation Flanders (FWO Vlaanderen). GJT is supported by a Fundamental Physics Fellowship. 

\appendix

\section{Classical winding solutions for QM on the group}
\label{app:qmgroup}
\begin{figure}[h]
\centering
\includegraphics[width=0.35\textwidth]{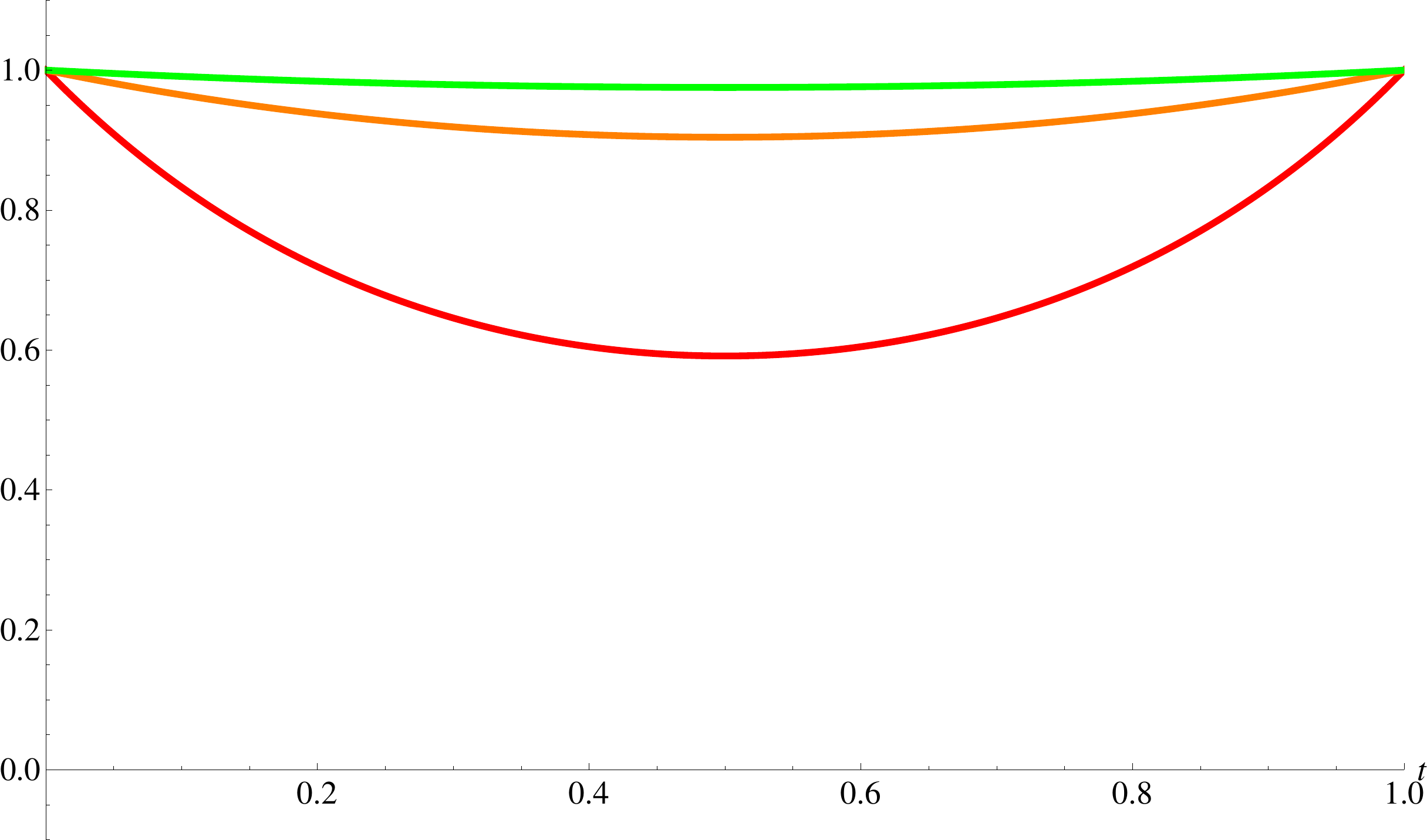}
\caption{SU(2) two-point correlator \eqref{sutwo} at $J=1$, $\mu=0$ and $\beta=1$. Red: $K=1$, Orange: $K=5$, Green: $K=20$.}
\label{WZWclassical}
\end{figure}

We provide a brief pedestrian analysis of the classical equations of motion associated to the particle on the group Lagrangian \eqref{nonab} (setting $\mu=0$ here), the appearance of winding quantum numbers and solutions, and the semi-classical regime $K\gg \tau,\beta$ of correlation functions.
\\~\\
The classical equation of motion of QM on the group manifold $G$ is
\begin{equation}
\partial_\tau (g^{-1} g') = 0,
\end{equation}
which can be integrated once into $g' = i g P$, for a constant matrix $P$. The general solution is $g(\tau) = S e^{i P \tau}$. Using the gauge equivalence $g \sim g G$, we can set $S = \mathbf{1}$ to impose $g(0)= \mathbf{1}$. This fixes the gauge of the solution. So the solution is 
\begin{equation}
g(\tau) = e^{i P \tau}.
\end{equation}
The monodromy of this solution is $g(\tau+\beta) = g(\tau) e^{i P \beta}$. Hence for zero monodromy, we have to solve $\mathbf{1} = e^{i P \beta}$ for the matrix $P \in \mathfrak{g}$. For compact groups, the matrix $P$ can be chosen to be hermitian, and hence diagonalizable, $\mathbf{1} = S^{-1} e^{i D \beta} S$ which implies
\begin{equation}
D = \text{diag}(2\pi m_1 /\beta, 2\pi m_2 / \beta \hdots), \qquad m_i \in \mathbb{Z}.
\end{equation}
The numbers $m_i$ are not all independent, but are linked by the fact that $D$ is in the Cartan subalgebra. This generalizes the U(1) winding quantum number $m$ in \eqref{eq:partu1} to one for each Cartan generator. In particular, in the unwound case, we just have the identity matrix solution:
\begin{equation}
g(\tau) = \mathbf{1}.
\end{equation}
The winding solutions on the other hand have on-shell action of the form:
\begin{equation}
S_{\text{on-shell}} = -\frac{K}{2} \int_{0}^{\beta} d\tau \, \text{Tr}\left[(g^{-1}g')^2\right] = \frac{K}{2}\beta\,  \text{Tr}P^2 =  \frac{K}{2}\beta \left(\frac{2\pi}{\beta}\right)^2 \sum_i m_i^2.
\end{equation}
Winding solutions are hence irrelevant for the semi-classical large $K$ limit, and the two-point correlator, asymptotes to 1 since
\begin{equation}
\frac{\text{Tr} \, G_{R,M}(\tau) e^{-\beta H}}{\text{Tr} e^{-\beta H}} \underset{\text{large }K}{\approx} G^{\text{on-shell}}_{R,M}(\tau) = 1.
\end{equation}
 Indeed, this can be checked numerically in Figure \ref{WZWclassical}.

\section{Orbits and Defects for compact BF theories}
\label{app:compact}
In this appendix we relate coadjoint orbits of the centrally-extended loop group $\widehat{LG}$ with defects in the BF disk.

\subsection{From 2d coadjoint orbits to monodromies}
We first relate the coadjoint orbits to monodromies. We start with the 2d $\widehat{LG}$ coadjoint orbit action:
\begin{equation}
\label{chiwzw}
S_{\chi \text{WZW}} = \frac{k}{4\pi}\int d \tau d \phi \Tr\left( \left(\gphi+\lambda\frac{4\pi}{k}\right)\gtau-\left(\gphi+\lambda\frac{4\pi}{k} \right)^2\right) + k\Gamma_{WZ},
\end{equation}
where the highest weight $\lambda = \bm{\lambda} \cdot \bm{H}$ labeling an integrable representations of $\widehat{LG}$.\footnote{In order for $e^{iS}$ to be single-valued, $\lambda$ has to be discretized as a weight.} This action can be viewed as a twisted chiral Wess-Zumino-Witten action. \\
The path integral with this action computes the Kac-Moody character $\chi_{\hat{\lambda}}(q)$. 
\\~\\
Performing a dimensional reduction along the $\phi$-direction results in:
\begin{equation}
S_{\text{KK}} = \frac{k}{4\pi}\int d \tau d \phi \Tr\left(\left(\lambda\frac{4\pi}{k}\right)\gtau - \left(\lambda\frac{4\pi}{k} \right)^2\right).
\end{equation}
Within path-integrals, this procedure can be achieved by taking the double-scaling limit $k\to+\infty$, $\int d\phi \to0$, with their product $k\int d\phi = K$, a fixed constant. Similarly, we require $\lambda \to +\infty$, keeping fixed the ratio
\begin{equation}
\label{dsc}
\Lambda \equiv \frac{4\pi}{k}\lambda.
\end{equation}
We rewrite 
\begin{equation}
S_{\text{KK}} = \frac{K}{4\pi}\int d \tau \Tr\left( \Lambda\gtau - \Lambda^2\right),
\end{equation}
which is the Alekseev-Shatashvili or Kirillov-Kostant action for the underlying group $G$.\footnote{This path-integral computes the character $\text{Tr}_\lambda(\mathbf{1}) \equiv (\text{dim }\lambda) e^{\frac{K}{4\pi}T \Tr \Lambda^2}$ in representation $\lambda$, weighted with a $\Lambda^2$ ``Hamiltonian''.}
\\~\\
For our purposes, we instead dimensionally reduce \eqref{chiwzw} along the $\tau$-direction and find:
\begin{equation}
\label{sqm}
S_{\text{QM}} = -\frac{K}{4\pi}\int d \phi \Tr(\gphi+\Lambda)^2,
\end{equation}
a particle-on-a-group action with ``twist'' $\Lambda$. \\
Notice that the quantum particle on the group $G$ is found in the $\phi$-direction. We can absorb the second term of \eqref{sqm} into a monodromy of $g$ by defining:
\begin{equation}
\tilde{g} =  g U(\phi), \quad U(\phi) = e^{\Lambda \phi},
\end{equation}
to find the action
\begin{equation}
S_{\text{QM}} = -\frac{K}{4\pi}\int d \phi \Tr(\tilde{g}^{-1}\partial_\phi \tilde{g})^2,
\end{equation}
with $\tilde{g}(\phi+2\pi) = \tilde{g}(\phi) M $ and monodromy matrix
\begin{equation}
M \equiv U(2\pi) = e^{2\pi \Lambda},
\end{equation}
where $\Lambda$ spans the entire Cartan subalgebra. Hence this argument can be reversed: given any monodromy matrix $M$, one can find a discrete representation label $\lambda$ that is associated to it by \eqref{dsc}, in the $k\to \infty$ limit.

\subsection{From defects to monodromies}
\label{app:defmono}
\begin{figure}[h]
\centering
\includegraphics[width=0.13\textwidth]{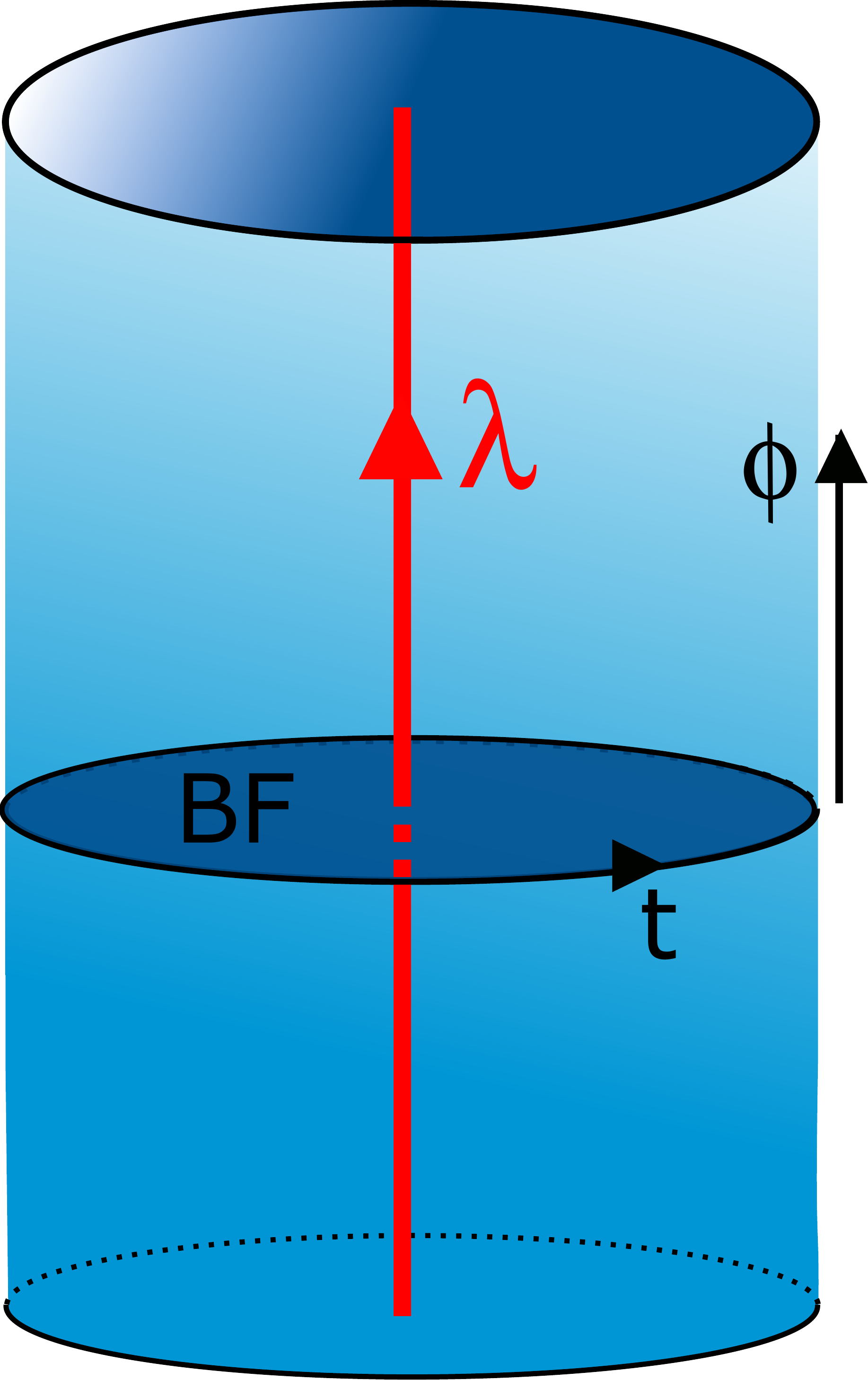}
\caption{Vertical Wilson loop in Chern-Simons theory piercing the BF disk.}
\label{vertical}
\end{figure}

Next, we classify defects in BF theory and relate these to the same monodromy matrices. Dimensionally reducing Chern-Simons along the angular $\phi$-direction leads to the 2d BF model plus boundary contribution:
\begin{equation}
\label{BF}
S_{\text{BF}} = \frac{K}{2\pi}\left(\int \text{Tr}\chi F + \frac{1}{2}\oint d\tau \, \text{Tr}\chi A_0\right).
\end{equation}
Consider a vertical Wilson loop in an irrep $\lambda$ in Chern-Simons theory (Figure \ref{vertical}).\footnote{Note the labeling of $\tau$ and $\phi$ in this figure. The coordinate $\tau$ will be identified with the thermal time of the resulting 2d BF theory.}
This line pierces the BF disk in a puncture, and these are the defects we consider here. The Wilson line is described by the operator:
\begin{equation}
\text{Tr}_\lambda \mathcal{P} e^{i \eta\int d\phi \, A_\phi},
\end{equation}
where $A_\phi \equiv \chi$.\footnote{A constant $\eta$ has been introduced here. One can view this as a rescaling of $\chi$ to find a finite limit.}
We can rewrite this expression using the well-known trick \cite{Moore:1989yh}:
\begin{equation}
\text{Tr}_\lambda \mathcal{P} e^{i \eta \int \chi} = \int \left[\mathcal{D}w\right]e^{i \int d\phi \text{Tr}\big( \lambda w^{-1} (\partial_\tau - i \eta \chi ) w\big)},
\end{equation}
in terms of an auxiliary group element $w$. Taking the double-scaling limit where $\int d\phi \to0$ and $\eta \to +\infty$, keeping their product fixed at some constant $\kappa$, we can write:\footnote{Equivalently, one can view this as an infinitesimal Hamiltonian propagation amplitude $e^{-i\epsilon H}$ with $H = \lambda w^{-1} i \eta \chi  w$. Care has to be taken to obtain the correct symplectic measure $d\mu(w)$ in this process.}
\begin{equation}
\text{Tr}_\lambda e^{i \kappa \chi} = \int d\mu(w) e^{\kappa \text{Tr}\big( \lambda w^{-1} \chi w\big)} = \int d\mu(w) e^{\left\langle \kappa \chi, w \lambda w^{-1}\right\rangle},
\end{equation}
which is the Kirillov character formula. The BF amplitude we want to compute:
\begin{equation}
\int \left[\mathcal{D}\chi\right]\left[\mathcal{D}A_\mu\right]\text{Tr}_\lambda e^{i \kappa \chi} e^{i S_{\text{BF}}},
\end{equation}
then projects onto the space:
\begin{equation}
\label{sourceeom}
F(\mathbf{x}) = \kappa \frac{2\pi}{K}w\lambda w^{-1} \delta(\mathbf{x}-\mathbf{y}),
\end{equation}
solved by 
\begin{equation}
A_{0} = g^{-1}\partial_\tau g + \frac{\kappa}{K} w \lambda w^{-1}.
\end{equation}
Since $w$ is a constant group element, we set $g\to gw^{-1}$ and the $w$-dependence cancels in the resulting action:
\begin{equation}
S = \frac{K}{4\pi}\int d\tau \text{Tr}A_0^2 =   \frac{K}{4\pi}\int d\tau \text{Tr}\left(g^{-1}\partial_\tau g  + \frac{\kappa}{K}\lambda\right)^2,
\end{equation}
which is the twisted particle on a group action with representation label 
\begin{equation}
\label{dsc2}
\Lambda \equiv \frac{\kappa}{K} \lambda,
\end{equation}
to be identified with \eqref{sqm}.
\\~\\
An alternative way to write the solution of \eqref{sourceeom} is by writing
\begin{equation}
A_{0} = \tilde{g}^{-1}\partial_\tau \tilde{g}, \qquad \tilde{g} =  g U(\tau), \quad U(\tau) = e^{\frac{\kappa}{K} w \lambda w^{-1} \tau}.
\end{equation}
The solution is an untwisted particle-on-a-group action, but where the group element has the monodromy:\footnote{The right-invariance $g \to g G$ allows the removal of the $w$-variable here as well.} $
g(\tau + 2\pi) = g(\tau) U(2\pi)$. Conversely, because of the right-invariance $g\to g G$, the monodromy matrix $U(2\pi)$ is a conjugacy class element and is hence in the maximal torus $T$ of the group $G$. And any such matrix can be parametrized as $U = \prod_{i} e^{ 2\pi \Lambda_i}$, where the product runs over all highest weights $\lambda_i$. By \eqref{dsc2}, one can associate an irreducible representation label $\lambda$ to it that signals the defect operator $\text{Tr}_R e^{i \kappa \chi}$ that is to be used. 
\\~\\
Summarizing, we find that defects within BF with gauge group $G$ correspond to coadjoint orbits of the loop group $\widehat{LG}$.

\section{Expectation values of functions of the area $\hat{A}$}
\label{app: f(A)}
It is interesting to generalize the discussion of section \ref{sect:areaop} to include arbitrary functions of the area operator $\hat{A}$, in particular $\log \hat{A}$ and negative powers are of interest to understand the significance of corrections to the Bekenstein-Hawking entropy in this context. \\
The $n$-th power of the horizon area operator is readily written down as:
\begin{equation}
\left\langle A^n\right\rangle = \left(\sqrt{\frac{4C}{\beta}}\pi\right)^n (-i)^n 2^{-n/2}H_n\left(i\pi \sqrt{\frac{2C}{\beta}}\right),
\end{equation}
in terms of (physics) Hermite polynomials $H_n(x)$. Using the asymptotics $2^{-n/2}H_n(x) \approx (\sqrt{2}x)^n$, we indeed find the large $C$-behavior: $\left\langle \hat{A}^n\right\rangle \approx \left(\frac{4\pi^2C}{\beta}\right)^n$. Also negative values of $n$ can be considered, e.g.:
\begin{equation}
\left\langle A^{-1}\right\rangle = - \sqrt{\frac{\beta}{2C\pi}} \frac{i}{2} e^{-\pi^2 2C/\beta}\text{erf}\left(i\pi \sqrt{\frac{2C}{\beta}}\right),
\end{equation}
found by integrating $\exp(\alpha \Phi)$ from $-\infty$ to $\alpha$. Since $\Phi$ is positive between $k$-eigenstates, after setting $\alpha=0$ in the end, this indeed leads to $1/\Phi$. This procedure is equivalent to integrating the Hermite generating function, leading to ``negative-order'' Hermite functions. At large $C$, one finds $\left\langle A^{-1}\right\rangle  \approx \frac{\beta}{4\pi^2C}$. \\
Finally, one can get to the log of the area operator as well. Using the contour representation of the Hermite polynomial, analytically continuing this and taking the derivative, one obtains:
\begin{equation}
\left\langle \log A\right\rangle = \ln \sqrt{\frac{4C}{\beta}}\pi - \gamma +i \frac{\pi}{2} - \left[\int_{\epsilon}^{+\infty}\frac{dt}{t} e^{-t^2/2+ i t \sqrt{4C/\beta}\pi} + \ln \epsilon\right],
\end{equation}
with $\gamma$ the Euler-Mascheroni constant. One proves that at large $C$, one has the asymptotic expansion:
\begin{equation}
\int_{\epsilon}^{+\infty}\frac{dt}{t} e^{-t^2/2+ i t \sqrt{4C/\beta}\pi} + \ln \epsilon \approx - \ln \sqrt{\frac{4C}{\beta}}\pi - \gamma + i \frac{\pi}{2},
\end{equation}
leading to the correct semi-classics:
\begin{equation}
\left\langle \ln A\right\rangle \approx 2\ln \sqrt{\frac{4C}{\beta}}\pi = \ln \frac{4 \pi^2C}{\beta}.
\end{equation}
All of these operators can be implemented within sectors of the diagram by inverse Laplace transforming from $\beta$ to $k^2$, analogously as we did for $\left\langle A\right\rangle$ itself. We leave the details to the reader.

\section{Degenerate Bilocal Correlators}
\label{app:deg}
Bilocal operators are represented in Liouville theory as a local Liouville primary operator between two boundary states \cite{Mertens:2017mtv}. Here we consider instead a degenerate primary operator between the branes. We focus on the Schwarzian two-point function.
\\~\\
A Liouville primary $e^{2\alpha \frac{\phi}{b}}$ with $2\alpha = Q - \frac{m}{b} -nb$ for $m,n \in \mathbb{N}$ describes Virasoro degenerate primaries. Setting $m > 1$ gives an ill-defined limit $b \to 0$, so the interesting degenerates are found by setting $n = 2\ell$ for a half-integer $\ell$, with Liouville operators:
\begin{equation}
\label{liouop}
V_{2\ell,1} = e^{-2\ell \phi}, \quad \ell = \frac{1}{2}, 1, \frac{3}{2}, \hdots
\end{equation}
We consider the simplest degenerate operator with $\ell = 1/2$. Within the Liouville evaluation of the Schwarzian correlators, we would consider the sphere three-point function:
\begin{equation}
\left\langle V_Q V_{2,1} V_P \right\rangle.
\end{equation}
This can be directly evaluated using the degenerate fusion algebra:
\begin{equation}
\label{fusalg}
V_{2,1} V_{P} = V_{P + i \frac{b}{2}} + V_{P - i \frac{b}{2}}.
\end{equation}
Since we set $P = b k$, the shift in $k$ is finite as $b\to 0$. 

Equivalently, in group theory language, the matrix element of the constrained finite-dimensional $\sltr$ irrep is simply \cite{Blommaert:2018oro}
\begin{equation}
\label{mele}
R(\phi) = e^{- 2\ell \phi}, \quad \ell = \frac{1}{2}, 1, \frac{3}{2}, \hdots
\end{equation}
matching the Liouville insertion \eqref{liouop}. 

\begin{figure}[h]
\centering
\includegraphics[width=0.55\textwidth]{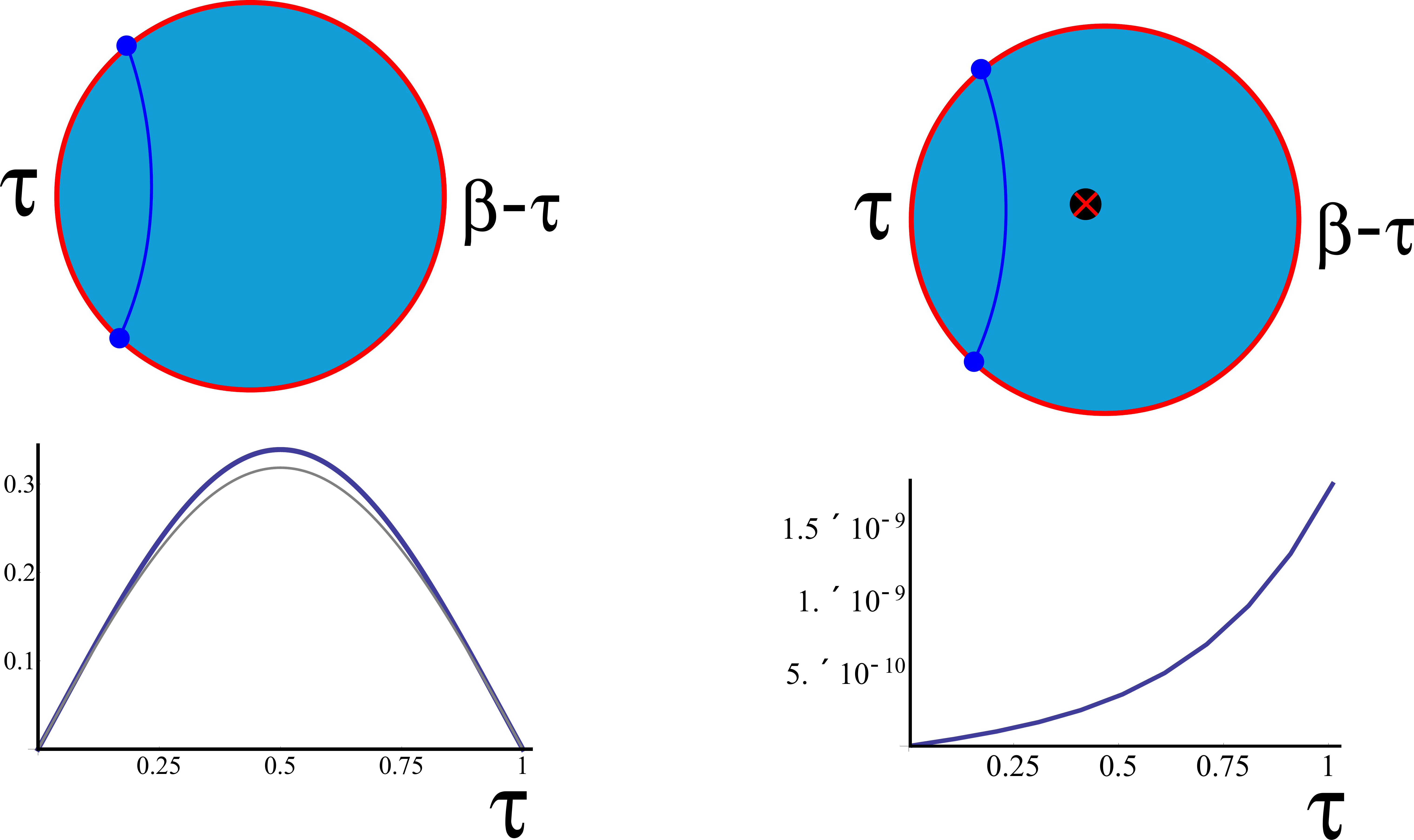}
\caption{Left: Bilocal correlator \eqref{bildegen} (blue) and its semi-classical approximation (gray), with $\beta=1$ and $C=1/2$. Right: Bilocal correlator \eqref{bildegendef} in the presence of a defect, with $\lambda =1$, $\beta=1$ and $C=1/2$.}
\label{finiteRep}
\end{figure}

Using
\begin{equation}
\label{besselprop}
\frac{K_{\alpha}(x)}{x} = \frac{1}{2\alpha}\left( K_{\alpha+1}(x) - K_{\alpha-1}(x)\right),
\end{equation}
which is the 1d analogue of the fusion algebra \eqref{fusalg}, one can evaluate the vertex function as:
\begin{equation}
\int d\phi K_{2i k_1}(e^{\phi}) K_{2i k_2}(e^{\phi}) e^{-\phi} = \frac{1}{32 i k_1 k_2 \sinh(2\pi k_2)}\left(\delta(k_1-k_2 + i/2) - \delta(k_1-k_2 - i/2)\right).
\end{equation}
This can be read as the minisuperspace evaluation of the Liouville amplitude, or as the definition of the 3j-symbol as the integral of three matrix elements. 
The bilocal correlation function can be written as:
\begin{align}
\label{bildegen}
\frac{1}{Z} &\int d\mu(k_1) d\mu(k_2) \frac{e^{-\tau \frac{k_1^2}{2C} - (\beta-\tau) \frac{k_2^2}{2C}}}{i k_1 k_2 \sinh(2\pi k_2)}\left(\delta(k_1-k_2 + i/2) - \delta(k_1-k_2 - i/2)\right) \nonumber \\
&= \frac{2C}{Z}\int_{0}^{+\infty} dk k \sinh(2\pi k) \frac{\sin\left( k \frac{\tau}{2C}\right)}{k} e^{-\frac{\beta}{2C} k^2} e^{\frac{\tau}{8C}} \nonumber \\
& = \left(\frac{\beta}{\pi} \sin\frac{\pi}{\beta}\tau \right) \,\, e^{\frac{\tau}{8C}\left( 1 - \frac{\tau}{\beta}\right)}.
\end{align}
Somewhat surprisingly, an explicit expression is obtained in this case, unlike the usual bilocal correlators as in \eqref{sch2pt}. In the second line we shifted the integration contour and evaluated the delta-functions. In the last line, the Gaussian integrals are performed. The first factor on the rhs is the semi-classical regime. The second factor disappears as $C \to \infty$, and represents the quantum corrections. Its influence is largest halfway the interval at $\tau =\beta/2$. \\
The correlator is zero at both endpoints at $\tau = 0$ and $\tau = \beta$ (Figure \ref{finiteRep}). 
As before, one can readily insert a defect in the bulk. E.g. inserting a hyperbolic defect in the sector exterior to the bilocal leads to the correlator:
\begin{align}
\label{bildegendef}
\frac{2C}{Z}\int_{0}^{+\infty} dk \cos(2\pi \lambda k) \frac{\sin\left( k \frac{\tau}{2C}\right)} {k}e^{-\frac{\beta}{2C} k^2} e^{\frac{\tau}{8C}}.
\end{align}
This set-up can be readily generalized to other finite-dimensional representations $\ell > 1/2$ by consecutive uses of the fusion algebra \eqref{besselprop} and to multiple bilocal insertions.
\\~\\
The bilocal operator in Schwarzian variables is 
\begin{equation}
\label{bil}
\left(\frac{\frac{\beta}{\pi} \sin \frac{\pi}{\beta}(f_1-f_2)}{\sqrt{f'_1 f'_2}}\right)^{2\ell},
\end{equation}
and its semiclassical limit is indeed $\frac{\beta}{\pi} \sin \frac{\pi}{\beta}\tau$ for $\ell=1/2$. From a matter CFT perspective, this operator corresponds to taking non-unitary matter primaries. This is reflected in the fact that the finite-dimensional irreps of $\sltr$ are also non-unitary.
\\~\\
As a degenerate operator in 2d CFT, it satisfies the null-state condition, and it is interesting to check this directly in the 1d Schwarzian context. Writing 
$\mathcal{O}_{-1/2}(\tau_1,\tau_2) = \frac{\frac{\beta}{\pi} \sin \left( \frac{\pi}{\beta}(f_1-f_2) \right)}{\sqrt{f'_1 f'_2}}$,
one readily checks explicitly that
\begin{equation}
\partial_{t_1}^2 \mathcal{O}_{-1/2}(\tau_1,\tau_2) \equiv  - \frac{1}{2} \left\{\tan \frac{\pi}{\beta}f, \tau_1\right\} \mathcal{O}_{-1/2}(\tau_1,\tau_2) = -\frac{1}{2C}T(\tau_1) \mathcal{O}_{-1/2}(\tau_1,\tau_2).
\end{equation}
This is the null state condition of the $V_{2,1}$ primary operator in 2d Virasoro CFT. Applying this formula to \eqref{bildegen}, and dropping a prefactor of $1/4C^2$, for the lhs one finds the insertion for each of the two terms:
\begin{equation}
(k_2^2-(k_2 \mp \frac{i}{2})^2)^2 = -k^2 \pm \frac{ik }{2} + \frac{1}{16},
\end{equation}
whereas the rhs is evaluated by taking the average of the energy $k_1^2/2C$ and $k_2^2/2C$ as:
\begin{equation}
-\frac{k_2^2+(k_2 \mp \frac{i}{2})^2}{2} = -k^2 \pm \frac{ik }{2} + \frac{1}{8}.
\end{equation}
These match save for a discrepancy in the zero-point energy. The latter mismatch can be viewed as a renormalization issue of the composite operator $:T(\tau_1) \mathcal{O}_{-1/2}(\tau_1,\tau_2):$. The same feature also appears in the standard Schwarzian bilocal correlators \cite{Mertens:2017mtv} for the higher terms in the series expansion of $\mathcal{O}_\ell(\tau_1,\tau_2)$ around $\tau_1=\tau_2$ .
\\~\\
Let us prove that \eqref{bil} is indeed the form of the bilocal operator that one finds using group-theoretic techniques \cite{Blommaert:2018oro}. In the finite-dimensional spin-$\ell$ representation, the $\mathfrak{sl}(2,\mathbb{R})$ generators take the form:
\begin{equation}
\scriptsize
J^- = i\left(\begin{array}{cccc}
0 & 0 & \hdots & 0\\
-1 & 0 & \hdots & 0\\
0 & -2 & \hdots  & 0 \\
\vdots & \vdots & \hdots & \vdots \\
0 & \hdots & -2 \ell & 0
\end{array}\right) , \, J^+ = i\left(\begin{array}{ccccc}
0 & -2\ell & 0 & 0\hdots & 0\\
0 & 0 & -2\ell +1 & \hdots & 0\\
\vdots & \vdots & \hdots & \vdots & \vdots \\
0 & 0 & 0 & \hdots & -1 \\
0 & 0 & 0 & 0 & 0
\end{array}\right), 
\, J^0 = i\left(\begin{array}{cccc}
\ell & 0 & \hdots & 0\\
0 & \ell-1 & \hdots & 0\\
0 & 0 & \hdots  & 0 \\
\vdots & \vdots & \hdots & \vdots \\
0 & 0 & \hdots & -\ell
\end{array}\right).
\end{equation}
With this representation of the generators, and the Gauss decomposition of the group element  $g^{-1}$:
\begin{equation}
\label{gauss}
g^{-1} = e^{i \gamma_- J^-} e^{2i \phi J^0} e^{i \gamma_+ J^+}, \qquad \gamma_- = F, \, e^{2\phi} = F', \, \gamma_+ = - \frac{1}{2} \frac{F''}{F'}, \quad \left\{F,\tau\right\} = T(\tau),
\end{equation}
one can check by direct computation that the $(+\ell,-\ell)$ component of the representation matrix of the Wilson line equals:
\begin{equation}
\left[R_\ell(g_2 \, g_1^{-1})\right]_{+\ell, -\ell} = \left(\frac{F_1-F_2}{\sqrt{F'_1 F'_2}}\right)^{2\ell}.
\end{equation}
This matrix element is computed in the $J^0$-basis. The bra and ket of this matrix element are respectively highest and lowest weight states of the representation, and as such are respectively zero-eigenstates of $J^-$ and $J^+$ respectively. Thus we can write for the mixed parabolic matrix element in the spin-$\ell$ representation:
\begin{equation}
\left\langle J^+ = 0\right| R_{\ell}(g_2 \, g_1^{-1}) \left|J^- = 0\right\rangle = \left(\frac{F_1-F_2}{\sqrt{F'_1 F'_2}}\right)^{2\ell}, \quad F = \tan \frac{\pi}{\beta} f.
\end{equation}
Evaluating the lhs directly in the Gauss decomposition \eqref{gauss}, one writes:
\begin{equation}
\left\langle J^+ = 0\right| R_{\ell}(g) \left|J^- = 0\right\rangle \,\, = \,\, \left\langle J^+ = 0\right| e^{-2i \phi J^0} \left|J^- = 0\right\rangle \,\, =\,\,  e^{-2 \ell \phi},
\end{equation}
in terms of the Cartan element $\phi$, in agreement with \eqref{mele}.

\end{document}